\documentclass[prx,twocolumn]{revtex4}
\usepackage{graphicx}
\usepackage{dcolumn}
\usepackage{bm}
\usepackage{multirow}
\usepackage{amsmath,amssymb}
\usepackage{threeparttable}

\begin{document}

\title{Emergence of electronic modes and triplet pairing from spin-1 antiferromagnetic insulators in the Kanamori-Hubbard model}

\author{Masanori Kohno}
\email{KOHNO.Masanori@nims.go.jp}
\affiliation{Research Center for Materials Nanoarchitectonics, National Institute for Materials Science, Tsukuba 305-0003, Japan}

\date{\today}

\begin{abstract}
The electronic band structures of conventional band insulators generally remain unchanged under doping via chemical-potential shifts or increasing temperature. However, in Mott and Kondo insulators described by models such as the Hubbard and Kondo lattice models, distinct electronic modes with momentum-shifted magnetic dispersion relations have been shown to emerge from band edges into the gap. Here, by investigating the two-orbital Kanamori-Hubbard model, this study elucidates how and why the band structures of spin-1 antiferromagnetic insulators change with doping and temperature, revealing emergent spectral features characteristic of orbitally degenerate systems. The nature of the emergent electronic modes is clarified using weak-hopping effective theory, selection-rule analysis, and numerical calculations. In a typical parameter regime, the doping-induced modes in each orbital reflect distinct types of spin modes, while the temperature-induced modes primarily reflect the low-energy spin mode in both orbitals; in both cases, the emergent modes exhibit momentum-shifted spin-mode dispersion relations. The momentum regimes of these emergent modes can differ between orbitals if signs of the hopping parameters differ. For small inter-orbital repulsion, onsite spin-triplet pairs can form after doping, leading to a pair-breaking gap, while the temperature-induced electronic modes continue to reflect the low-energy spin mode without a pair-breaking gap. Moreover, the emergence and characteristics of electronic modes in nonequilibrium states by spin or charge perturbations are elucidated. These findings provide a comprehensive understanding of how the interplay among the spin, charge, and orbital degrees of freedom in large-spin antiferromagnetic insulators leads to emergent electronic modes, opening avenues for exploring unconventional quantum phenomena in strongly correlated systems. 
\end{abstract}

\maketitle

\section{Introduction}
In orbitally degenerate materials, Hund's coupling tends to align the spins of electrons in different orbitals in parallel, leading to Mott insulators that can be effectively regarded as antiferromagnets with spin larger than $\frac{1}{2}$. 
The Mott transition from such a large-spin antiferromagnetic insulator may exhibit distinctive features compared to that from a canonical spin-$\frac{1}{2}$ insulator. 
\par
One of the characteristics of orbitally degenerate systems is the orbital-selective Mott transition (OSMT) \cite{AnisimovOSMT,GeorgesOSMT}. This phenomenon may resemble conventional insulator--metal transitions in multiorbital band insulators; nevertheless, the nature remains elusive. 
In single-orbital or orbital-hybridized systems described by the Hubbard model (HM), Kondo lattice model (KLM), and periodic Anderson model (PAM), recent theoretical studies have shown that electronic modes are induced from the band edges into the band gap by doping or increasing temperature in Mott and Kondo insulators, which exhibit magnetic dispersion relations shifted by the Fermi momenta \cite{Kohno1DHub,Kohno2DHub,KohnoDIS,KohnoRPP,KohnoHubLadder,KohnoKLM,KohnoGW,KohnoAF,KohnoSpin,Kohno1DtJ,Kohno2DtJ,KohnoMottT,KohnoTinduced,KohnoDrivenMott}. These constitute essential features of strongly correlated insulators, which are absent in band insulators. 
However, in the case of orbitally degenerate systems, it is still unclear how such essential features---especially the emergence of electronic modes reflecting spin excitations in an orbital-dependent manner---are realized. In addition, the temperature-induced change in the band structure \cite{KohnoMottT,KohnoTinduced} and the emergence of electronic modes in spin- or charge-perturbed nonequilibrium states \cite{KohnoDrivenMott} remain unclear in orbitally degenerate systems. 
\par
Another topic in orbitally degenerate systems is the possibility of superconductivity associated with Hund's coupling. Recently, pairing in multiorbital systems has attracted particular attention, especially since the discovery of nickelate superconductors \cite{nickelateSuper}. In orbitally degenerate systems, in contrast to single-orbital systems, pairing can also occur between different orbitals and, as a result of Hund's coupling, spin-triplet pairing may emerge. 
\par
Addressing the above issues, this study clarifies the following features: the emergence of electronic modes originating from spin excitations upon doping; their orbital-specific correspondence with distinct types of spin excitations; the emergence of electronic modes reflecting spin excitations at nonzero temperatures; the emergence of electronic modes in nonequilibrium states under spin or charge perturbations; and the formation of onsite spin-triplet pairs mediated by Hund's coupling. 
\par
The underlying mechanisms of these features and the dispersion relations of the emergent modes are theoretically explained by an effective theory for weak inter-site hopping together with a selection-rule analysis. The validity and generality of the results in one and two dimensions are demonstrated using the non-Abelian dynamical density-matrix renormalization group (DDMRG) method and cluster perturbation theory (CPT) for the Kanamori-Hubbard model (KHM). The cases of electron doping and different signs of inter-site hopping parameters for different orbitals are also systematically examined based on symmetry transformations. 
\par
Specifically, this study shows that electronic modes exhibiting momentum-shifted spin-mode dispersion relations emerge within the band gap upon doping large-spin antiferromagnetic insulators, reflecting spin excitations in an orbital-dependent manner. In the doped orbital, a low-energy electronic mode emerges, as in canonical spin-1/2 Mott insulators. These emergent modes constitute an essential feature of the OSMT, in clear contrast to conventional insulator--metal transitions in multiorbital band insulators. A concise overview of this feature is also presented in a separate letter \cite{KohnoKHPRL}. 
\par
Electronic modes are also shown to be induced at nonzero temperatures. These modes emerge in both orbitals, in contrast to the doping-induced modes. The temperature-driven changes in the band structure are also a general characteristic of strongly correlated insulators, regardless of spatial dimensions, spin length, or model details. 
\par
In nonequilibrium states generated by spin or charge perturbations, electronic modes can also be induced. The dispersion relations of these modes can be expressed as combinations of unperturbed electronic and spin-mode dispersion relations. 
This contrasts with noninteracting systems, in which the band structure remains unchanged in any eigenstate. 
\par
Moreover, in the regime of small inter-orbital repulsion, onsite spin-triplet pairs can be formed upon doping. These triplet pairs can coexist with low-energy spin excitations, in contrast to onsite singlet pairs as in the attractive HM. The electronic excitation exhibits a pair-breaking gap, whereas the spin excitation remains (almost) gapless. Superconductivity is expected to occur in dimensions higher than one in specific parameter regimes. 
\par
This study contributes to a comprehensive understanding of how and why electronic modes emerge depending on the orbital, via the interplay among spin, charge, and orbital degrees of freedom in large-spin antiferromagnetic insulators described by the KHM, under doping, nonzero temperature, and spin or charge perturbations. 
\section{Model and methods}
\subsection{Model, spectral function, and dynamical spin structure factors} 
\subsubsection{Kanamori-Hubbard model (KHM)}
In this paper, we consider the two-orbital KHM defined by the following Hamiltonian \cite{Kanamori}: 
\begin{align}
{\cal H}=&-\sum_{\langle i,j\rangle,m,\sigma}t_m\left(c_{i,\sigma}^{m\dagger}c_{j,\sigma}^m+{\rm H.c.}\right)\nonumber\\
&+U\sum_{i,m}\left(n_{i,\uparrow}^m-\frac{1}{2}\right)\left(n_{i,\downarrow}^m-\frac{1}{2}\right)\nonumber\\
&+\left(U^{\prime}-\frac{J_{\rm H}}{2}\right)\sum_i\left(n_i^1-1\right)\left(n_i^2-1\right)\nonumber\\
&+J_{\rm H}\sum_i{\bm \eta}_i^1\cdot{\bm \eta}_i^2-J_{\rm H}\sum_i{\bm S}_i^1\cdot{\bm S}_i^2\nonumber\\
&-\mu\sum_{i,m}n_i^m-H\sum_{i,m}S_i^{m,z},
\label{eq:Ham}
\end{align}
where $\langle i,j\rangle$ means that sites $i$ and $j$ are nearest neighbors on a chain and on a square lattice for the one-dimensional (1D) KHM and two-dimensional (2D) KHM, respectively. 
Here, $c_{i,\sigma}^{m\dagger}$ and $n_{i,\sigma}$ denote the creation and number operators of an electron with spin $\sigma(=\uparrow,\downarrow)$ on the $m$th orbital ($m=1,2$) at site $i$, respectively. The number operator, spin operator, and $\eta$ operator on the $m$th orbital at site $i$ are denoted by $n_i^m$, ${\bm S}_i^m$, and ${\bm \eta}_i^m$, respectively; 
\begin{equation}
\begin{array}{ll}
n_{i,\sigma}^m=c_{i,\sigma}^{m\dagger}c_{i,\sigma}^m,&n_i^m=\sum_{\sigma}n_{i,\sigma}^m,\\
S_i^{m,z}=\frac{1}{2}\left(n_{i,\uparrow}^m-n_{i,\downarrow}^m\right),&\eta_i^{m,z}=\frac{1}{2}\left(n_i^m-1\right),\\
S^{m,x}_i=\frac{1}{2}\left(S^{m,+}_i+S^{m,-}_i\right),&\eta^{m,x}_i=\frac{1}{2}\left(\eta^{m,+}_i+\eta^{m,-}_i\right),\\
S^{m,y}_i=\frac{1}{2{\rm i}}\left(S^{m,+}_i-S^{m,-}_i\right),&\eta^{m,y}_i=\frac{1}{2{\rm i}}\left(\eta^{m,+}_i-\eta^{m,-}_i\right),\\
S_i^{m,+}=c_{i,\uparrow}^{m\dagger}c_{i,\downarrow}^{m},&\eta_i^{m,+}=(-1)^ic_{i,\uparrow}^{m\dagger}c_{i,\downarrow}^{m\dagger},\\
S_i^{m,-}=c_{i,\downarrow}^{m\dagger}c_{i,\uparrow}^{m},&\eta_i^{m,-}=(-1)^ic_{i,\downarrow}^{m}c_{i,\uparrow}^m,
\end{array}
\label{eq:spinEta}
\end{equation}
where $(-1)^i=+1$ and $-1$ for sites $i$ on the A and B sublattices, respectively. 
In Eq. (\ref{eq:Ham}), the atomic orbitals are assumed to be degenerate, and inter-site hopping between different types of atomic orbitals is neglected. We assume orbital rotational invariance, which leads to $U=U^{\prime}+J_{\rm H}$. We also assume $t_1>0$ unless otherwise mentioned. 
\par
The numbers of sites and electrons are denoted by $N_{\rm s}$ and $N_{\rm e}$, respectively. 
At half filling, $N_{\rm e}=2N_{\rm s}$. The number of holes introduced into the half-filled system is denoted by $N_{\rm h}$; 
$N_{\rm h}=2N_{\rm s}-N_{\rm e}$. The hole doping concentration $\delta$ is defined as $\delta=\frac{N_{\rm h}}{N_{\rm s}}$. 
The total spin quantum number and the $z$ component of the spin in units of $\hbar$ are denoted by $S$ and $S^z$, respectively. 
For an electron, $S=\frac{1}{2}$, and its $z$ component, denoted by $s^z$, takes the values $+\frac{1}{2}$ and $-\frac{1}{2}$ for $\sigma=\uparrow$ and $\downarrow$, respectively. 
\subsubsection{Spectral function} 
The spectral function of an electron with momentum ${\bm k}$ for the $m$th orbital from state $|X\rangle$ at zero temperature is defined as 
\begin{align}
\label{eq:Akwspin}
A_X^m({\bm k},\omega)=\frac{1}{2}\sum_{\sigma}A_{X}^{m,\sigma}({\bm k},\omega),\\
\label{eq:AkwGF}
A_{X}^{m,\sigma}({\bm k},\omega)=-\frac{1}{\pi}\Im G_{X}^{m,\sigma}({\bm k},\omega),
\end{align}
using the following Green's function: 
\begin{align}
G_{X}^{m,\sigma}({\bm k},\omega)&=-\frac{\rm i}{Z_X}\int_0^{\infty}dt e^{{\rm i}\omega t-\epsilon t}\langle X|\{c_{{\bm k},\sigma}^m(t),c_{{\bm k},\sigma}^{m\dagger}\}|X\rangle,\nonumber\\
Z_X&=\langle X|X\rangle,\quad c_{{\bm k},\sigma}^m(t)=e^{{\rm i}{\cal H}t}c_{{\bm k},\sigma}^me^{-{\rm i}{\cal H}t}
\end{align}
in the limit of $\epsilon\rightarrow +0$, where $c_{{\bm k},\sigma}^{m\dagger}$ is defined as the Fourier transform of $c_{i,\sigma}^{m\dagger}$, 
\begin{equation}
\label{eq:cdk}
c_{{\bm k},\sigma}^{m\dagger}=\frac{1}{\sqrt{N_{\rm s}}}\sum_i e^{{\rm i}{\bm k}\cdot{\bm r}_i}c_{i,\sigma}^{m\dagger}. 
\end{equation}
For an eigenstate of the Hamiltonian $|X\rangle$ with energy $E_X$, the spectral function can be expressed as 
\begin{equation}
A_{X}^{m,\sigma}({\bm k},\omega)=\frac{1}{Z_X}\sum_n\left[
\begin{array}{r}
|\langle n|c_{{\bm k},\sigma}^{m\dagger}|X\rangle|^2\delta(\omega-E_n+E_X)\\
+|\langle n|c_{{\bm k},\sigma}^m|X\rangle|^2\delta(\omega+E_n-E_X)
\end{array}\right],
\label{eq:Akw}
\end{equation}
where $|n\rangle$ denotes the $n$th eigenstate of the Hamiltonian with energy $E_n$. 
The eigenstates are assumed to be normalized to unity. 
The momentum distribution function for the $m$th orbital is defined as 
\begin{equation}
n_{X}^m({\bm k})=\int_{-\infty}^{0}d\omega A_{X}^m({\bm k},\omega).
\end{equation}
\par
At temperature $T$, the spectral function is defined as 
\begin{align}
\label{eq:AkwspinT}
A_T^m({\bm k},\omega)=\frac{1}{2}\sum_{\sigma}A_{T}^{m,\sigma}({\bm k},\omega),\\
\label{eq:AkwGFT}
A_{T}^{m,\sigma}({\bm k},\omega)=-\frac{1}{\pi}\Im G_{T}^{m,\sigma}({\bm k},\omega),
\end{align}
using the following Green's function: 
\begin{align}
G_{T}^{m,\sigma}({\bm k},\omega)&=-{\rm i}\int_0^{\infty}dte^{{\rm i}\omega t-\epsilon t}
\frac{\sum_l{}_l\langle{\rm T}|\{c_{{\bm k},\sigma}^m(t),c_{{\bm k},\sigma}^{m\dagger}\}|{\rm T}\rangle_l}{\sum_{l}{}_l\langle{\rm T}|{\rm T}\rangle_l},\nonumber\\
|{\rm T}\rangle_l&=e^{-\frac{\beta{\cal H}}{2}}|{\rm I}\rangle_l,
\label{eq:thermalState}
\end{align}
where $\beta$ denotes the inverse temperature, $\beta=\frac{1}{T}$, and the Boltzmann constant $k_{\rm B}$ is set to unity. 
The spectral function can be expressed as 
\begin{align}
A_{T}^{m,\sigma}({\bm k},\omega)=&\frac{1}{Z_T}\sum_{l,n}e^{-\beta E_l}\left[
\begin{array}{r}
|\langle n|c_{{\bm k},\sigma}^{m\dagger}|l\rangle|^2\delta(\omega-E_n+E_l)\\
+|\langle n|c_{{\bm k},\sigma}^m|l\rangle|^2\delta(\omega+E_n-E_l)
\end{array}\right]\nonumber\\
=&\frac{1}{Z_T}\sum_{l,n}\left(e^{-\beta E_n}+e^{-\beta E_l}\right)|\langle n|c_{{\bm k},\sigma}^{m\dagger}|l\rangle|^2\nonumber\\
&\quad\quad\quad\times\delta(\omega-E_n+E_l)
\label{eq:AkwT}
\end{align}
with $|\langle l|c_{{\bm k},\sigma}^{m\dagger}|n\rangle|^2=|\langle n|c_{{\bm k},\sigma}^m|l\rangle|^2$, where $Z_T=\sum_l e^{-\beta E_l}$. 
\par
The electronic density of states is defined as 
\begin{equation}
\rho^m(\omega)=\frac{1}{N_{\rm s}}\sum_{\bm k}A_Y^m({\bm k},\omega),
\end{equation}
where the summation is taken over the Brillouin zone, for $Y=T$ at temperature $T$ [Eq. (\ref{eq:AkwspinT})] and $X$ at $T=0$ [Eq. (\ref{eq:Akwspin})]. 
\subsubsection{Dynamical spin structure factors} 
The dynamical spin structure factors from the ground state at half filling are defined as 
\begin{align}
\label{eq:Skwpm}
S^{\pm}({\bm k},\omega)&=\sum_{n,\alpha}|\langle n|S^{\pm,\alpha}_{\bm k}|{\rm GS}\rangle|^2\delta(\omega-E_n+E_{\rm GS}),\\
\label{eq:Tkwpm}
T^{\pm}({\bm k},\omega)&=\sum_{n,\alpha}|\langle n|T^{\pm,\alpha}_{\bm k}|{\rm GS}\rangle|^2\delta(\omega-E_n+E_{\rm GS})
\end{align}
for $\alpha=x,y,z$, where $|{\rm GS}\rangle$ denotes the ground state at half filling with energy $E_{\rm GS}$, 
and $S^{\pm,\alpha}_{\bm k}$ and $T^{\pm,\alpha}_{\bm k}$ are given by the Fourier transforms of $S_i^{\pm,\alpha}$ and $T_i^{\pm,\alpha}$, respectively. Here, 
\begin{equation}
S_i^{\pm,\alpha}=\frac{1}{\sqrt{2}}\left(S^{1,\alpha}_i\pm S^{2,\alpha}_i\right)
\label{eq:Sipma}
\end{equation}
with the spin operator $S^{m,\alpha}_i$ on the $m$th orbital at site $i$ [Eq. (\ref{eq:spinEta})]. 
The inter-orbital spin operators $T^{\pm,\alpha}_i$ at site $i$ are defined as 
\begin{equation}
\begin{array}{ll}
T^{\pm,x}_i&=\frac{1}{2}\left(T^{\pm,+}_i+T^{\pm,-}_i\right),\\
T^{\pm,y}_i&=\frac{1}{2{\rm i}}\left(T^{\pm,+}_i-T^{\pm,-}_i\right),\\
T^{\pm,+}_i&=\frac{1}{\sqrt{2}}\left(c^{1\dagger}_{i,\uparrow}c^2_{i,\downarrow}\pm c^{2\dagger}_{i,\uparrow}c^1_{i,\downarrow}\right),\\
T^{\pm,-}_i&=\frac{1}{\sqrt{2}}\left(c^{1\dagger}_{i,\downarrow}c^2_{i,\uparrow}\pm c^{2\dagger}_{i,\downarrow}c^1_{i,\uparrow}\right),\\
T^{\pm,z}_i&=\frac{1}{\sqrt{2}}\sum_{\sigma}s^z\left(c^{1\dagger}_{i,\sigma}c^2_{i,\sigma}\pm c^{2\dagger}_{i,\sigma}c^1_{i,\sigma}\right).
\end{array}
\label{eq:Ti}
\end{equation}
These operators transform as SU(2) operators in spin space, since $(-T^{\pm,+}_i,\sqrt{2}T^{\pm,z}_i,T^{\pm,-}_i)$ behave as rank-1 spherical tensor operators with respect to the spin ${\bm S}_i(={\bm S}^1_i+{\bm S}^2_i)$. 
\par
The following spin operators are also defined: 
\begin{equation}
S^{\pm,+}_i=S^{\pm,x}_i+{\rm i}S^{\pm,y}_i,\quad S^{\pm,-}_i=S^{\pm,x}_i-{\rm i}S^{\pm,y}_i,
\label{eq:Sipmpm}
\end{equation}
and the Fourier transforms of $S^{\pm,\gamma}_i$ and $T^{\pm,\gamma}_i$ are denoted by $S^{\pm,\gamma}_{\bm k}$ and $T^{\pm,\gamma}_{\bm k}$, respectively, for $\gamma=+,-,z$. 
\subsection{Methods} 
To calculate the spectral functions and dynamical spin structure factors from the ground state in the 1D KHM, the DDMRG method \cite{DDMRG} in the spin-SU(2) non-Abelian basis \cite{nonAbelianHub} was used on 60-site clusters under open boundary conditions, keeping 240 eigenstates [SU(2)-multiplets] of the density matrix \cite{KohnoDIS,Kohno1DtJ,Kohno2DtJ,KohnoHubLadder,KohnoKLM,KohnoGW,KohnoMottT,KohnoTinduced,KohnoDrivenMott,KohnoRPP}. The operators in momentum space were calculated as 
\begin{equation}
{\cal O}_k=\sqrt{\frac{2}{N_{\rm s}+1}}\sum_j\sin(kr_j){\cal O}_j
\end{equation}
with $k=\frac{\pi l}{N_{\rm s}+1}$ for $l=1,\cdots,N_{\rm s}$, where ${\cal O}_k$ and ${\cal O}_j$ denote the operators in momentum space and real space, respectively \cite{DDMRGAkw}. 
The chemical potential $\mu$ of doped systems with the number of electrons $N_{\rm e}={\bar N_{\rm e}}$ was calculated as $[E_{\rm GS}^{\mu=0}({\bar N_{\rm e}}+2)-E_{\rm GS}^{\mu=0}({\bar N_{\rm e}}-2)]/4$, where $E_{\rm GS}^{\mu=0}({\bar N_{\rm e}}\pm 2)$ denotes the ground-state energy at $N_{\rm e}={\bar N_{\rm e}}\pm 2$ and $\mu=0$. At half filling, $\mu$ was set to zero. 
\par
Other numerical results presented in this paper were obtained using the CPT \cite{CPTPRL,CPTPRB}. The cluster Green's functions were calculated under open boundary conditions on 8-site clusters at $T=0$ and on 6-site clusters at $T>0$ in 1D, and on $3\times 2$-site clusters in 2D (see Appendix \ref{sec:sizeEffects} for a discussion of the finite-size effects). 
These cluster Green's functions were used to calculate the Green's functions in momentum space following the standard CPT procedures \cite{CPTPRL,CPTPRB}. 
The symmetrized spectral function is defined as ${\bar A}_Y^m({\bm k},\omega)=[A_Y^m(k_x,k_y,\omega)+A_Y^m(k_y,k_x,\omega)]/2$. 
The real-space Green's functions within a cluster at intermediate density were calculated as a weighted average of those at the lower and higher densities in clusters with an even number of electrons so that the averaged density matches the targeted value \cite{CPTHanke}. 
The Fermi level ($\omega=0$) of doped systems was set such that the integrated spectral weight below the Fermi level is consistent with the electron density. In the pairing regime, the Fermi level was set to the center of the pair-breaking gap. At half filling, $\mu$ was set to zero. 
\par
In the CPT calculation of the spectral functions from states generated by spin fluctuations, the single-mode approximation for spin excitation was used. Specifically, Eq. (\ref{eq:Akw}) with $|X\rangle=\sum_i e^{{\rm i}{\bm q}\cdot{\bm r}_i}{\cal X}_i|{\rm GS}\rangle$ for ${\cal X}_i=S^{\pm,z}_i$ and $T^{\pm,z}_i$ was used, with $E_X=\langle X|{\cal H}|X\rangle/\langle X|X\rangle$, under the assumption that these states can effectively represent the dominant states in the dynamical spin structure factors \cite{KohnoDrivenMott}. 
\par
In the CPT calculation at nonzero temperature, the low-temperature Lanczos method \cite{LTLM}, which is equivalent to the thermal-pure-quantum-state method with the Lanczos algorithm \cite{TPQS,TPQSLanczos}, was used \cite{KohnoMottT,KohnoTinduced}. Typically 10 thermal states were generated from random vectors $|{\rm I}\rangle_l$ in Eq. (\ref{eq:thermalState}) in each sector specified by the numbers of up spins ($N_{\uparrow}$) and down spins ($N_{\downarrow}$). The orthonormal states obtained via QR decomposition from the random vectors were used as initial block states in the block Lanczos method. Typically 600 eigenstates obtained with the block Lanczos method were used as intermediate eigenstates in each $(N_{\uparrow},N_{\downarrow})$ sector. All $(N_{\uparrow},N_{\downarrow})$ sectors were included. 
\par
In this paper, we consider systems with an even number of sites and assume that the ground state for an even number of electrons is a spin-singlet state in the absence of magnetic fields. In the numerical results presented in this paper, Gaussian broadening with a standard deviation of $0.1t_1$ was used after deconvolution from Lorentzian broadening of a half width at half maximum of $0.1t_1$. 
\par
The spectral function, momentum distribution function, and dynamical spin structure factors of the ground state at hole density $\delta$ are represented as $A_{\delta}^m({\bm k},\omega)$, $n_{\delta}^m({\bm k})$, $S_{\delta}^{\pm}({\bm k},\omega)$, and $T_{\delta}^{\pm}({\bm k},\omega)$; specifically, $A_{\delta=0}^m({\bm k},\omega)=A_{\rm GS}^m({\bm k},\omega)$, $S_{\delta=0}^{\pm}({\bm k},\omega)=S^{\pm}({\bm k},\omega)$, and $T_{\delta=0}^{\pm}({\bm k},\omega)=T^{\pm}({\bm k},\omega)$. Similarly, the spectral function of the ground state at magnetization density $\frac{S^z}{N_{\rm s}}$ for spin $\sigma$ is represented as $A_{\frac{S^z}{N_{\rm s}}}^{m,\sigma}({\bm k},\omega)$. 
\section{Spectral features} 
\label{sec:spectralfeatures}
\subsection{Half-filled systems at zero temperature} 
\label{sec:haffilngT0}
\begin{figure*} 
\centering
\includegraphics[width=\linewidth]{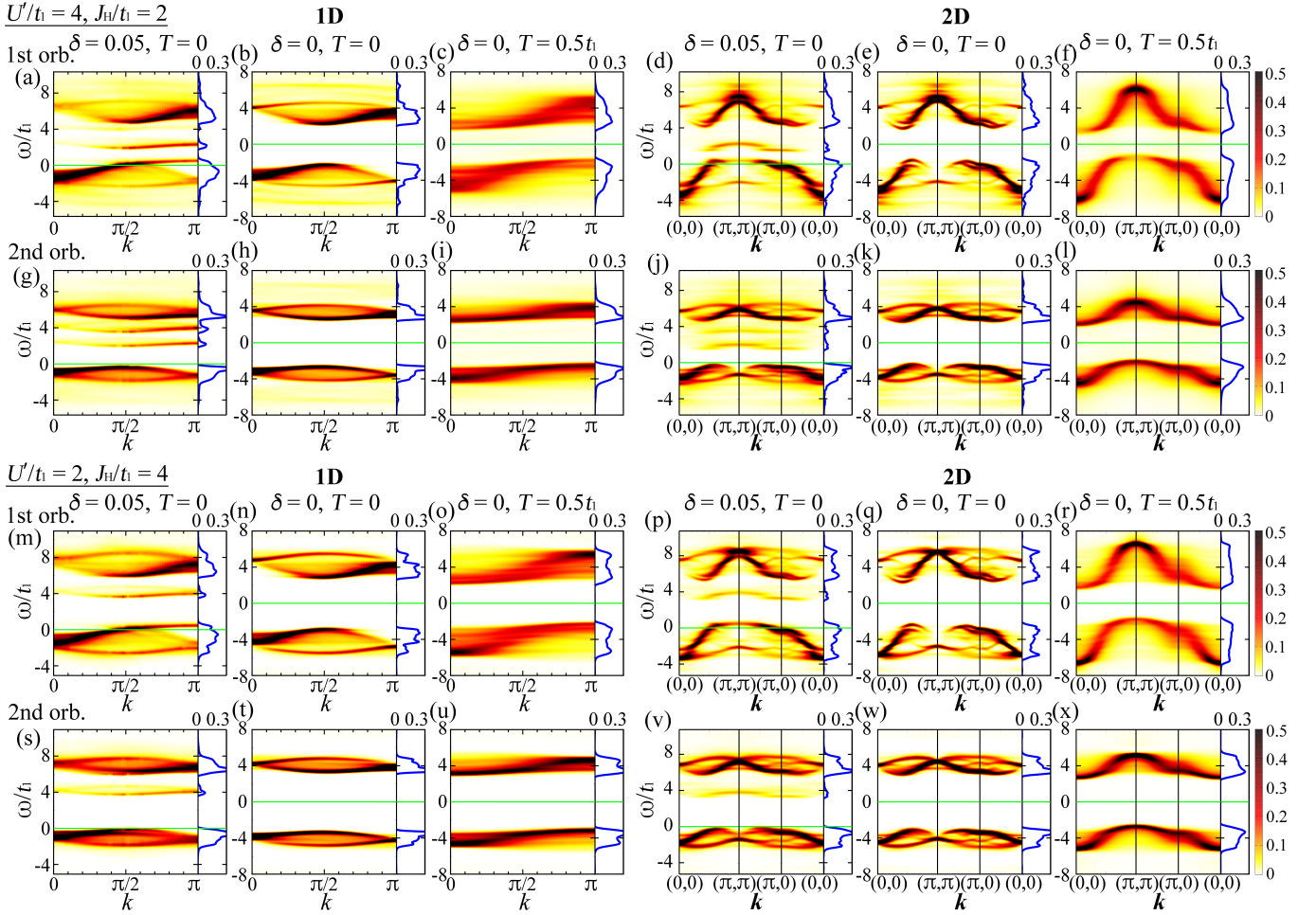}
\caption{Spectral functions for $U^{\prime}/t_1=4$ and $J_{\rm H}/t_1=2$ [(a)--(l)], and $U^{\prime}/t_1=2$ and $J_{\rm H}/t_1=4$ [(m)--(x)], with $U/t_1=6$ and $t_2/t_1=0.5$ for the 1st orbital ($m=1$) [(a)--(f), (m)--(r)] and 2nd orbital ($m=2$) [(g)--(l), (s)--(x)]. 
(a)--(c), (g)--(i), (m)--(o), (s)--(u) $A_{\delta=0.05}^m(k,\omega)t_1$ [(a), (g), (m), (s)], $A_{\delta=0}^m(k,\omega)t_1$ [(b), (h), (n), (t)], $A_{T=0.5t_1}^m(k,\omega)t_1$ [(c), (i), (o), (u)] in the 1D KHM. 
(d)--(f), (j)--(l), (p)--(r), (v)--(x) ${\bar A}_{\delta=0.05}^m({\bm k},\omega)t_1$ [(d), (j), (p), (v)], ${\bar A}_{\delta=0}^m({\bm k},\omega)t_1$ [(e), (k), (q), (w)], ${\bar A}_{T=0.5t_1}^m({\bm k},\omega)t_1$ [(f), (l), (r), (x)] in the 2D KHM. The blue curves on the right-hand side show $\rho^m(\omega)t_1$ calculated from the spectral functions on the left. 
The green lines indicate $\omega=0$.}
\label{fig:Akw}
\end{figure*}
\begin{figure*} 
\centering
\includegraphics[width=\linewidth]{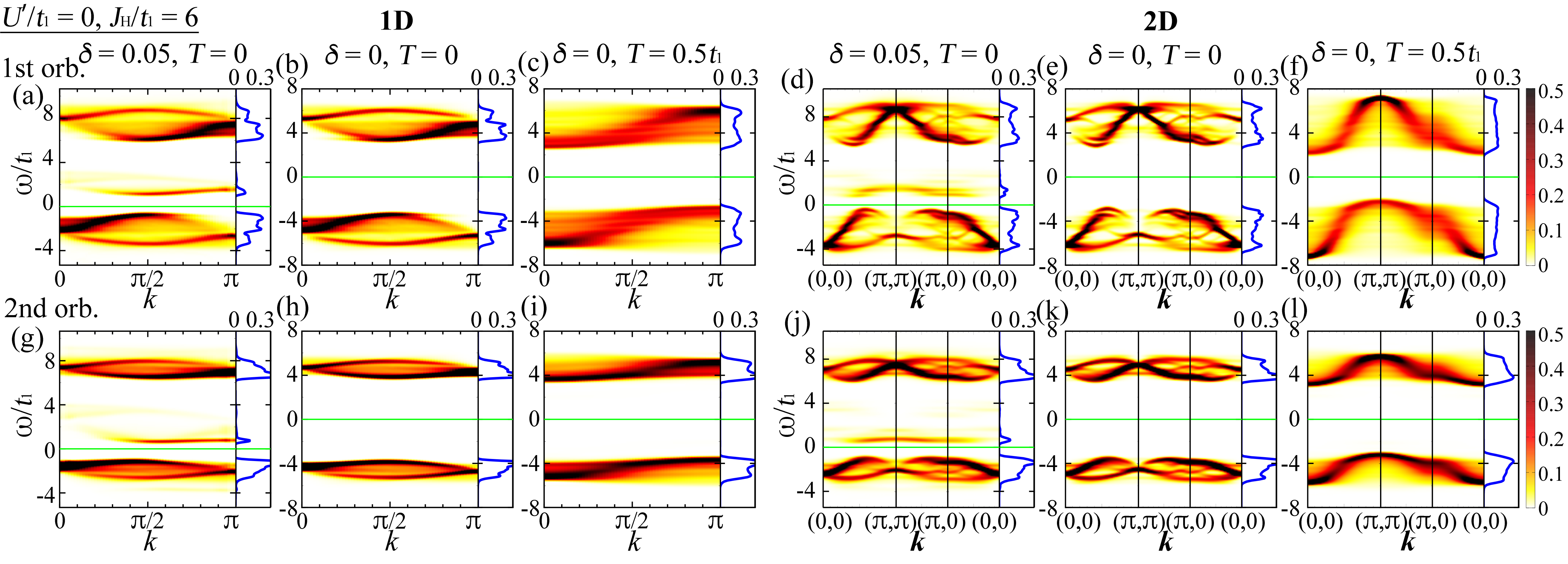}
\caption{Spectral functions for $U^{\prime}/t_1=0$, $J_{\rm H}/t_1=6$, $U/t_1=6$, and $t_2/t_1=0.5$ for the 1st orbital ($m=1$) [(a)--(f)] and 2nd orbital ($m=2$) [(g)--(l)]. 
(a)--(c), (g)--(i) $A_{\delta=0.05}^m(k,\omega)t_1$ [(a), (g)], $A_{\delta=0}^m(k,\omega)t_1$ [(b), (h)], $A_{T=0.5t_1}^m(k,\omega)t_1$ [(c), (i)] in the 1D KHM. 
(d)--(f), (j)--(l) ${\bar A}_{\delta=0.05}^m({\bm k},\omega)t_1$ [(d), (j)], ${\bar A}_{\delta=0}^m({\bm k},\omega)t_1$ [(e), (k)], ${\bar A}_{T=0.5t_1}^m({\bm k},\omega)t_1$ [(f), (l)] in the 2D KHM. The blue curves on the right-hand side show $\rho^m(\omega)t_1$ calculated from the spectral functions on the left. 
The green lines indicate $\omega=0$.}
\label{fig:AkwUp0}
\end{figure*}
At half filling and zero temperature, there are two bands for each orbital: the upper Hubbard band (UHB) for $\omega>0$ and the lower Hubbard band (LHB) for $\omega<0$ [Figs. \ref{fig:Akw}(b), \ref{fig:Akw}(e), \ref{fig:Akw}(h), \ref{fig:Akw}(k), \ref{fig:Akw}(n), \ref{fig:Akw}(q), \ref{fig:Akw}(t), \ref{fig:Akw}(w), \ref{fig:AkwUp0}(b), \ref{fig:AkwUp0}(e), \ref{fig:AkwUp0}(h), and \ref{fig:AkwUp0}(k)]. The Fermi level is located in the band gap, and the system is a Mott insulator. 
\par
The overall spectral features in each orbital are analogous to those of the HM \cite{Kohno1DHub,Kohno2DHub,KohnoRPP,Essler,TakahashiBook}. 
In 1D, within the LHB, the low-$|\omega|$ branch for $|k|<\frac{\pi}{2}$ corresponds to the spinon mode, and the fan-shaped mode corresponds to the holon mode. The UHB is symmetric to the LHB with respect to $k=\frac{\pi}{2}$ and $\omega=0$ [Figs. \ref{fig:Akw}(b), \ref{fig:Akw}(h), \ref{fig:Akw}(n), \ref{fig:Akw}(t), \ref{fig:AkwUp0}(b), and \ref{fig:AkwUp0}(h)]. 
In 2D, within the LHB, the low-$|\omega|$ branch along ${\bm k}=(0,0)$--$(\frac{\pi}{2},\frac{\pi}{2})$ can be interpreted as originating from the 1D spinon mode modified by inter-chain hopping \cite{Kohno2DHub,KohnoRPP}, or as a manifestation of the low-energy spin mode in the one-hole-doped system \cite{KohnoMottT}, and the mode with a bandwidth of approximately $4t_m$ for the $m$th orbital may be regarded as originating from the 1D holon mode \cite{Kohno2DHub,KohnoRPP}. The UHB is symmetric to the LHB with respect to ${\bm k}=(\frac{\pi}{2},\frac{\pi}{2})$ and $\omega=0$ [Figs. \ref{fig:Akw}(e), \ref{fig:Akw}(k), \ref{fig:Akw}(q), \ref{fig:Akw}(w), \ref{fig:AkwUp0}(e), and \ref{fig:AkwUp0}(k)]. 
\subsection{Hole-doped systems at zero temperature} 
In hole-doped systems with large $U^{\prime}/t_1$, an electronic mode is induced in the band gap from the vicinity of the top of the LHB of the orbital whose upper edge is higher in $\omega$ than that of the other orbital, exhibiting essentially gapless (metallic) dispersion relation [Figs. \ref{fig:Akw}(a), \ref{fig:Akw}(d), \ref{fig:Akw}(m), and \ref{fig:Akw}(p)], as found in the HM \cite{Kohno1DHub,Kohno2DHub,KohnoDrivenMott,KohnoRPP,Kohno1DtJ,Kohno2DtJ,KohnoMottT,KohnoGW,KohnoSpin,KohnoAF}. 
\par
Moreover, additional electronic modes are induced in the band gap, which are distinct from the emergent mode near the top of the LHB. 
The emergent electronic modes differ depending on the orbital; the low-$\omega$ emergent mode appears only in the orbital from which an electron is removed (i.e., the orbital whose top of the LHB is higher in $\omega$ than the other) [Figs. \ref{fig:Akw}(a), \ref{fig:Akw}(d), \ref{fig:Akw}(m), and \ref{fig:Akw}(p)], whereas the high-$\omega$ emergent mode just below the UHB primarily appears in the other orbital [Figs. \ref{fig:Akw}(g) and \ref{fig:Akw}(j)]. The intermediate-$\omega$ emergent modes appear in both orbitals [Figs. \ref{fig:Akw}(a), \ref{fig:Akw}(d), \ref{fig:Akw}(g), \ref{fig:Akw}(j), \ref{fig:Akw}(m), \ref{fig:Akw}(p), \ref{fig:Akw}(s), and \ref{fig:Akw}(v)]. 
\par
On the other hand, for small $U^{\prime}/t_1$, an electronic mode emerges within the band gap for each orbital, separated from the LHB of each orbital by a gap [Figs. \ref{fig:AkwUp0}(a), \ref{fig:AkwUp0}(d), \ref{fig:AkwUp0}(g), and \ref{fig:AkwUp0}(j)]. The Fermi level lies within this gap, resulting in gapped single-particle electronic excitations, as in a superconducting state. 
\subsection{Half-filled systems at nonzero temperature}
Regardless of the values of $U^{\prime}$, electronic modes emerge in the band gap from both the bottom of the UHB and the top of the LHB for both orbitals as the temperature increases [Figs. \ref{fig:Akw}(c), \ref{fig:Akw}(f), \ref{fig:Akw}(i), \ref{fig:Akw}(l), \ref{fig:Akw}(o), \ref{fig:Akw}(r), \ref{fig:Akw}(u), \ref{fig:Akw}(x), \ref{fig:AkwUp0}(c), \ref{fig:AkwUp0}(f), \ref{fig:AkwUp0}(i), and \ref{fig:AkwUp0}(l)], as in the case of the HM \cite{KohnoMottT}. The emergence of the electronic modes alters the band structure; specifically, the momentum of the top of the LHB changes from $\frac{\bm \pi}{2}$ to ${\bm \pi}$, and that of the bottom of the UHB from $\frac{\bm \pi}{2}$ to ${\bm 0}$. Unlike the hole-doped case, the emergent electronic modes far away from the band edges are essentially negligible at least up to the temperature shown in Figs. \ref{fig:Akw}(c), \ref{fig:Akw}(f), \ref{fig:Akw}(i), \ref{fig:Akw}(l), \ref{fig:Akw}(o), \ref{fig:Akw}(r), \ref{fig:Akw}(u), \ref{fig:Akw}(x), \ref{fig:AkwUp0}(c), \ref{fig:AkwUp0}(f), \ref{fig:AkwUp0}(i), and \ref{fig:AkwUp0}(l). 
\par
These spectral features obtained in both the hole-doped and nonzero-temperature cases can be collectively and consistently explained, as described below. 
\section{Theoretical interpretation}
\subsection{States in a system without inter-site hopping}
\subsubsection{Ground states at half filling}
\label{sec:GSt0}
To explain the spectral features of the KHM in the strong interaction regime, we first consider the case of $t_1=t_2=0$. Hereafter, $t$ refers collectively to $t_1$ and $t_2$. The system is decoupled into isolated sites. The eigenstates of the Hamiltonian can be simply obtained as direct product states of the local eigenstates of individual sites. 
Eigenstates at a single site are listed in Table \ref{tbl:states}. 
\begin{table*}
\caption{States in the single-site KHM}
\begin{tabular}{cccccc}\hline
Symbols&States&$N_{\rm e}$&$S$&$S^z$&Energies\\\hline
$|{\rm V}\rangle$&$|0,0\rangle$&0&0&0&$E_{\rm V}=\frac{U}{2}+U^{\prime}-\frac{J_{\rm H}}{4}$\\
$|{\rm Z}\rangle$&$|\uparrow\downarrow,\uparrow\downarrow\rangle$&4&0&0&$E_{\rm Z}=\frac{U}{2}+U^{\prime}-\frac{J_{\rm H}}{4}-4\mu$\\
$|{\rm D}^+\rangle$&$\frac{1}{\sqrt{2}}\left(|\uparrow\downarrow,0\rangle+|0,\uparrow\downarrow\rangle\right)$&2&0&0&$E_{{\rm D}^+}=\frac{U}{2}-U^{\prime}+\frac{3J_{\rm H}}{4}-2\mu$\\
$|{\rm D}^-\rangle$&$\frac{1}{\sqrt{2}}\left(|\uparrow\downarrow,0\rangle-|0,\uparrow\downarrow\rangle\right)$&2&0&0&$E_{{\rm D}^-}=\frac{U}{2}-U^{\prime}-\frac{J_{\rm H}}{4}-2\mu$\\
$|{\rm S}\rangle$&$\frac{1}{\sqrt{2}}\left(|\uparrow,\downarrow\rangle-|\downarrow,\uparrow\rangle\right)$&2&0&0&$E_{{\rm S}}=-\frac{U}{2}+\frac{3J_{\rm H}}{4}-2\mu$\\
$|{\rm T}^0\rangle$&$\frac{1}{\sqrt{2}}\left(|\uparrow,\downarrow\rangle+|\downarrow,\uparrow\rangle\right)$&2&1&0&$E_{{\rm T}^0}=-\frac{U}{2}-\frac{J_{\rm H}}{4}-2\mu$\\
$|{\rm T}^{2s^z}\rangle$&$|\sigma,\sigma\rangle$&2&1&$2s^z$&$E_{{\rm T}^{2s^z}}=-\frac{U}{2}-\frac{J_{\rm H}}{4}-2\mu-2s^zH$\\
$|{\rm R}^1_{\sigma}\rangle$&$|\sigma,0\rangle$&1&$\frac{1}{2}$&$s^z$&$E_{{\rm R}_{\sigma}}=-\mu-s^zH$\\
$|{\rm R}^2_{\sigma}\rangle$&$|0,\sigma\rangle$&1&$\frac{1}{2}$&$s^z$&$E_{{\rm R}_{\sigma}}=-\mu-s^zH$\\
$|{\rm P}^1_{\sigma}\rangle$&$|\sigma,\uparrow\downarrow\rangle$&3&$\frac{1}{2}$&$s^z$&$E_{{\rm P}_{\sigma}}=-3\mu-s^zH$\\
$|{\rm P}^2_{\sigma}\rangle$&$|\uparrow\downarrow,\sigma\rangle$&3&$\frac{1}{2}$&$s^z$&$E_{{\rm P}_{\sigma}}=-3\mu-s^zH$\\\hline
\end{tabular}
\par
$|\alpha^1,\alpha^2\rangle$ denotes the state with $|\alpha^m\rangle$ in the $m$th orbital ($m=1,2$).\\ 
$s^z=+\frac{1}{2}$ and $-\frac{1}{2}$ for $\sigma=\uparrow$ and $\downarrow$, respectively.\\
Energies of $|{\rm R}^1_{\sigma}\rangle$ and $|{\rm R}^2_{\sigma}\rangle$ are degenerate, as are those of \\
$|{\rm P}^1_{\sigma}\rangle$ and $|{\rm P}^2_{\sigma}\rangle$, due to atomic orbital degeneracy.
\label{tbl:states}
\end{table*}
\par
The ground states at half filling are described as 
\begin{equation}
|{\rm GS}\rangle_{N_{\rm h}=0}^{t=0}=\prod_i|{\rm T}^{\gamma_i}\rangle_i,
\end{equation}
where $\gamma_i=1,-1$, or $0$ at $H=0$, and $\prod_i$ denotes the direct product over all sites. Because the spin-triplet states at each site are triply degenerate, the ground states have $3^{N_{\rm s}}$-fold degeneracy. The undoped ground-state energy is $E_{\rm T}N_{\rm s}$. Here, $E_{\rm T}=E_{{\rm T}^{\gamma}}$ for $\gamma=1,-1,0$ at $H=0$. 
\subsubsection{Spin-triplet pairing}
\label{sec:tripletPairing}
The two-hole-doped ground state, i.e., the lowest-energy state after removing two electrons, $|{\rm GS}\rangle_{N_{\rm h}=2}^{t=0}$, depends on the parameters. 
The energies of 
\begin{equation}
\begin{array}{ccc}
|{\rm V}\rangle_i\prod_{j(\ne i)}|{\rm T}^{\gamma_j}\rangle_j&\text{and}&
|{\rm R}^m_{\sigma}\rangle_i|{\rm R}^{m^{\prime}}_{\sigma^{\prime}}\rangle_{j(\ne i)}\prod_{l(\ne i,j)}|{\rm T}^{\gamma_l}\rangle_l
\end{array}
\label{eq:2holeStates}
\end{equation}
are 
\begin{equation}
\begin{array}{ll}
E_{\rm V}+E_{\rm T}(N_{\rm s}-1)=U^{\prime}-\frac{J_{\rm H}}{2}-2\mu+E_{\rm T}(N_{\rm s}-2)&\text{and}\nonumber\\
E_{{\rm R}_{\sigma}}+E_{{\rm R}_{\sigma^{\prime}}}+E_{\rm T}(N_{\rm s}-2)=-2\mu+E_{\rm T}(N_{\rm s}-2),&
\end{array}
\end{equation}
respectively. 
Hence, $|{\rm GS}\rangle_{N_{\rm h}=2}^{t=0}$ is the former for $U^{\prime}<\frac{J_{\rm H}}{2}$, 
and the latter for $U^{\prime}>\frac{J_{\rm H}}{2}$. 
This implies that, in doped systems with an even number of electrons for $U^{\prime}<\frac{J_{\rm H}}{2}$, electron pairs are more stable than separated single electrons; electrons on a site move together as a spin-triplet pair to neighboring sites with effective hopping of $\mathcal{O}(t^2)$ upon introducing weak inter-site electronic hopping ($t$), whereas electrons behave as single particles for $U^{\prime}>\frac{J_{\rm H}}{2}$. 
\par
When electrons form spin-triplet pairs between different orbitals on a site ($U^{\prime}<\frac{J_{\rm H}}{2}$), single-particle electronic excitation has an energy gap because breaking a pair costs nonzero energy, although spin excitations remain almost unchanged from the undoped case. 
This feature persists regardless of the lattice geometry or spatial dimensions, since the pair formation is an on-site property like that of the attractive HM. 
In dimensions higher than one, triplet superconductivity is expected to occur in specific parameter regimes. 
\par
Although pairing occurs for $U^{\prime}<\frac{J_{\rm H}}{2}$ at $t=0$, if the values of $\omega$ at the top of the LHBs (bottom of the UHBs) differ between orbitals, electron pairs tend to break up into single electrons in hole-doped (electron-doped) systems with $t\ne 0$. As shown in Figs. \ref{fig:Akw}(m) and \ref{fig:Akw}(p), electrons are removed from the top of the LHB of the orbital with the highest upper edge even though the states in Eq. (\ref{eq:2holeStates}) are degenerate at $t=0$ for $U^{\prime}=\frac{J_{\rm H}}{2}$. At the site where an electron is removed from that orbital, an electron remains in the other orbital, which behaves as a single particle. This indicates that differences in inter-site hopping between orbitals tend to suppress electron pairing, and a more negative value of $U^{\prime}-\frac{J_{\rm H}}{2}$ is required to maintain electron pairing. 
\par
Unfortunately, the parameter regime for the electron pairing is difficult to realize in conventional materials, because 
\begin{align}
U=&\int d{\bm r}\int d{\bm r}^{\prime}|\phi^m({\bm r})|^2v(|{\bm r}-{\bm r}^{\prime}|)|\phi^m({\bm r}^{\prime})|^2,\nonumber\\
U^{\prime}=&\int d{\bm r}\int d{\bm r}^{\prime}|\phi^1({\bm r})|^2v(|{\bm r}-{\bm r}^{\prime}|)|\phi^2({\bm r}^{\prime})|^2,\nonumber\\
\frac{J_{\rm H}}{2}=&\int d{\bm r}\int d{\bm r}^{\prime}\phi^{1*}({\bm r})\phi^2({\bm r})v(|{\bm r}-{\bm r}^{\prime}|)\phi^{2*}({\bm r}^{\prime})\phi^1({\bm r}^{\prime})
\end{align}
in the derivation of the parameters of the KHM from atomic orbitals \cite{Kanamori}, where $\phi^m({\bm r})$ denotes the wavefunction of the $m$th orbital ($m=1,2$) on a site, and $v(|{\bm r}-{\bm r}^{\prime}|)$ denotes the Coulomb potential. The integrands of $U$ and $U^{\prime}$ are always nonnegative, whereas that of $J_{\rm H}$ can take complex or negative values, leading to possible cancellations upon integration. 
\par
To realize the spin-triplet superconductivity, additional effects that strongly suppress $U^{\prime}$ or strongly enhance $J_{\rm H}$ are required in materials. Alternatively, this type of superconductivity may be realized in artificial systems, such as cold atoms, if they are engineered so that $U^{\prime}\ll\frac{J_{\rm H}}{2}$ and $|t_1|\approx |t_2|$. 
\subsubsection{Excited states from the undoped ground state}
\label{sec:excUndopedt0}
\begin{figure*} 
\centering
\includegraphics[width=\linewidth]{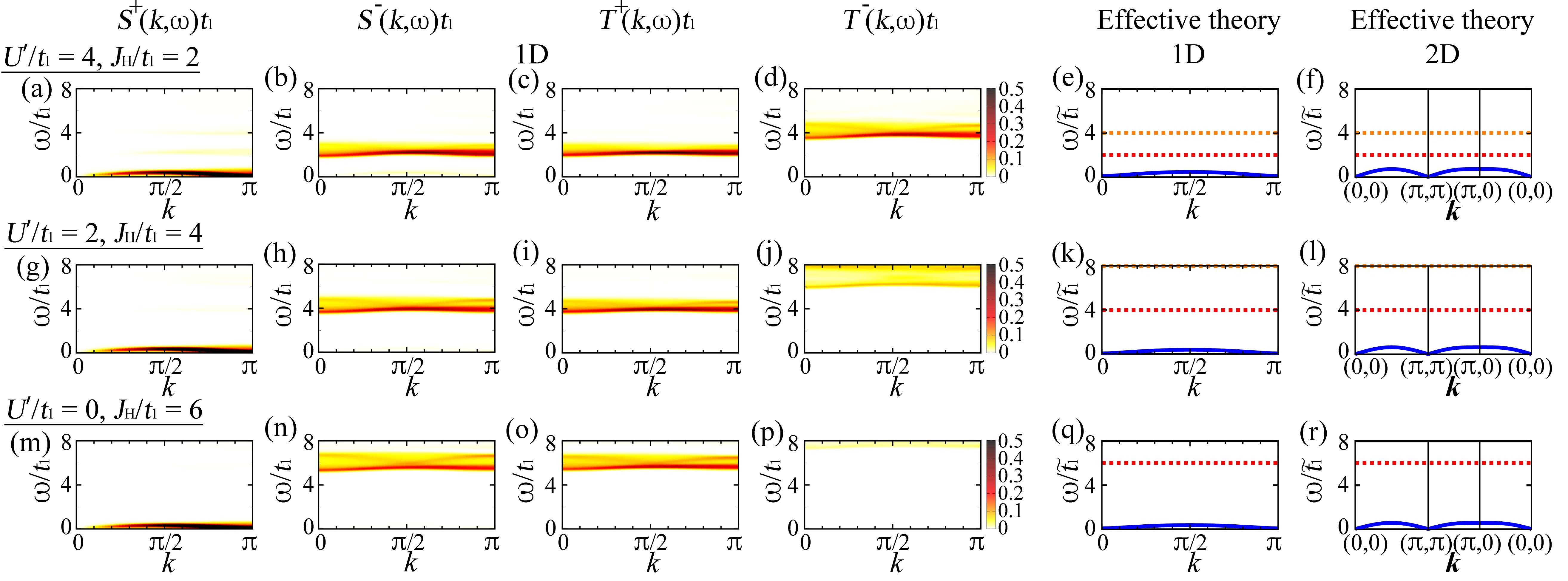}
\caption{Spin excitations at half filling for $U^{\prime}/t_1=4$ and $J_{\rm H}/t_1=2$ [(a)--(f)]; $U^{\prime}/t_1=2$ and $J_{\rm H}/t_1=4$ [(g)--(l)]; and $U^{\prime}/t_1=0$ and $J_{\rm H}/t_1=6$ [(m)--(r)], with $U/t_1=6$. 
(a), (g), (m) $S^+(k,\omega)t_1$; (b), (h), (n) $S^-(k,\omega)t_1$; (c), (i), (o) $T^+(k,\omega)t_1$; and (d), (j), (p) $T^-(k,\omega)t_1$, in the 1D KHM for $t_2/t_1=0.5$. 
(e), (f), (k), (l), (q), (r) The dispersion relations of the low-energy spin mode in the effective theory [$\omega=e_{\bm k}^{S^+,d}$; Eq. (\ref{eq:dispSb})] (solid blue curves) in the 1D KHM [(e), (k), (q)] and 2D KHM [(f), (l), (r)] for $t_2/t_1=0.5$ with ${\tilde t}_1=t_1$. The dispersion relations $\omega=\Delta E_{\rm T}^{{\rm D}^+}$ (dashed orange lines), $\omega=\Delta E_{\rm T}^{{\rm S}}$ and $\Delta E_{\rm T}^{{\rm D}^-}$ (dashed red lines) at $t_1=t_2=0$ are shown for $U^{\prime}/{\tilde t}_1=4$ and $J_{\rm H}/{\tilde t}_1=2$ [(e), (f)]; $U^{\prime}/{\tilde t}_1=2$ and $J_{\rm H}/{\tilde t}_1=4$ [(k), (l)]; and $U^{\prime}/{\tilde t}_1=0$ and $J_{\rm H}/{\tilde t}_1=6$ [(q), (r)] with ${\tilde t}_1>0$.}
\label{fig:Skw}
\end{figure*}
The excited state obtained by applying an operator ${\cal X}_{\bm k}$ to the undoped ground state $|{\rm GS}\rangle_{N_{\rm h}=0}^{t=0}$ is given by 
\begin{align}
{\cal X}_{\bm k}|{\rm GS}\rangle_{N_{\rm h}=0}^{t=0}
&=\frac{1}{\sqrt{N_{\rm s}}}\sum_ie^{{\rm i}{\bm k}\cdot{\bm r}_i}{\cal X}_i|{\rm T}^{\gamma_i}\rangle_i\prod_{j(\ne i)}|{\rm T}^{\gamma_j}\rangle_j\nonumber\\
&=\frac{1}{\sqrt{N_{\rm s}}}\sum_ie^{{\rm i}{\bm k}\cdot{\bm r}_i}\sum_n\alpha_{{\rm T}^{{\gamma}_i}}^{{\rm X}^n}|{\rm X}^n\rangle_i\prod_{j(\ne i)}|{\rm T}^{\gamma_j}\rangle_j,
\label{eq:exStatest0}
\end{align}
where $\alpha_{\rm Y}^{{\rm X}^n}$ is a c-number defined as 
\begin{equation}
{\cal X}_i|{\rm Y}\rangle_i=\sum_n\alpha_{\rm Y}^{{\rm X}^n}|{\rm X}^n\rangle_i.
\end{equation}
For ${\cal X}_{\bm k}=c_{{\bm k},\sigma}^m$, $e^{{\rm i}{\bm k}\cdot{\bm r}_i}$ is replaced with $e^{-{\rm i}{\bm k}\cdot{\bm r}_i}$ in Eq. (\ref{eq:exStatest0}). 
\par
For example, spin excited states generated by $S^{-,\gamma}_{\bm k}$ for $\gamma=+,-,z$ are given by 
\begin{equation}
S^{-,\gamma}_{\bm k}|{\rm GS}\rangle_{N_{\rm h}=0}^{t=0}=\frac{1}{\sqrt{N_{\rm s}}}\sum_ie^{{\rm i}{\bm k}\cdot{\bm r}_i}\alpha_{{\rm T}^{{\gamma}_i}}^{\rm S}|{\rm S}\rangle_i\prod_{j(\ne i)}|{\rm T}^{\gamma_j}\rangle_j.
\label{eq:Snt0}
\end{equation}
The energy of this state at $t_1=t_2=0$ is $E_{\rm S}+E_{\rm T}(N_{\rm s}-1)$, unless $\alpha_{{\rm T}^{{\gamma}_i}}^{\rm S}=0$ for all $i$. 
The excitation energy from the ground state is $\Delta E_{\rm T}^{\rm S}=E_{\rm S}-E_{\rm T}$.
Thus, spectral weights appear in $S^-({\bm k},\omega)$ at $\omega=\Delta E_{\rm T}^{\rm S}$ at all momenta; the dispersion relation is flat and independent of ${\bm k}$ [dashed red curves in Figs. \ref{fig:Skw}(e), \ref{fig:Skw}(f), \ref{fig:Skw}(k), \ref{fig:Skw}(l), \ref{fig:Skw}(q), and \ref{fig:Skw}(r)]. 
\par
Similarly, excited states generated by $T^{\pm,\gamma}_{\bm k}$ for $\gamma=+,-,z$ are given by 
\begin{equation}
T_{\bm k}^{\pm,\gamma}|{\rm GS}\rangle_{N_{\rm h}=0}^{t=0}=\frac{1}{\sqrt{N_{\rm s}}}\sum_ie^{{\rm i}{\bm k}\cdot{\bm r}_i}\alpha_{{\rm T}^{{\gamma}_i}}^{{\rm D}^{\mp}}|{\rm D}^{\mp}\rangle_i\prod_{j(\ne i)}|{\rm T}^{\gamma_j}\rangle_j.
\label{eq:Tpmt0}
\end{equation}
The energy of this state at $t_1=t_2=0$ is $E_{{\rm D}^{\mp}}+E_{\rm T}(N_{\rm s}-1)$, unless $\alpha_{{\rm T}^{{\gamma}_i}}^{{\rm D}^{\mp}}=0$ for all $i$. 
The excitation energy from the ground state is $\Delta E_{\rm T}^{{\rm D}^{\mp}}=E_{{\rm D}^{\mp}}-E_{\rm T}$.
Thus, spectral weights appear in $T^{\pm}({\bm k},\omega)$ at $\omega=\Delta E_{\rm T}^{{\rm D}^{\mp}}$ at all momenta; the dispersion relation is flat and independent of ${\bm k}$ [dashed orange and red curves in Figs. \ref{fig:Skw}(e), \ref{fig:Skw}(f), \ref{fig:Skw}(k), \ref{fig:Skw}(l), \ref{fig:Skw}(q), and \ref{fig:Skw}(r)]. Here, $\Delta E_{\rm T}^{{\rm D}^-}=\Delta E_{\rm T}^{\rm S}$ under the orbital rotational symmetry condition $U=U^{\prime}+J_{\rm H}$ (Table \ref{tbl:states}). 
\subsection{Excitations from the ground state in the weak inter-site hopping regime} 
\subsubsection{Excitations from the undoped ground state} 
When inter-site hopping is introduced ($t_1, t_2\ne 0$), the degeneracy of the $t=0$ ground states at half filling is lifted by the second-order perturbation with respect to $t$ in the small-$|t|$ regime ($|t|\ll U,J_{\rm H}$). The lowest-energy state is selected as the ground state, which is essentially the same as the ground state of the spin-1 Heisenberg model because the effective model is described by 
\begin{equation}
{\cal H}_{\rm eff}=J_{\rm eff}\sum_{\langle i,j\rangle}{\bm S}_i\cdot{\bm S}_j,
\end{equation}
where ${\bm S}_i$ denotes the spin-1 operator at site $i$, and $J_{\rm eff}=\frac{t_1^2+t_2^2}{U+J_{\rm H}/2}$. 
The dispersion relations become dispersive; the bandwidth of bosonic excitations is of ${\cal O}(t^2)$. 
\par
It is known that the spin-1 Heisenberg model in one dimension has a spin gap \cite{HaldaneGap} and those in higher dimensions exhibit a gapless spin-wave mode (magnon) \cite{AndersonSW}. 
\par
Thus, the dispersion relations of the low-energy spin excitations can be effectively expressed as 
\begin{equation}
\begin{array}{lll}
e^{S^+,{d=1}}_{\bm k}=J_{\rm eff}\sqrt{\Delta^2+v_{\rm s}^2(1-\gamma_{\bm k}^2)}&\text{for}&d=1,\\
e^{S^+,{d>1}}_{\bm k}=2dJ_{\rm eff}\sqrt{1-\gamma_{\bm k}^2}&\text{for}&d>1
\label{eq:dispSb}
\end{array}
\end{equation}
with spin exchange coupling $J_{\rm eff}$, where
\begin{equation}
\gamma_{\bm k}=\frac{1}{d}\sum_{i=1}^d \cos k_i
\label{eq:gammak}
\end{equation}
for $k_1=k_x$, $k_2=k_y$, and $k_3=k_z$ in $d$ dimensions [solid blue curves in Figs. \ref{fig:Skw}(e), \ref{fig:Skw}(f), \ref{fig:Skw}(k), \ref{fig:Skw}(l), \ref{fig:Skw}(q), and \ref{fig:Skw}(r)]. 
In Figs. \ref{fig:Skw}(e), \ref{fig:Skw}(k), and \ref{fig:Skw}(q), $\Delta\approx0.4105$ \cite{HaldaneGapDMRG} and $v_{\rm s}\approx2.49$ \cite{velocity1DS1Heis} were used. 
The effective dispersion relation [Eq. (\ref{eq:dispSb})] is consistent with the numerical results [Figs. \ref{fig:Skw}(a), \ref{fig:Skw}(g), and \ref{fig:Skw}(m)]. 
\par
The dispersion relations of the modes of $S^-({\bm k},\omega)$ and $T^{\pm}({\bm k},\omega)$ also become dispersive as shown in Figs. \ref{fig:Skw}(b)--\ref{fig:Skw}(d), \ref{fig:Skw}(h), \ref{fig:Skw}(i), \ref{fig:Skw}(n), and \ref{fig:Skw}(o). 
Note that the high-energy mode is absorbed into the continuum for $U^{\prime}/t_1=2$ and $J_{\rm H}/t_1=4$ [Fig. \ref{fig:Skw}(j)]; it lies above the plotted $\omega$ range for $U^{\prime}/t_1=0$ and $J_{\rm H}/t_1=6$ [Fig. \ref{fig:Skw}(p)]. 
\par
The ground state at half filling can be effectively expressed as a linear combination of the $t=0$ ground states as 
\begin{equation}
|{\rm GS}\rangle_{N_{\rm h}=0}=\sum_l\psi({\cal C}^l)\prod_i|{\rm T}^{\gamma_i^l}\rangle_i,
\end{equation}
where ${\cal C}^l$ denotes the $l$th configuration of the spin-triplet states over all sites, e.g., ${\cal C}^l
=({\rm T}^{1},{\rm T}^{0},\cdots, {\rm T}^{-1})$. 
Hereafter, we assume that the ground state is a spin-singlet state with zero momentum. 
\par
The excited state generated by an operator ${\cal X}_{\bm k}$ is obtained as
\begin{align}
{\cal X}_{\bm k}|{\rm GS}\rangle_{N_{\rm h}=0}
&=\frac{1}{\sqrt{N_{\rm s}}}\sum_{i,l}e^{{\rm i}{\bm k}\cdot{\bm r}_i}\psi({\cal C}^l){\cal X}_i|{\rm T}^{\gamma_i^l}\rangle_i\prod_{j(\ne i)}|{\rm T}^{\gamma_j^l}\rangle_j\nonumber\\
&=\frac{1}{\sqrt{N_{\rm s}}}\sum_{i,l}e^{{\rm i}{\bm k}\cdot{\bm r}_i}\psi({\cal C}^l)\sum_n\alpha_{{\rm T}^{{\gamma}_i^l}}^{{\rm X}^n}|{\rm X}^n\rangle_i\prod_{j(\ne i)}|{\rm T}^{\gamma_j^l}\rangle_j,
\label{eq:exStates}
\end{align}
as in the case of $t=0$ [Eq. (\ref{eq:exStatest0})]. For ${\cal X}_{\bm k}=c_{{\bm k},\sigma}^m$, $e^{{\rm i}{\bm k}\cdot{\bm r}_i}$ is replaced with $e^{-{\rm i}{\bm k}\cdot{\bm r}_i}$ in Eq. (\ref{eq:exStates}). 
\par
The spectral functions and dynamical spin structure factors are obtained by calculating the matrix elements $\langle i|{\cal X}_{\bm k}|{\rm GS}\rangle_{N_{\rm h}=0}$ with eigenstates $|i\rangle$ [Eqs. (\ref{eq:Akw}), (\ref{eq:Skwpm}), and (\ref{eq:Tkwpm})]. Because the excited state can be expressed as a linear combination of the eigenstates of the Hamiltonian with the same quantum numbers, the spectral weights generally spread over a wide $\omega$ range in strongly interacting systems. However, the characteristic spectral feature can be explained in terms of the dominant mode. The eigenstate of the dominant mode is denoted by $|{\cal X}\rangle_{\bm k}$, and its excitation energy from the ground state is denoted by $e^{\cal X}_{\bm k}$: 
\begin{align}
&|{\cal X}\rangle_{\bm k}\overset{\text{dominant}}{\longleftarrow}{\cal X}_{\bm k}|{\rm GS}\rangle_{N_{\rm h}=0},\nonumber\\
&({\cal H}-E_{{\rm GS}_{N_{\rm h}=0}})|{\cal X}\rangle_{\bm k}=e^{\cal X}_{\bm k}|{\cal X}\rangle_{\bm k},
\label{eq:dominantState}
\end{align}
where the ground-state energy of $N_{\rm h}=n$ is denoted by $E_{{\rm GS}_{N_{\rm h}=n}}$. 
\par
Specifically, the dispersion relations of the spin modes can be expressed as follows: 
$\omega=e^{S^+}_{\bm k}$ for the low-energy spin mode in Figs. \ref{fig:Skw}(a), \ref{fig:Skw}(g), and \ref{fig:Skw}(m); 
$\omega=e^{S^-}_{\bm k}$ for the intermediate-energy mode in Figs. \ref{fig:Skw}(b), \ref{fig:Skw}(h), and \ref{fig:Skw}(n); $\omega=e^{T^+}_{\bm k}$ for the intermediate-energy mode in Figs. \ref{fig:Skw}(c), \ref{fig:Skw}(i), and \ref{fig:Skw}(o); and 
$\omega=e^{T^-}_{\bm k}$ for the high-energy mode in Fig. \ref{fig:Skw}(d). 
Here, the spin indices are omitted due to the SU(2) symmetry. 
Similarly, notations without spin indices are used below unless otherwise required for clarity. 
\subsubsection{Electronic excited states from the one-hole-doped ground state} 
\label{sec:holedoping}
We consider the low-hole-doping limit in the parameter regime where electrons behave as single particles without pairing (Sec. \ref{sec:tripletPairing}). 
The one-hole-doped ground state, defined as the lowest-energy eigenstate with the same quantum numbers as the state obtained by removing an electron with momentum ${\bm k}_{\rm F}^-$ from the first orbital in the undoped ground state, can be effectively expressed as 
\begin{equation}
|{\rm GS}\rangle_{N_{\rm h}=1}=\frac{1}{\sqrt{N_{\rm s}}}\sum_{i,l}e^{-{\rm i}{\bm k}_{\rm F}^-\cdot{\bm r}_i}{\tilde \psi}({\cal C}^l)|{\rm R}^2\rangle_i\prod_{j(\ne i)}|{\rm T}^{\gamma_j^l}\rangle_j,
\label{eq:GS1h}
\end{equation}
where the top of the LHB for the first orbital is assumed to be higher in $\omega$ than that of the second orbital, and ${\bm k}_{\rm F}^-$ denotes the Fermi momentum in the low-hole-doping limit, i.e., the momentum at the top of the LHB for the first orbital. The weight of the configuration ${\cal C}^l$ is denoted by ${\tilde \psi}({\cal C}^l)$; ${\cal C}^l$ represents the $l$th configuration over all sites, with the state at site $i$ being ${\rm R}^2$. 
Although $|{\rm GS}\rangle_{N_{\rm h}=1}$ is not exactly the same as $c_{{\bm k}_{\rm F}^-}^1|{\rm GS}\rangle_{N_{\rm h}=0}$, their overlap is significant. In fact, the spectral weight at the top of the LHB for the first orbital ($|c^1\rangle_{{\bm k}_{\rm F}^-}$ [Eq. (\ref{eq:dominantState})]) [Figs. \ref{fig:Akw}(b), \ref{fig:Akw}(e), \ref{fig:Akw}(n), and \ref{fig:Akw}(q)] at half filling, which corresponds to the one-hole-doped ground state ($|{\rm GS}\rangle_{N_{\rm h}=1}$), is dominant. 
\par
By adding or removing an electron with momentum ${\bm k}$ on the $m$th orbital, the electronic states are obtained as 
\begin{align}
&c_{\bm k}^{m\dagger}|{\rm GS}\rangle_{N_{\rm h}=1}\nonumber\\
&=\frac{1}{N_{\rm s}}\sum_{i,j(\ne i),l}e^{{\rm i}({\bm k}\cdot{\bm r}_j-{\bm k}_{\rm F}^-\cdot{\bm r}_i)}\nonumber\\
&\quad\quad\quad\quad\quad\times{\tilde \psi}({\cal C}^l)c_j^{m\dagger}|{\rm T}^{\gamma_j^l}\rangle_j|{\rm R}^2\rangle_i\prod_{n(\ne i,j)}|{\rm T}^{\gamma_n^l}\rangle_n\nonumber\\
\label{eq:addition}
&+\frac{1}{N_{\rm s}}\sum_{i,l}e^{{\rm i}({\bm k}-{\bm k}_{\rm F}^-)\cdot{\bm r}_i}{\tilde \psi}({\cal C}^l)c_i^{m\dagger}|{\rm R}^2\rangle_i\prod_{j(\ne i)}|{\rm T}^{\gamma_j^l}\rangle_j,\\
&c_{\bm k}^m|{\rm GS}\rangle_{N_{\rm h}=1}\nonumber\\
&=\frac{1}{N_{\rm s}}\sum_{i,j(\ne i),l}e^{-{\rm i}({\bm k}\cdot{\bm r}_j+{\bm k}_{\rm F}^-\cdot{\bm r}_i)}\nonumber\\
&\quad\quad\quad\quad\quad\times{\tilde \psi}({\cal C}^l)c_j^m|{\rm T}^{\gamma_j^l}\rangle_j|{\rm R}^2\rangle_i\prod_{n(\ne i,j)}|{\rm T}^{\gamma_n^l}\rangle_n\nonumber\\
\label{eq:removal}
&+\frac{1}{N_{\rm s}}\sum_{i,l}e^{-{\rm i}({\bm k}+{\bm k}_{\rm F}^-)\cdot{\bm r}_i}{\tilde \psi}({\cal C}^l)c_i^m|{\rm R}^2\rangle_i\prod_{j(\ne i)}|{\rm T}^{\gamma_j^l}\rangle_j.
\end{align}
The first terms in Eqs. (\ref{eq:addition}) and (\ref{eq:removal}) exhibit essentially the same spectral features as those of the undoped system, i.e., the UHB and LHB. The excitation energies are primarily determined by $c_j^{m(\dagger)}|{\rm T}^{\gamma_j^l}\rangle_j$ as in the case of the undoped systems. 
\par
For $m=1$, the second term of Eq. (\ref{eq:addition}) is rewritten as 
\begin{align}
&\frac{1}{N_{\rm s}}\sum_{i,l}e^{{\rm i}({\bm k}-{\bm k}_{\rm F}^-)\cdot{\bm r}_i}{\tilde \psi}({\cal C}^l)\alpha_{{\rm R}^2}^{\rm T}|{\rm T}\rangle_i\prod_{j(\ne i)}|{\rm T}^{\gamma_j^l}\rangle_j\nonumber\\
+&\frac{1}{N_{\rm s}}\sum_{i,l}e^{{\rm i}({\bm k}-{\bm k}_{\rm F}^-)\cdot{\bm r}_i}{\tilde \psi}({\cal C}^l)\alpha_{{\rm R}^2}^{\rm S}|{\rm S}\rangle_i\prod_{j(\ne i)}|{\rm T}^{\gamma_j^l}\rangle_j.
\label{eq:DISTS}
\end{align}
\par
The first term of Eq. (\ref{eq:DISTS}) can overlap with $|{\rm GS}\rangle_{N_{\rm h}=0}$ (when ${\bm k}={\bm k}_{\rm F}^-$), which contributes to the top of the LHB for the first orbital. This term can also overlap with $S^+_{{\bm k}-{\bm k}_{\rm F}^-}|{\rm GS}\rangle_{N_{\rm h}=0}$, particularly with the dominant eigenstate $|S^+\rangle_{{\bm k}-{\bm k}_{\rm F}^-}$ [Eq. (\ref{eq:dominantState})], which contributes to the electronic mode originating from the spin mode in $S^+({\bm k},\omega)$ in the low-$\omega$ regime. 
According to Eq. (\ref{eq:Akw}), the dispersion relation of the electronic mode is expressed as 
\begin{equation}
\omega=e^{S^+}_{{\bm k}-{\bm k}_{\rm F}^-}.
\label{eq:wSp}
\end{equation}
Here, the chemical potential $\mu$ is adjusted so that the Fermi level is located at the top of the LHB for the first orbital, i.e., $E_{{\rm GS}_{N_{\rm h}=1}}=E_{{\rm GS}_{N_{\rm h}=0}}$. 
\par
Because nonzero components of $S_i^{-}|{\rm T}\rangle_i$ are proportional to $|{\rm S}\rangle_i$ [Eqs. (\ref{eq:Sipma}) and (\ref{eq:Sipmpm}); Table \ref{tbl:states}], the second term of Eq. (\ref{eq:DISTS}) can overlap with $S^-_{{\bm k}-{\bm k}_{\rm F}^-}|{\rm GS}\rangle_{N_{\rm h}=0}$, particularly with the dominant eigenstate $|S^-\rangle_{{\bm k}-{\bm k}_{\rm F}^-}$ [Eq. (\ref{eq:dominantState})], which contributes to the electronic mode originating from the spin mode in $S^-({\bm k},\omega)$ in the intermediate-$\omega$ regime. 
According to Eq. (\ref{eq:Akw}), the dispersion relation of the electronic mode is expressed as 
\begin{equation}
\omega=e^{S^-}_{{\bm k}-{\bm k}_{\rm F}^-}.
\label{eq:wSn}
\end{equation}
\par
For $m=2$, the second term of Eq. (\ref{eq:addition}) is rewritten as 
\begin{align}
&\frac{1}{N_{\rm s}}\sum_{i,l}e^{{\rm i}({\bm k}-{\bm k}_{\rm F}^-)\cdot{\bm r}_i}{\tilde \psi}({\cal C}^l)\alpha_{{\rm R}^2}^{{\rm D}^{-}}|{\rm D}^{-}\rangle_i\prod_{j(\ne i)}|{\rm T}^{\gamma_j^l}\rangle_j\nonumber\\
+&\frac{1}{N_{\rm s}}\sum_{i,l}e^{{\rm i}({\bm k}-{\bm k}_{\rm F}^-)\cdot{\bm r}_i}{\tilde \psi}({\cal C}^l)\alpha_{{\rm R}^2}^{{\rm D}^{+}}|{\rm D}^{+}\rangle_i\prod_{j(\ne i)}|{\rm T}^{\gamma_j^l}\rangle_j.
\label{eq:DISD}
\end{align}
\par
Because nonzero components of $T_i^{\pm}|{\rm T}\rangle_i$ are proportional to $|{\rm D}^{\mp}\rangle_i$ [Eq. (\ref{eq:Ti}); Table \ref{tbl:states}], the first term of Eq. (\ref{eq:DISD}) can overlap with $T^{+}_{{\bm k}-{\bm k}_{\rm F}^-}|{\rm GS}\rangle_{N_{\rm h}=0}$, particularly with the dominant eigenstate $|T^{+}\rangle_{{\bm k}-{\bm k}_{\rm F}^-}$ [Eq. (\ref{eq:dominantState})], which contributes to the electronic mode originating from the spin mode in $T^{+}({\bm k},\omega)$ in the intermediate-$\omega$ regime. 
According to Eq. (\ref{eq:Akw}), the dispersion relation of the electronic mode is expressed as 
\begin{equation}
\omega=e^{T^{+}}_{{\bm k}-{\bm k}_{\rm F}^-}.
\label{eq:wTp}
\end{equation}
\par
The second term of Eq. (\ref{eq:DISD}) can overlap with $T^{-}_{{\bm k}-{\bm k}_{\rm F}^-}|{\rm GS}\rangle_{N_{\rm h}=0}$, particularly with the dominant eigenstate $|T^{-}\rangle_{{\bm k}-{\bm k}_{\rm F}^-}$ [Eq. (\ref{eq:dominantState})], which contributes to the electronic mode originating from the spin mode in $T^{-}({\bm k},\omega)$ in the high-$\omega$ regime. 
According to Eq. (\ref{eq:Akw}), the dispersion relation of the electronic mode is expressed as 
\begin{equation}
\omega=e^{T^{-}}_{{\bm k}-{\bm k}_{\rm F}^-}.
\label{eq:wTn}
\end{equation}
\par
\begin{figure} 
\centering
\includegraphics[width=\linewidth]{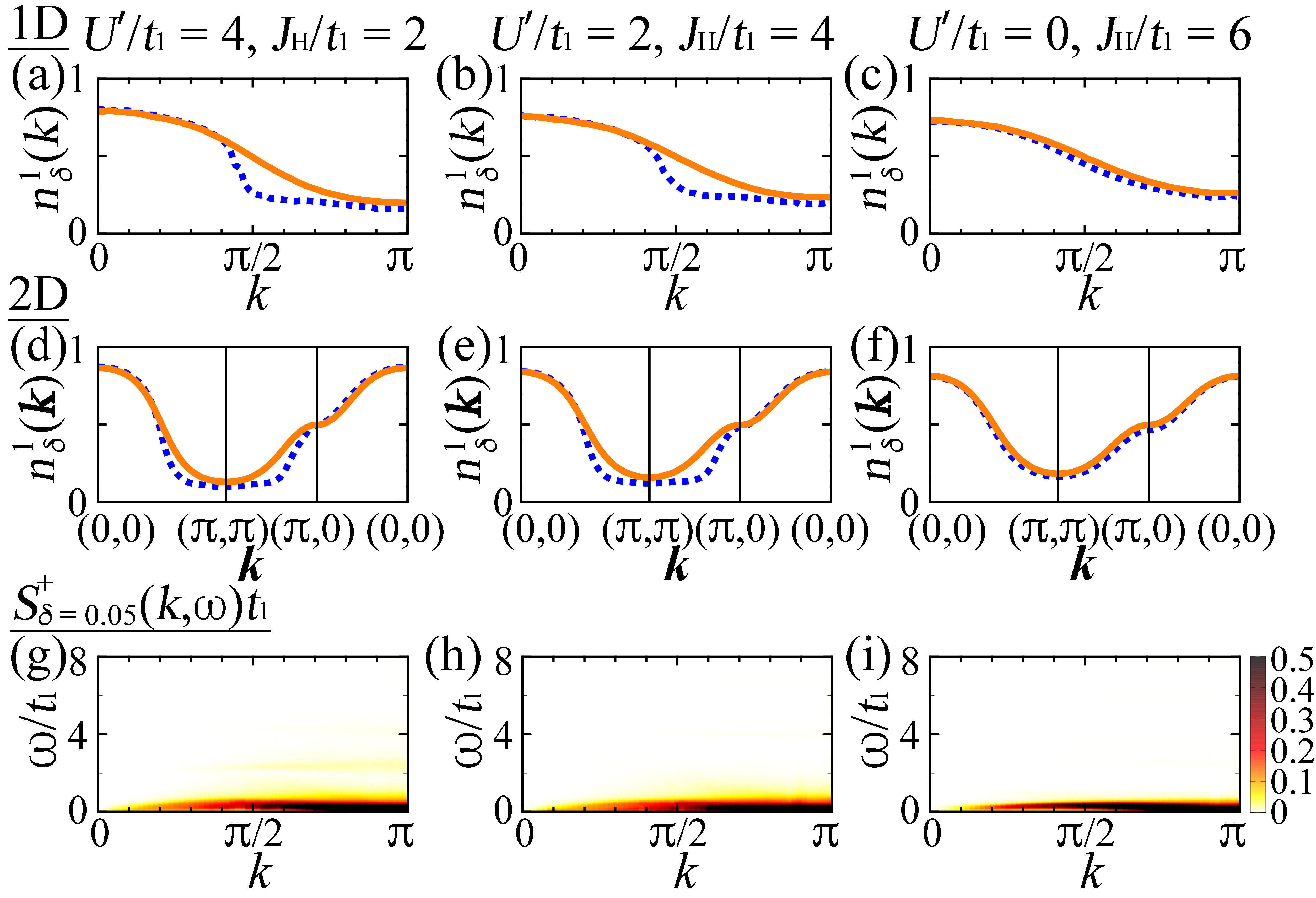}
\caption{Momentum distribution functions in the 1D KHM [(a)--(c)] and 2D KHM [(d)--(f)], and dynamical spin structure factors at $\delta=0.05$ in the 1D KHM [(g)--(i)], for $U/t_1=6$ and $t_2/t_1=0.5$. 
(a)--(f) $n_{\delta=0}^1({\bm k})$ (solid orange curves) and $n_{\delta=0.05}^1({\bm k})$ (dashed blue curves) for $U^{\prime}/t_1=4$ and $J_{\rm H}/t_1=2$ [(a), (d)]; $U^{\prime}/t_1=2$ and $J_{\rm H}/t_1=4$ [(b), (e)]; and $U^{\prime}/t_1=0$ and $J_{\rm H}/t_1=6$ [(c), (f)]. 
(g)--(i) $S_{\delta=0.05}^+(k,\omega)t_1$ for $U^{\prime}/t_1=4$ and $J_{\rm H}/t_1=2$ [(g)]; $U^{\prime}/t_1=2$ and $J_{\rm H}/t_1=4$ [(h)]; and $U^{\prime}/t_1=0$ and $J_{\rm H}/t_1=6$ [(i)].}
\label{fig:nkSk}
\end{figure}

In hole-doped systems, because the LHB of the orbital whose band top is higher in $\omega$ remains mostly filled with electrons in the momentum regime inside the Fermi sea ($|k|<\frac{\pi}{2}$ in 1D; $|k_x\pm k_y|<\pi$ in 2D, for $\delta\rightarrow +0$) [Figs. \ref{fig:nkSk}(a)--\ref{fig:nkSk}(f)], the emergent electron-addition modes primarily appear in the momentum regime outside the Fermi sea of the orbital ($|k|>\frac{\pi}{2}$ in 1D; $|k_x\pm k_y|>\pi$ in 2D, for $\delta\rightarrow +0$). 
If the sign of the hopping parameter for the other orbital is the same as that for this orbital (i.e., the momentum regime of the Fermi sea is the same), the emergent electron-addition modes primarily appear in the momentum regime outside the Fermi sea for the other orbital as well. 
\begin{figure} 
\centering
\includegraphics[width=\linewidth]{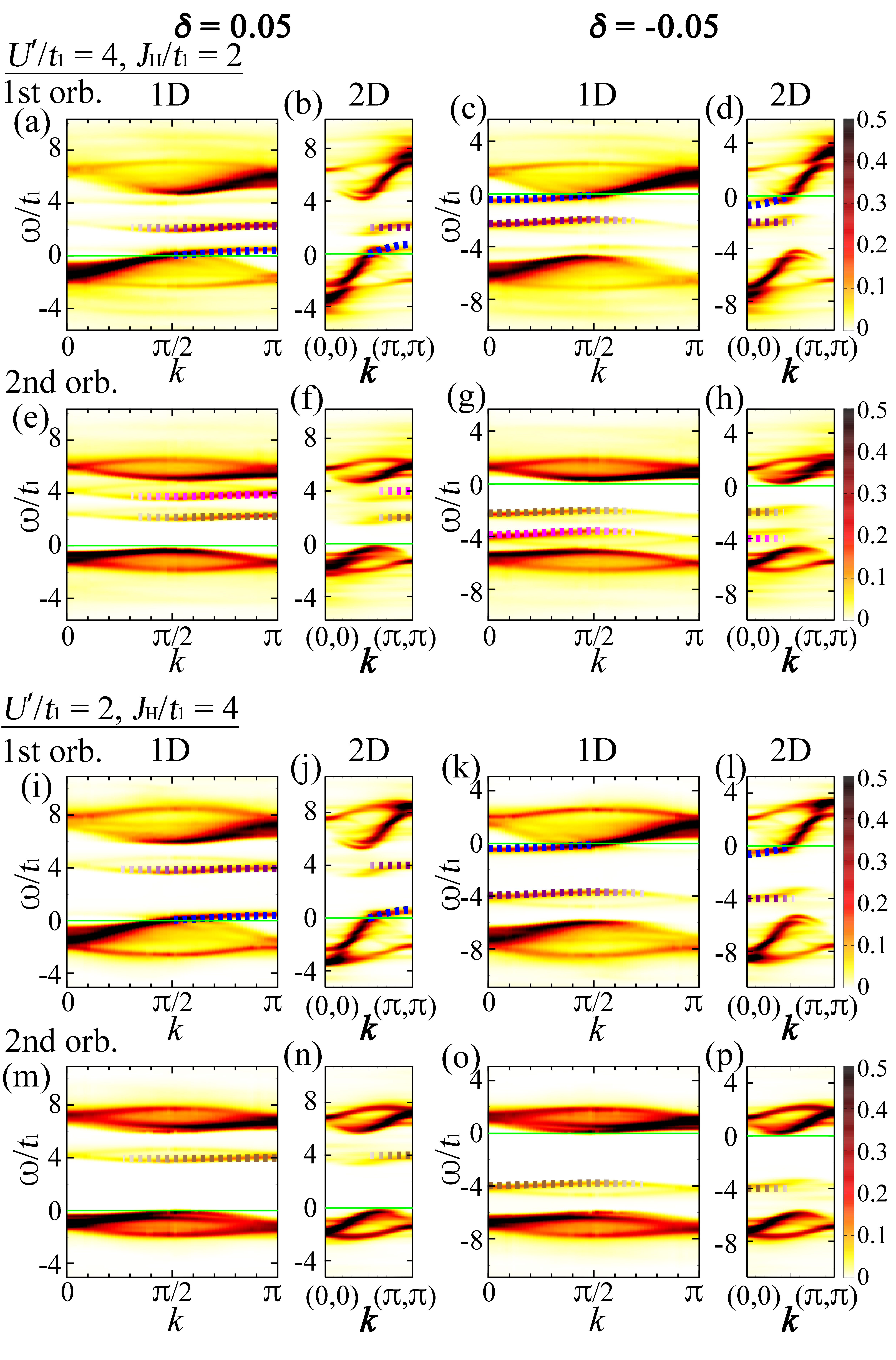}
\caption{Comparisons between the dispersion relations of the doping-induced electronic modes in the effective theory and the spectral functions for $U^{\prime}/t_1=4$ and $J_{\rm H}/t_1=2$ [(a)--(h)]; $U^{\prime}/t_1=2$ and $J_{\rm H}/t_1=4$ [(i)--(p)] 
in the 1st orbital ($m=1$) [(a)--(d), (i)--(l)] and 2nd orbital ($m=2$) [(e)--(h), (m)--(p)], with $U/t_1=6$ and $t_2/t_1=0.5$. 
(a), (e), (i), (m) $A_{\delta=0.05}^m(k,\omega)t_1$; (c), (g), (k), (o) $A_{\delta=-0.05}^m(k,\omega)t_1$ for the 1D KHM. 
(b), (f), (j), (n) ${\bar A}_{\delta=0.05}^m({\bm k},\omega)t_1$; (d), (h), (l), (p) ${\bar A}_{\delta=-0.05}^m({\bm k},\omega)t_1$ along ${\bm k}=(0,0)$--$(\pi,\pi)$ for the 2D KHM. The dispersion relations of the doping-induced modes are $\omega=e_{{\bm k}-{\bm k}_{\rm F}^{-}}^{S^-}$ (dashed purple lines), $\omega=e_{{\bm k}-{\bm k}_{\rm F}^{-}}^{S^+}$ (dashed blue lines) [(a), (b), (i), (j)]; $\omega=-e_{-{\bm k}+{\bm k}_{\rm F}^{+}}^{S^-}$ (dashed purple lines), $\omega=-e_{-{\bm k}+{\bm k}_{\rm F}^{+}}^{S^+}$ (dashed blue lines) [(c), (d), (k), (l)]; $\omega=e_{{\bm k}-{\bm k}_{\rm F}^{-}}^{T^-}$ (dashed magenta lines), $\omega=e_{{\bm k}-{\bm k}_{\rm F}^{-}}^{T^+}$ (dashed brown lines) [(e), (f), (m), (n)]; and $\omega=-e_{-{\bm k}+{\bm k}_{\rm F}^{+}}^{T^-}$ (dashed magenta lines), $\omega=-e_{-{\bm k}+{\bm k}_{\rm F}^{+}}^{T^+}$ (dashed brown lines) [(g), (h), (o), (p)], plotted in the relevant momentum regimes. The excitation energies of the spin modes are extracted from Figs. \ref{fig:Skw}(a)--\ref{fig:Skw}(d) and \ref{fig:Skw}(g)--\ref{fig:Skw}(i) for 1D and approximated by $e_{\bm k}^{S^+,d=2}$ [Eq. (\ref{eq:dispSb})], $\Delta E_{\rm T}^{\rm S}$, and $\Delta E_{\rm T}^{{\rm D}^\pm}$ for 2D. The Fermi momenta are approximated as ${\bm k}_{\rm F}^{\pm}=\frac{\bm \pi}{2}$. The green lines indicate $\omega=0$.}
\label{fig:compDoping}
\end{figure}
\par
In Figs. \ref{fig:compDoping}(a), \ref{fig:compDoping}(b), \ref{fig:compDoping}(e), \ref{fig:compDoping}(f), \ref{fig:compDoping}(i), \ref{fig:compDoping}(j), \ref{fig:compDoping}(m), and \ref{fig:compDoping}(n), 
the dispersion relations of the hole-doping-induced modes, as obtained from the effective theory [Eqs. (\ref{eq:wSp}), (\ref{eq:wSn}), (\ref{eq:wTp}), and (\ref{eq:wTn})], are compared with the spectral functions obtained by numerical calculations [Figs. \ref{fig:Akw}(a), \ref{fig:Akw}(d), \ref{fig:Akw}(g), \ref{fig:Akw}(j), \ref{fig:Akw}(m), \ref{fig:Akw}(p), \ref{fig:Akw}(s), and \ref{fig:Akw}(v)]. 
In these figures, the dispersion relations of the emergent modes within the band gap are obtained using the excitation energies of the spin modes, which are extracted from Figs. \ref{fig:Skw}(a)--\ref{fig:Skw}(d) and \ref{fig:Skw}(g)--\ref{fig:Skw}(i) for 1D, and approximated by $e_{\bm k}^{S^+,d=2}$ [Eq. (\ref{eq:dispSb})], $\Delta E_{\rm T}^{\rm S}$, and $\Delta E_{\rm T}^{{\rm D}^\pm}$ for 2D. 
The above theoretical analysis reasonably well explains the emergent electronic modes found in the numerical results, indicating that the doping-induced electronic modes originate from the spin modes at half filling. The emergent electronic modes can be interpreted as the spin modes whose dispersion relations are shifted by the Fermi momentum. 
Detailed analyses for 1D are shown in Appendix \ref{sec:dopingEffects} [Figs. \ref{fig:dopingDpn}(a)--\ref{fig:dopingDpn}(g)], where it is demonstrated that the excitation energies of the emergent modes at $k=\pi$ approach those of the corresponding spin modes at $k=\frac{\pi}{2}$ as doping concentration decreases. 
\par
Furthermore, the theoretical analysis elucidates the orbital-selective emergence of the electronic modes. The electronic mode reflecting the low-energy spin excitation $|S^+\rangle_{\bm k}$ appears only in the doped orbital [dashed blue lines in Figs. \ref{fig:compDoping}(a), \ref{fig:compDoping}(b), \ref{fig:compDoping}(i), and \ref{fig:compDoping}(j); Eq. (\ref{eq:wSp})], whereas that reflecting the high-energy spin excitation $|T^-\rangle_{\bm k}$ appears only in the undoped orbital [dashed orange lines in Figs. \ref{fig:compDoping}(e) and \ref{fig:compDoping}(f); Eq. (\ref{eq:wTn})]. 
The electronic modes in the intermediate-$\omega$ regime in the doped and undoped orbitals can be identified as those originating from the spin excitations $|S^-\rangle_{\bm k}$ [dashed purple lines in Figs. \ref{fig:compDoping}(a), \ref{fig:compDoping}(b), \ref{fig:compDoping}(i), and \ref{fig:compDoping}(j); Eq. (\ref{eq:wSn})] and $|T^+\rangle_{\bm k}$ [dashed red lines in Figs. \ref{fig:compDoping}(e), \ref{fig:compDoping}(f), \ref{fig:compDoping}(m), and \ref{fig:compDoping}(n); Eq. (\ref{eq:wTp})], respectively. 
\par
The emergence of electronic states with intermediate or high excitation energies in the band gap has been reported. Specifically, the emergence of electronic states due to excitation from atomic configuration $|{\rm P}_{\sigma}^2\rangle$ to $|{\rm S}\rangle$ (Table \ref{tbl:states}) was reported in electron-doped systems, and the emergent band is called the Hund band in Ref. \cite{Hundband}. In Refs. \cite{HDDMRG,dopeHDDMFT,undopeHDDMFT}, in-gap electronic states due to excitation from $|{\rm R}_{\sigma}^2\rangle$ to $|0,\uparrow\downarrow\rangle$ were reported and are called holon-doublon (HD) states. In Ref. \cite{splitHDJDMFT}, splitting of the band of HD states by the Hund’s coupling was reported. However, these studies have mainly focused on density-of-states calculations, and the identification has been within the atomic limit. 
\par
The present study goes beyond these approaches by explicitly identifying the orbital-selective emergent dispersive modes reflecting distinct spin excitations. 
This study elucidates, in a unified manner, their origins, mechanisms, direct connections to the spin modes, and dispersion relations, particularly noting the shift in momentum equal to the Fermi momentum, in line with manifestations of spin excitations as electronic modes \cite{Kohno1DHub,Kohno2DHub,KohnoDIS,KohnoRPP,KohnoHubLadder,KohnoKLM,KohnoGW,KohnoAF,KohnoSpin,Kohno1DtJ,Kohno2DtJ,KohnoMottT,KohnoTinduced,KohnoDrivenMott}. 
\subsubsection{Electronic excited states from the one-electron-doped ground state} 
\label{sec:eledoping}
The electron-doped case can be explained in parallel with the hole-doped case. The one-electron-doped ground state can be expressed as 
\begin{equation}
|{\rm GS}\rangle_{N_{\rm h}=-1}=\frac{1}{\sqrt{N_{\rm s}}}\sum_{i,l}e^{{\rm i}{\bm k}_{\rm F}^+\cdot{\bm r}_i}{\tilde \psi}({\cal C}^l)|{\rm P}^2\rangle_i\prod_{j(\ne i)}|{\rm T}^{\gamma_j^l}\rangle_j,
\label{eq:GS1e}
\end{equation}
where the bottom of the UHB for the first orbital is assumed to be lower than that of the second orbital, and ${\bm k}_{\rm F}^+$ denotes the Fermi momentum in the low-electron-doping limit, i.e., the momentum at the bottom of the UHB for the first orbital. Here, the weight of the configuration ${\cal C}^l$ is denoted by ${\tilde \psi}({\cal C}^l)$; ${\cal C}^l$ represents the $l$th configuration over all sites, with the state at site $i$ being ${\rm P}^2$. 
\par
By adding or removing an electron with momentum ${\bm k}$ on the $m$th orbital, the electronic states are obtained as 
\begin{align}
&c_{\bm k}^{m\dagger}|{\rm GS}\rangle_{N_{\rm h}=-1}\nonumber\\
&=\frac{1}{N_{\rm s}}\sum_{i,j(\ne i),l}e^{{\rm i}({\bm k}\cdot{\bm r}_j+{\bm k}_{\rm F}^+\cdot{\bm r}_i)}\nonumber\\
&\quad\quad\quad\quad\quad\times{\tilde \psi}({\cal C}^l)c_j^{m\dagger}|{\rm T}^{\gamma_j^l}\rangle_j|{\rm P}^2\rangle_i\prod_{n(\ne i,j)}|{\rm T}^{\gamma_n^l}\rangle_n\nonumber\\
\label{eq:addition_e}
&+\frac{1}{N_{\rm s}}\sum_{i,l}e^{{\rm i}({\bm k}+{\bm k}_{\rm F}^+)\cdot{\bm r}_i}{\tilde \psi}({\cal C}^l)c_i^{m\dagger}|{\rm P}^2\rangle_i\prod_{j(\ne i)}|{\rm T}^{\gamma_j^l}\rangle_j,\\
&c_{\bm k}^m|{\rm GS}\rangle_{N_{\rm h}=-1}\nonumber\\
&=\frac{1}{N_{\rm s}}\sum_{i,j(\ne i),l}e^{{\rm i}(-{\bm k}\cdot{\bm r}_j+{\bm k}_{\rm F}^+\cdot{\bm r}_i)}\nonumber\\
&\quad\quad\quad\quad\quad\times{\tilde \psi}({\cal C}^l)c_j^m|{\rm T}^{\gamma_j^l}\rangle_j|{\rm P}^2\rangle_i\prod_{n(\ne i,j)}|{\rm T}^{\gamma_n^l}\rangle_n\nonumber\\
\label{eq:removal_e}
&+\frac{1}{N_{\rm s}}\sum_{i,l}e^{{\rm i}(-{\bm k}+{\bm k}_{\rm F}^+)\cdot{\bm r}_i}{\tilde \psi}({\cal C}^l)c_i^m|{\rm P}^2\rangle_i\prod_{j(\ne i)}|{\rm T}^{\gamma_j^l}\rangle_j.
\end{align}
The first terms in Eqs. (\ref{eq:addition_e}) and (\ref{eq:removal_e}) exhibit essentially the same spectral features as those of the undoped system, i.e., the UHB and LHB. The excitation energies are primarily determined by $c_j^{m(\dagger)}|{\rm T}^{\gamma_j^l}\rangle_j$ as in the case of the undoped systems. 
\par
For $m=1$, the second term of Eq. (\ref{eq:removal_e}) is rewritten as 
\begin{align}
&\frac{1}{N_{\rm s}}\sum_{i,l}e^{{\rm i}(-{\bm k}+{\bm k}_{\rm F}^+)\cdot{\bm r}_i}{\tilde \psi}({\cal C}^l)\alpha_{{\rm P}^2}^{\rm T}|{\rm T}\rangle_i\prod_{j(\ne i)}|{\rm T}^{\gamma_j^l}\rangle_j\nonumber\\
+&\frac{1}{N_{\rm s}}\sum_{i,l}e^{{\rm i}(-{\bm k}+{\bm k}_{\rm F}^+)\cdot{\bm r}_i}{\tilde \psi}({\cal C}^l)\alpha_{{\rm P}^2}^{\rm S}|{\rm S}\rangle_i\prod_{j(\ne i)}|{\rm T}^{\gamma_j^l}\rangle_j.
\label{eq:DISTS_e}
\end{align}
\par
The first term of Eq. (\ref{eq:DISTS_e}) can overlap with $|{\rm GS}\rangle_{N_{\rm h}=0}$ (when ${\bm k}={\bm k}_{\rm F}^+$), which contributes to the bottom of the UHB for the first orbital. This term can also overlap with $S^+_{-{\bm k}+{\bm k}_{\rm F}^+}|{\rm GS}\rangle_{N_{\rm h}=0}$, particularly with the dominant eigenstate $|S^+\rangle_{-{\bm k}+{\bm k}_{\rm F}^+}$ [Eq. (\ref{eq:dominantState})], which contributes to the electronic mode originating from the spin mode in $S^+({\bm k},\omega)$ in the low-$|\omega|$ regime. 
According to Eq. (\ref{eq:Akw}), the dispersion relation of the electronic mode is expressed as 
\begin{equation}
\omega=-e^{S^+}_{-{\bm k}+{\bm k}_{\rm F}^+}.
\label{eq:wSp_e}
\end{equation}
Here, the chemical potential $\mu$ is adjusted so that the Fermi level is located at the bottom of the UHB for the first orbital, i.e., $E_{{\rm GS}_{N_{\rm h}=-1}}=E_{{\rm GS}_{N_{\rm h}=0}}$. 
\par
Because nonzero components of $S_i^-|{\rm T}\rangle_i$ are proportional to $|{\rm S}\rangle_i$ [Eqs. (\ref{eq:Sipma}) and (\ref{eq:Sipmpm}); Table \ref{tbl:states}], the second term of Eq. (\ref{eq:DISTS_e}) can overlap with $S^-_{-{\bm k}+{\bm k}_{\rm F}^+}|{\rm GS}\rangle_{N_{\rm h}=0}$, particularly with the dominant eigenstate $|S^-\rangle_{-{\bm k}+{\bm k}_{\rm F}^+}$ [Eq. (\ref{eq:dominantState})], which contributes to the electronic mode originating from the spin mode in $S^-({\bm k},\omega)$ in the intermediate-$|\omega|$ regime. 
According to Eq. (\ref{eq:Akw}), the dispersion relation of the electronic mode is expressed as 
\begin{equation}
\omega=-e^{S^-}_{-{\bm k}+{\bm k}_{\rm F}^+}.
\label{eq:wSn_e}
\end{equation}
\par
For $m=2$, the second term of Eq. (\ref{eq:removal_e}) is rewritten as 
\begin{align}
&\frac{1}{N_{\rm s}}\sum_{i,l}e^{{\rm i}(-{\bm k}+{\bm k}_{\rm F}^+)\cdot{\bm r}_i}{\tilde \psi}({\cal C}^l)\alpha_{{\rm P}^2}^{{\rm D}^{-}}|{\rm D}^{-}\rangle_i\prod_{j(\ne i)}|{\rm T}^{\gamma_j^l}\rangle_j\nonumber\\
+&\frac{1}{N_{\rm s}}\sum_{i,l}e^{{\rm i}(-{\bm k}+{\bm k}_{\rm F}^+)\cdot{\bm r}_i}{\tilde \psi}({\cal C}^l)\alpha_{{\rm P}^2}^{{\rm D}^{+}}|{\rm D}^{+}\rangle_i\prod_{j(\ne i)}|{\rm T}^{\gamma_j^l}\rangle_j.
\label{eq:DISD_e}
\end{align}
\par
Because nonzero components of $T_i^{\pm}|{\rm T}\rangle_i$ are proportional to $|{\rm D}^{\mp}\rangle_i$ [Eq. (\ref{eq:Ti}); Table \ref{tbl:states}], the first term of Eq. (\ref{eq:DISD_e}) can overlap with $T^{+}_{-{\bm k}+{\bm k}_{\rm F}^+}|{\rm GS}\rangle_{N_{\rm h}=0}$, particularly with the dominant eigenstate $|T^{+}\rangle_{-{\bm k}+{\bm k}_{\rm F}^+}$ [Eq. (\ref{eq:dominantState})], which contributes to the electronic mode originating from the spin mode in $T^{+}({\bm k},\omega)$ in the intermediate-$|\omega|$ regime. 
According to Eq. (\ref{eq:Akw}), the dispersion relation of the electronic mode is expressed as 
\begin{equation}
\omega=-e^{T^{+}}_{-{\bm k}+{\bm k}_{\rm F}^+}.
\label{eq:wTp_e}
\end{equation}
\par
The second term of Eq. (\ref{eq:DISD_e}) can overlap with $T^{-}_{-{\bm k}+{\bm k}_{\rm F}^+}|{\rm GS}\rangle_{N_{\rm h}=0}$, particularly with the dominant eigenstate $|T^{-}\rangle_{-{\bm k}+{\bm k}_{\rm F}^+}$ [Eq. (\ref{eq:dominantState})], which contributes to the electronic mode originating from the spin mode in $T^{-}({\bm k},\omega)$ in the high-$|\omega|$ regime. 
According to Eq. (\ref{eq:Akw}), the dispersion relation of the electronic mode is expressed as 
\begin{equation}
\omega=-e^{T^{-}}_{-{\bm k}+{\bm k}_{\rm F}^+}.
\label{eq:wTn_e}
\end{equation}
\par
In electron-doped systems, because the UHB of the orbital whose band bottom is lower in $\omega$ remains mostly empty in the momentum regime outside the Fermi sea ($|k|>\frac{\pi}{2}$ in 1D; $|k_x\pm k_y|>\pi$ in 2D, for $\delta\rightarrow -0$), the emergent electron-removal modes primarily appear in the momentum regime inside the Fermi sea of the orbital ($|k|<\frac{\pi}{2}$ in 1D; $|k_x\pm k_y|<\pi$ in 2D, for $\delta\rightarrow -0$). 
If the sign of the hopping parameter for the other orbital is the same as that for this orbital (i.e., the momentum regime of the Fermi sea is the same), the emergent electron-removal modes primarily appear in the momentum regime inside the Fermi sea for the other orbital as well. 
\subsubsection{Particle-hole transformation} 
The spectral function of the electron-doped state $|X\rangle$ with $N_{\rm h}=-n$ can be obtained from that of the corresponding hole-doped state $|X^{\prime}\rangle$ with $N_{\rm h}=n$ as 
\begin{equation}
A_X^m({\bm k},\omega)=A_{X^{\prime}}^m({\bm \pi}-{\bm k},-\omega)
\label{eq:AkwPH}
\end{equation}
on a bipartite lattice in the KHM for $H=0$, via the particle-hole transformation, which is derived in the same way as for the HM \cite{KohnoDrivenMott}. 
\par
The particle-hole transformation \cite{TakahashiBook,Essler} is given by 
\begin{equation}
c_{i,\sigma}^m\rightarrow (-1)^ic_{i,\sigma}^{m\dagger}
\label{eq:phTrans}
\end{equation}
for all $\sigma$, orbitals $m$, and sites $i$. 
Under this transformation, the Hamiltonian of the KHM [Eq. (\ref{eq:Ham})], which is represented by ${\cal H}(t_1,t_2,U,U^{\prime},J_{\rm H},\mu,H)$, transforms into ${\cal H}(t_1,t_2,U,U^{\prime},J_{\rm H},-\mu,-H)-4\mu N_{\rm s}$ on a bipartite lattice. States with $N_{\rm h}=-n$ are transformed into those with $N_{\rm h}=n$. 
Since Eq. (\ref{eq:phTrans}) leads to 
\begin{equation}
c_{{\bm k},\sigma}^m\rightarrow c_{{\bm \pi}-{\bm k},\sigma}^{m\dagger},
\label{eq:phTransk}
\end{equation}
the spectral function of the electron-doped state $|X\rangle$ [Eq. (\ref{eq:Akw})] is transformed as 
\begin{align}
A_{X}^{m,\sigma}({\bm k},\omega)\rightarrow&\frac{1}{Z_{X^{\prime}}}\sum_{n}\left[
\begin{array}{r}
|\langle n|c_{{\bm \pi}-{\bm k},\sigma}^{m}|X^{\prime}\rangle|^2\delta(\omega-E_n+E_{X^{\prime}})\\
+|\langle n|c_{{\bm \pi}-{\bm k},\sigma}^{m\dagger}|X^{\prime}\rangle|^2\delta(\omega+E_n-E_{X^{\prime}})
\end{array}\right]\nonumber\\
&=A_{X^{\prime}}^{m,\sigma}({\bm \pi}-{\bm k},-\omega),
\end{align}
where $|X^{\prime}\rangle$ is the hole-doped state obtained from the electron-doped state $|X\rangle$ via the particle-hole transformation. 
\par
In Figs. \ref{fig:compDoping}(c), \ref{fig:compDoping}(d), \ref{fig:compDoping}(g), \ref{fig:compDoping}(h), \ref{fig:compDoping}(k), \ref{fig:compDoping}(l), \ref{fig:compDoping}(o), and \ref{fig:compDoping}(p), 
the dispersion relations of the electron-doping-induced modes, as obtained from the effective theory [Eqs. (\ref{eq:wSp_e}), (\ref{eq:wSn_e}), (\ref{eq:wTp_e}), and (\ref{eq:wTn_e})], are compared with the spectral functions obtained via the transformation of Eq. (\ref{eq:AkwPH}) from those of the hole-doped systems in Figs. \ref{fig:Akw}(a), \ref{fig:Akw}(d), \ref{fig:Akw}(g), \ref{fig:Akw}(j), \ref{fig:Akw}(m), \ref{fig:Akw}(p), \ref{fig:Akw}(s), and \ref{fig:Akw}(v). 
As in the hole-doped case, the above theoretical analysis successfully explains the emergent electronic modes in electron-doped systems as well, indicating that the doping-induced electronic modes originate from the spin modes at half filling. The emergent electronic modes exhibit spin-mode dispersion relations shifted by the Fermi momentum with $\omega$ inverted. 
The orbital-selective emergence of the electronic modes is also demonstrated. 
\subsubsection{Selection rules} 
The emergence of electronic modes reflecting spin excitations can generally be understood in terms of quantum numbers. 
We consider selection rules for electronic excitations between the exact eigenstates on a cluster with an even number of sites, in the low-doping limit, and in the parameter regime where electrons behave as single particles without pairing (Sec. \ref{sec:tripletPairing}). 
\par
The exact ground state with $N_{\rm h}=n$ is denoted by $|{\rm GS}\rangle_{N_{\rm h}=n}^{\rm exact}$, which is expressed in terms of the full site basis constructed from all possible local states shown in Table \ref{tbl:states}. The energy is denoted by $E_{{\rm GS}_{N_{\rm h}=n}}^{\rm exact}$. The ground state at half filling $|{\rm GS}\rangle_{N_{\rm h}=0}^{\rm exact}$ is assumed to have momentum ${\bm 0}$, $S=0$, and $N_{\rm h}=0$. The dominant eigenstate of ${\cal X}_{\bm k}|{\rm GS}\rangle_{N_{\rm h}=0}^{\rm exact}$ is denoted by $|{\cal X}\rangle_{\bm k}^{\rm exact}$, where ${\cal X}=S^{\pm}$ or $T^{\pm}$ (omitting the spin indices); these states have momentum ${\bm k}$, $S=1$, and $N_{\rm h}=0$. The excitation energies from $|{\rm GS}\rangle_{N_{\rm h}=0}^{\rm exact}$ are denoted by $e_{\bm k}^{S^{\pm,{\rm exact}}}$ and $e_{\bm k}^{T^{\pm,{\rm exact}}}$. 
\par
The one-hole-doped ground state $|{\rm GS}\rangle_{N_{\rm h}=1}^{\rm exact}$ has momentum $-{\bm k}_{\rm F}^-$, $S=\frac{1}{2}$, and $N_{\rm h}=1$, which are the same quantum numbers as those of the state obtained by removing an electron with momentum ${\bm k}_{\rm F}^-$ from the top of the LHB for the orbital with the highest upper edge in $\omega$. 
If an electron with momentum ${\bm k}$ is added to $|{\rm GS}\rangle_{N_{\rm h}=1}^{\rm exact}$, the resulting state has momentum ${\bm k}-{\bm k}_{\rm F}^-$, $S=0$ or $1$, and $N_{\rm h}=0$. This state can overlap with $|{\rm GS}\rangle_{N_{\rm h}=0}^{\rm exact}$, $|S^{\pm}\rangle_{{\bm k}-{\bm k}_{\rm F}^-}^{\rm exact}$, and $|T^{\pm}\rangle_{{\bm k}-{\bm k}_{\rm F}^-}^{\rm exact}$. Thus, according to Eq. (\ref{eq:Akw}), the spectral weights can emerge in the electron-addition spectrum along 
\begin{equation}
\omega=e^{S^{\pm,{\rm exact}}}_{{\bm k}-{\bm k}_{\rm F}^-}\quad\text{and}\quad
\omega=e^{T^{\pm,{\rm exact}}}_{{\bm k}-{\bm k}_{\rm F}^-}
\end{equation}
[corresponding to Eqs. (\ref{eq:wSp}), (\ref{eq:wSn}), (\ref{eq:wTp}), and (\ref{eq:wTn})] as well as at the top of the LHB at the Fermi level. Here, the chemical potential $\mu$ is adjusted so that the Fermi level is located at the top of the LHB of the orbital with the highest upper edge in $\omega$ (i.e., $E_{{\rm GS}_{N_{\rm h}=1}}^{\rm exact}=E_{{\rm GS}_{N_{\rm h}=0}}^{\rm exact}$). 
\par
Similarly, the one-electron-doped ground state $|{\rm GS}\rangle_{N_{\rm h}=-1}^{\rm exact}$ has momentum ${\bm k}_{\rm F}^+$, $S=\frac{1}{2}$, and $N_{\rm h}=-1$, which are the same quantum numbers as those of the state obtained by adding an electron with momentum ${\bm k}_{\rm F}^+$ to the bottom of the UHB for the orbital with the lowest lower edge in $\omega$. 
If an electron with momentum ${\bm k}$ is removed from $|{\rm GS}\rangle_{N_{\rm h}=-1}^{\rm exact}$, the resulting state has momentum $-{\bm k}+{\bm k}_{\rm F}^+$, $S=0$ or $1$, and $N_{\rm h}=0$. This state can overlap with $|{\rm GS}\rangle_{N_{\rm h}=0}^{\rm exact}$, $|S^{\pm}\rangle_{-{\bm k}+{\bm k}_{\rm F}^+}^{\rm exact}$, and $|T^{\pm}\rangle_{-{\bm k}+{\bm k}_{\rm F}^+}^{\rm exact}$. Thus, according to Eq. (\ref{eq:Akw}), the spectral weights can emerge in the electron-removal spectrum along 
\begin{equation}
\omega=-e^{S^{\pm,{\rm exact}}}_{-{\bm k}+{\bm k}_{\rm F}^+}\quad\text{and}\quad
\omega=-e^{T^{\pm,{\rm exact}}}_{-{\bm k}+{\bm k}_{\rm F}^+}
\end{equation}
[corresponding to Eqs. (\ref{eq:wSp_e}), (\ref{eq:wSn_e}), (\ref{eq:wTp_e}), and (\ref{eq:wTn_e})] as well as at the bottom of the UHB at the Fermi level. Here, the chemical potential $\mu$ is adjusted so that the Fermi level is located at the bottom of the UHB of the orbital with the lowest lower edge in $\omega$ (i.e., $E_{{\rm GS}_{N_{\rm h}=-1}}^{\rm exact}=E_{{\rm GS}_{N_{\rm h}=0}}^{\rm exact}$). 
\par
These results imply that electronic modes reflecting the spin excitations can emerge upon doping. These modes exhibit the spin-mode dispersion relations ($\pm e_{\pm{\bm k}}^{\cal X}$) shifted by the Fermi momenta ($\mp{\bm k}_{\rm F}^{\mp}$) for ${\cal X}=S^+, S^-, T^+$, and $T^-$. Here, the upper (lower) sign indicates the electron-addition (removal) excitation in the low-hole-doping (low-electron-doping) limit. 
\subsubsection{Spectral weights of the doping-induced electronic modes}
\label{sec:SWdoping}
In the $n$-hole-doped ($n$-electron-doped) ground state, there are essentially $n$ doped sites [Eqs. (\ref{eq:GS1h}) and (\ref{eq:GS1e})] in the low-doping regime ($n\ll N_{\rm s}$). 
The probability of adding (removing) an electron on the doped sites, described by the second term of Eq. (\ref{eq:addition}) \textbf{(}Eq. (\ref{eq:removal_e})\textbf{)}, becomes $n$ times as large as that in the one-hole-doped (one-electron-doped) case. Thus, the spectral weights of the emergent modes in the $n$-hole-doped ($n$-electron-doped) system are essentially $n$ times those of the one-hole-doped (one-electron-doped) system. This implies that the spectral weights of the emergent modes for both orbitals---even for the orbital that remains essentially undoped---increase in proportion to doping concentration $|\delta|$ in the low-doping regime. 
This behavior is confirmed by numerical calculations in Appendix \ref{sec:dopingEffects} [Figs. \ref{fig:dopingDpn}(h) and \ref{fig:dopingDpn}(i)]. 
\par
The spectral features remain almost unchanged in the low-doping regime, because the Fermi momenta and spin excitations remain almost unchanged [Figs. \ref{fig:Akw}(a), \ref{fig:Akw}(d), \ref{fig:Akw}(m), \ref{fig:Akw}(p), \ref{fig:nkSk}(g) and \ref{fig:nkSk}(h)]. As doping concentration increases, the emergent electronic modes gradually evolve into well-developed bands; in particular, the low-energy mode in the doped orbital forms a gapless mode carrying significant spectral weight which may be regarded as a quasiparticle of a Fermi liquid. 
\subsection{Electronic excitations at nonzero temperatures} 
\label{sec:nonzeroT}
\subsubsection{Origins and mechanisms of temperature-induced electronic modes} 
At nonzero temperatures, the thermal equilibrium state involves not only the ground state but also all the excited states, particularly those whose excitation energies are lower than the temperature. 
If the temperature is comparable to the low-energy spin-excitation energies, electronic excitations from the spin-excited states that are involved in the thermal equilibrium state can appear in the spectral function \cite{KohnoMottT,KohnoTinduced}. 
\par
The eigenstate of the low-energy spin mode with momentum ${\bm q}$, which has a significant overlap with $S^+_{\bm q}|{\rm GS}\rangle_{N_{\rm h}=0}$, can be effectively expressed as 
\begin{equation}
|S^+\rangle_{\bm q}=\frac{1}{\sqrt{N_{\rm s}}}\sum_{i,l}e^{{\rm i}{\bm q}\cdot{\bm r}_i}{\bar \psi}({\cal C}^l)|{\rm T}^{\gamma_i^l}\rangle_i\prod_{j(\ne i)}|{\rm T}^{\gamma_j^l}\rangle_j,
\label{eq:Spq}
\end{equation}
whose excitation energy from $|{\rm GS}\rangle_{N_{\rm h}=0}$ is $e_{\bm q}^{S^+}$ [Eq. (\ref{eq:dominantState})]. 
The electronic excited state obtained by removing an electron with momentum ${\bm k}$ from the $m$th orbital, $c_{\bm k}^m|S^+\rangle_{\bm q}$, involves the following state: 
\begin{equation}
\frac{1}{N_{\rm s}}\sum_{i,l}e^{{\rm i}({\bm q}-{\bm k})\cdot{\bm r}_i}{\bar \psi}({\cal C}^l)\alpha_{{\rm T}^{\gamma_i^l}}^{{\rm R}^{\bar m}}|{\rm R}^{\bar m}\rangle_i\prod_{j(\ne i)}|{\rm T}^{\gamma_j^l}\rangle_j,
\end{equation}
where ${\bar m}$ denotes the orbital opposite to $m$; specifically, ${\bar m}=2$ and $1$ for $m=1$ and $2$, respectively. 
This state can overlap with the eigenstate associated with the top of the LHB for the $m$th orbital at zero temperature, i.e., $|c^m\rangle_{{\bm k}_{\rm F}^-}$, whose excitation energy from $|{\rm GS}\rangle_{N_{\rm h}=0}$ is $e_{{\bm k}_{\rm F}^-}^{c^m}$ [Eq. (\ref{eq:dominantState})] if ${\bm q}={\bm k}-{\bm k}_{\rm F}^-$. 
Thus, in the electron-removal spectral function for the $m$th orbital [Eq. (\ref{eq:AkwT})], an electronic mode can emerge from the top of the LHB, exhibiting 
\begin{equation}
\omega=e_{{\bm k}-{\bm k}_{\rm F}^-}^{S^+}+\mu_-^m, 
\label{eq:wSp_rT}
\end{equation}
where $\mu_-^m$ denotes the value of $\omega$ at the top of the LHB for the $m$th orbital at zero temperature: $\mu_-^m=-e_{{\bm k}_{\rm F}^-}^{c^m}$. 
This indicates that an electronic mode reflecting the low-energy spin mode can emerge from the top of the LHB for each orbital, exhibiting the momentum-shifted spin-mode dispersion relation. 
\par
Similarly, the electronic excited state obtained by adding an electron with momentum ${\bm k}$ to the $m$th orbital, $c_{\bm k}^{m\dagger}|S^+\rangle_{\bm q}$, involves the following state: 
\begin{equation}
\frac{1}{N_{\rm s}}\sum_{i,l}e^{{\rm i}({\bm q}+{\bm k})\cdot{\bm r}_i}{\bar \psi}({\cal C}^l)\alpha_{{\rm T}^{\gamma_i^l}}^{{\rm P}^{\bar m}}|{\rm P}^{\bar m}\rangle_i\prod_{j(\ne i)}|{\rm T}^{\gamma_j^l}\rangle_j.
\end{equation}
This state can overlap with the eigenstate associated with the bottom of the UHB for the $m$th orbital at zero temperature, i.e., $|c^{m\dagger}\rangle_{{\bm k}_{\rm F}^+}$, whose excitation energy from $|{\rm GS}\rangle_{N_{\rm h}=0}$ is $e_{{\bm k}_{\rm F}^+}^{c^{m\dagger}}$ [Eq. (\ref{eq:dominantState})] if ${\bm q}=-{\bm k}+{\bm k}_{\rm F}^+$. 
Thus, in the electron-addition spectral function for the $m$th orbital [Eq. (\ref{eq:AkwT})], an electronic mode can emerge from the bottom of the UHB, 
exhibiting 
\begin{equation}
\omega=-e_{-{\bm k}+{\bm k}_{\rm F}^+}^{S^+}+\mu_+^m, 
\label{eq:wSp_aT}
\end{equation}
where $\mu_+^m$ denotes the value of $\omega$ at the bottom of the UHB for the $m$th orbital at zero temperature: $\mu_+^m=e_{{\bm k}_{\rm F}^+}^{c^{m\dagger}}$. 
This indicates that an electronic mode reflecting the low-energy spin mode can emerge from the bottom of the UHB for each orbital, exhibiting the momentum-shifted spin-mode dispersion relation inverted with respect to $\omega$. 
\par
If the temperature is well below the excitation energies of higher-energy spin modes such as those of $|S^-_{\bm k}\rangle$ and $|T_{\bm k}^{\pm}\rangle$, the number of their excited spins in the thermal equilibrium state should be sparse, making the spectral weights originating from these modes negligible or difficult to discern. 
Thus, at low temperatures, emergent electronic modes are basically those originating from the low-energy spin mode only, and the emergent modes appear in the spectral functions for both orbitals, exhibiting the low-energy spin-mode dispersion relation ($\pm e^{S^+}_{\pm{\bm k}}$) shifted by the Fermi momenta ($\mp{\bm k}_{\rm F}^{\mp}$) from the band edges ($\mu_{\mp}^m$) [Eqs. (\ref{eq:wSp_rT}) and (\ref{eq:wSp_aT})]. 
\par
Because the electronic excitation from $|S^+\rangle_{\bm k}$ to the lowest-energy electron-removed state for the $m$th orbital is the inverse of the excitation from the lowest-energy electron-removed state for the $m$th orbital to $|S^+\rangle_{\bm k}$ [Eq. (\ref{eq:AkwT})], the spectral weights of the temperature-induced modes primarily appear in the same momentum regimes as the doping-induced modes in the one-electron-removed system for the $m$th orbital, i.e., outside the Fermi sea of the $m$th orbital (Sec. \ref{sec:holedoping}). 
\par
Similarly, because the electronic excitation from $|S^+\rangle_{\bm k}$ to the lowest-energy electron-added state for the $m$th orbital is the inverse of the excitation from the lowest-energy electron-added state for the $m$th orbital to $|S^+\rangle_{\bm k}$ [Eq. (\ref{eq:AkwT})], the spectral weights of the temperature-induced modes primarily appear in the same momentum regimes as the doping-induced modes in the one-electron-added system for the $m$th orbital, i.e., inside the Fermi sea of the $m$th orbital (Sec. \ref{sec:eledoping}). 
\par
Note that these spectral features of the temperature-induced modes persist even in the parameter regime where electrons form spin-triplet pairs upon doping (Sec. \ref{sec:tripletPairing}). 
\par
\begin{figure} 
\centering
\includegraphics[width=\linewidth]{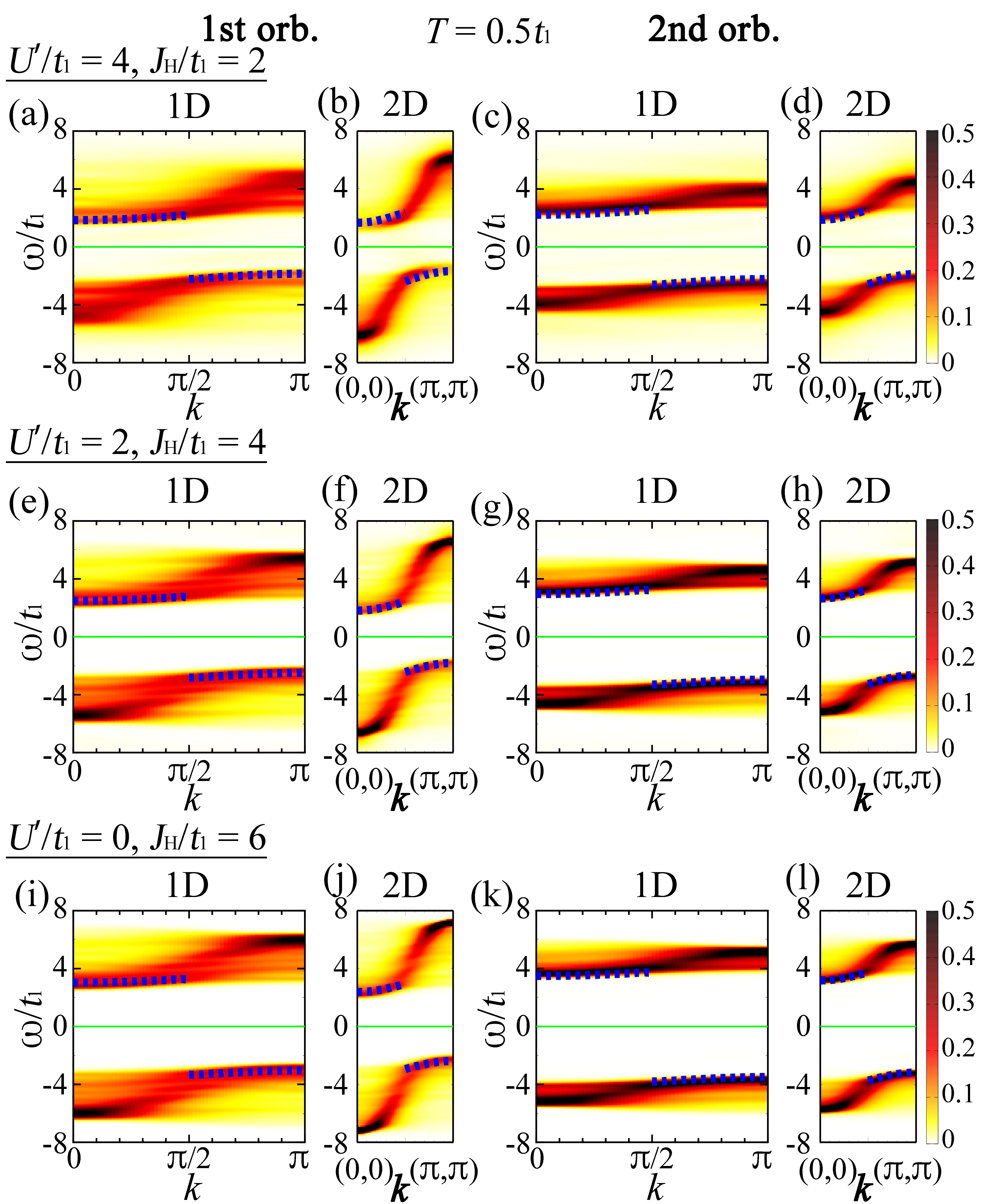}
\caption{Comparisons between the dispersion relations of the temperature-induced electronic modes in the effective theory and the spectral functions for $U^{\prime}/t_1=4$ and $J_{\rm H}/t_1=2$ [(a)--(d)]; $U^{\prime}/t_1=2$ and $J_{\rm H}/t_1=4$ [(e)--(h)]; and $U^{\prime}/t_1=0$ and $J_{\rm H}/t_1=6$ [(i)--(l)]. 
(a), (e), (i) $A_{T=0.5t_1}^1(k,\omega)t_1$; (c), (g), (k) $A_{T=0.5t_1}^2(k,\omega)t_1$ in the 1D KHM. 
(b), (f), (j) ${\bar A}_{T=0.5t_1}^1({\bm k},\omega)t_1$; (d), (h), (l) ${\bar A}_{T=0.5t_1}^2({\bm k},\omega)t_1$ along ${\bm k}=(0,0)$--$(\pi,\pi)$ in the 2D KHM. The dispersion relations of the temperature-induced modes are $\omega=\pm e_{\pm{\bm k}\mp{\bm k}_{\rm F}^{\mp}}^{S^+}+\mu_{\mp}^m$ (dashed blue curves), plotted in the relevant momentum regimes. The excitation energies of the spin modes are extracted from Figs. \ref{fig:Skw}(a), \ref{fig:Skw}(g), and \ref{fig:Skw}(m) for 1D and approximated by $e_{\bm k}^{S^+,d=2}$ [Eq. (\ref{eq:dispSb})] for 2D. The green lines indicate $\omega=0$.}
\label{fig:compTemp}
\end{figure}

In Fig. \ref{fig:compTemp}, the dispersion relations of the temperature-induced modes, as obtained from the effective theory [Eqs. (\ref{eq:wSp_rT}) and (\ref{eq:wSp_aT})], are compared with the spectral functions obtained by numerical calculations [Figs. \ref{fig:Akw}(c), \ref{fig:Akw}(f), \ref{fig:Akw}(i), \ref{fig:Akw}(l), \ref{fig:Akw}(o), \ref{fig:Akw}(r), \ref{fig:Akw}(u), \ref{fig:Akw}(x), \ref{fig:AkwUp0}(c), \ref{fig:AkwUp0}(f), \ref{fig:AkwUp0}(i), and \ref{fig:AkwUp0}(l)]. 
The above theoretical analysis reasonably well explains the emergent electronic modes found in the numerical results, indicating that, in both orbitals, the temperature-induced electronic modes originate from the low-energy spin mode $|S^+\rangle_{\bm k}$. The emergent electronic modes exhibit the low-energy spin-mode dispersion relation shifted by the Fermi momenta. 
The spectral features at lower nonzero temperature are discussed in Appendix \ref{sec:temperatureEffects}. 
\par
The present theoretical analysis [Eqs. (\ref{eq:wSp_rT}) and (\ref{eq:wSp_aT})] elucidates that the temperature-induced electronic modes, originating only from the low-energy spin excitation, appear in both orbitals even in the parameter regime for pairing (Fig. \ref{fig:compTemp}), in contrast to the doping-induced electronic modes, which reflect distinct spin excitations depending on the orbitals in the parameter regime without pairing (Secs. \ref{sec:holedoping} and \ref{sec:eledoping}). 
\subsubsection{Spectral weights of the temperature-induced electronic modes}
\label{sec:SWT}
The spectral weights of the emergent modes increase as the temperature increases, primarily because the number of excited spins involved in the thermal equilibrium state increases \cite{KohnoTinduced}. 
This implies that an increase in the components of the doped states is not necessary to enhance the spectral weights of the temperature-induced modes [Eq. (\ref{eq:AkwT})]; the tops of the LHBs or the bottoms of the UHBs can be far below or above the temperature scale, provided that spin-excitation energies are comparable to the temperature \cite{KohnoMottT,KohnoTinduced}. 
\par
The emergent electronic modes gradually evolve into well-developed bands as the temperature increases.
\subsubsection{Selection rules} 
The exact dominant eigenstate of the low-energy spin excitation $|S^+\rangle_{\bm q}^{\rm exact}$ [Eq. (\ref{eq:dominantState})] is 
involved in the thermal equilibrium state at nonzero temperature; this state has momentum ${\bm q}$, $S=1$, and $N_{\rm h}=0$. 
If an electron with momentum ${\bm k}$ is removed from the $m$th orbital in this state, the resulting state $c_{\bm k}^m|S^+\rangle_{\bm q}^{\rm exact}$ has momentum ${\bm q}-{\bm k}$, $S=\frac{1}{2}$ or $\frac{3}{2}$, and $N_{\rm h}=1$. 
This resulting state can overlap with the one-electron-removed lowest-energy state (i.e., the exact eigenstate at the top of the LHB) for the $m$th orbital $|c^m\rangle_{{\bm k}_{\rm F}^-}^{\rm exact}$ with momentum $-{\bm k}_{\rm F}^-$, $S=\frac{1}{2}$, and $N_{\rm h}=1$, provided that ${\bm q}={\bm k}-{\bm k}_{\rm F}^-$. Thus, according to Eq. (\ref{eq:AkwT}), an electronic mode can emerge for the $m$th orbital along 
\begin{equation}
\omega=e_{{\bm k}-{\bm k}_{\rm F}^-}^{S^{+,{\rm exact}}}+\mu_-^{m,{\rm exact}},
\label{eq:TindL}
\end{equation}
where $\mu_-^{m,{\rm exact}}(=-e_{{\bm k}_{\rm F}^-}^{c^{m,{\rm exact}}})$ corresponds to the $\omega$ value at the top of the LHB for the $m$th orbital at zero temperature. Equation (\ref{eq:TindL}) corresponds to Eq. (\ref{eq:wSp_rT}). 
\par
Similarly, if an electron with momentum ${\bm k}$ is added to the $m$th orbital in $|S^+\rangle_{\bm q}^{\rm exact}$, the resulting state $c_{\bm k}^{m\dagger}|S^+\rangle_{\bm q}^{\rm exact}$ has momentum ${\bm q}+{\bm k}$, $S=\frac{1}{2}$ or $\frac{3}{2}$, and $N_{\rm h}=-1$. 
This resulting state can overlap with the one-electron-added lowest-energy state (i.e., the exact eigenstate at the bottom of the UHB) for the $m$th orbital $|c^{m\dagger}\rangle_{{\bm k}_{\rm F}^+}^{\rm exact}$ with momentum ${\bm k}_{\rm F}^+$, $S=\frac{1}{2}$, and $N_{\rm h}=-1$, provided that ${\bm q}=-{\bm k}+{\bm k}_{\rm F}^+$. Thus, according to Eq. (\ref{eq:AkwT}), an electronic mode can emerge for the $m$th orbital along 
\begin{equation}
\omega=-e_{-{\bm k}+{\bm k}_{\rm F}^+}^{S^{+,{\rm exact}}}+\mu_+^{m,{\rm exact}},
\label{eq:TindU}
\end{equation}
where $\mu_+^{m,{\rm exact}}(=e_{{\bm k}_{\rm F}^+}^{c^{m\dagger,{\rm exact}}})$ corresponds to the $\omega$ value at the bottom of the UHB for the $m$th orbital at zero temperature. Equation (\ref{eq:TindU}) corresponds to Eq. (\ref{eq:wSp_aT}). 
\par
Thus, at nonzero temperature, electronic modes can emerge, exhibiting the low-energy spin-mode dispersion relation ($\pm e_{\pm{\bm k}}^{S^{+,{\rm exact}}}$) shifted by the Fermi momenta ($\mp{\bm k}_{\rm F}^{\mp}$) from the band edges ($\mu_{\mp}^m$) for both orbitals. 
\par
The above argument remains valid even in the parameter regime where electrons form spin-triplet pairs upon doping (Sec. \ref{sec:tripletPairing}). 
\subsection{Momentum regimes of the emergent modes depending on the signs of the hopping parameters} 
\label{sec:momentum}
The momentum regimes where the emergent modes primarily appear depend on the signs of the hopping parameters $t_1$ and $t_2$. This can be explained in terms of gauge transformation (Appendix \ref{sec:gaugeTrans}). 
\par
Under a $U(1)$ gauge transformation defined as 
\begin{equation}
c_{i,\sigma}^m\rightarrow (-1)^ic_{i,\sigma}^m
\label{eq:gaugeTrans}
\end{equation}
for all $\sigma$ and sites $i$ on a bipartite lattice, 
the spectral function of eigenstate $|X^{\prime}\rangle$ for ${\cal H}(-t_m)$ can be expressed using that of $|X\rangle$ (the eigenstate with the same energy as $|X^{\prime}\rangle$) for ${\cal H}(t_m)$ as 
\begin{equation}
A_{X^{\prime}}^m({\bm k},\omega)=A_X^m({\bm \pi}-{\bm k},\omega) 
\label{eq:Akwntm}
\end{equation}
in inversion symmetric systems; $A_X^m({\bm k},\omega)=A_X^m(-{\bm k},\omega)$. 
Here, ${\cal H}(t_m)$ is a shorthand notation for the Hamiltonian ${\cal H}(t_m,t_{\bar m},U,U^{\prime},J_{\rm H},\mu,H)$ [Eq. (\ref{eq:Ham})], since all the parameters remain unchanged except for the sign of $t_m$. 
\par
Thus, the gauge transformation, i.e., the sign reversal of $t_m$, maps the spectral function for the $m$th orbital at each momentum ${\bm k}$ to that of ${\bm \pi}-{\bm k}$, thus interchanging the regions around ${\bm k}={\bm 0}$ and ${\bm \pi}$. 
\par
\begin{figure}
\centering
\includegraphics[width=\linewidth]{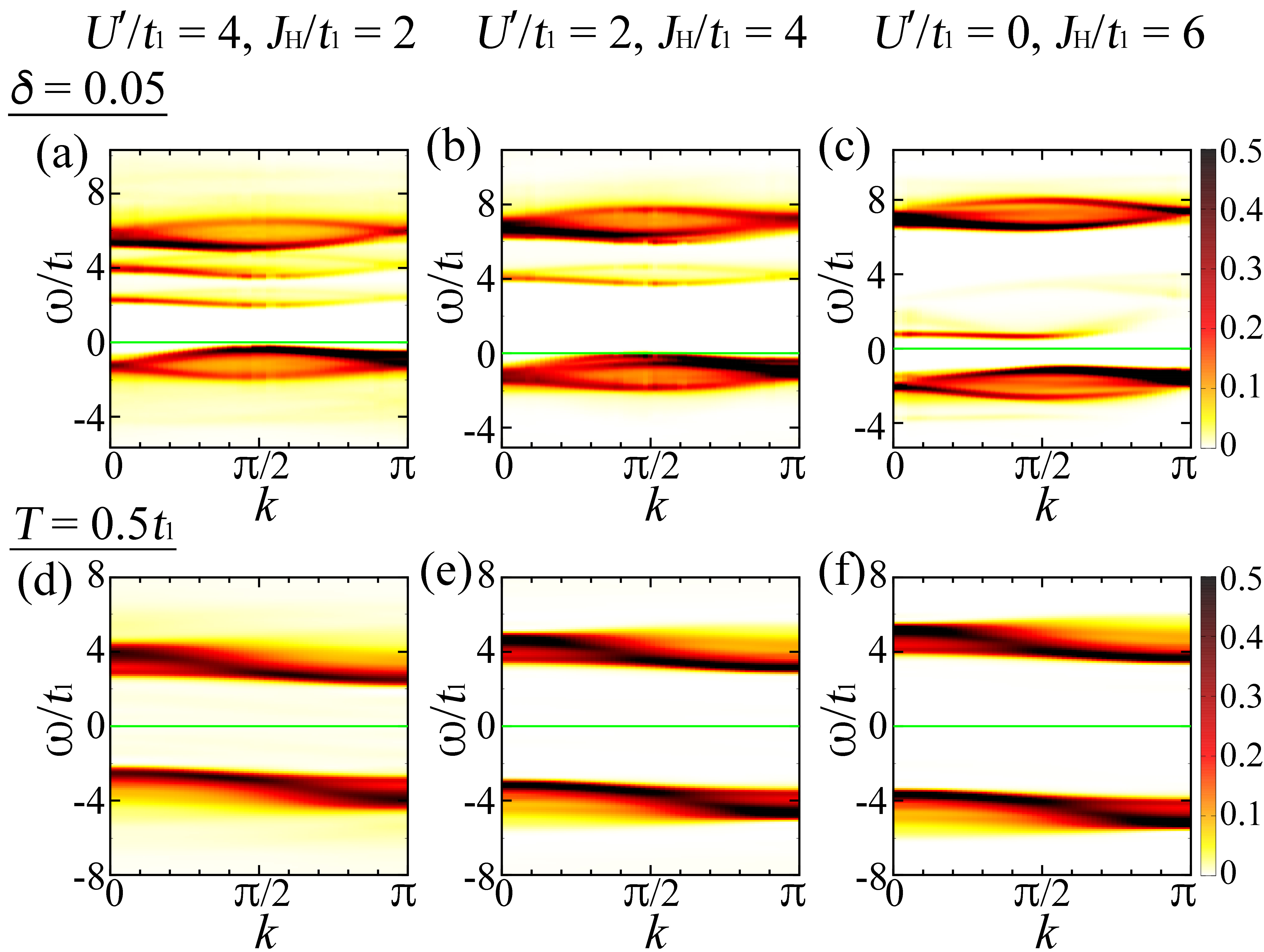}
\caption{Spectral functions for $t_2/t_1=-0.5$ and $U/t_1=6$ in the 1D KHM.
(a)--(c) $A_{\delta=0.05}^2(k,\omega)t_1$; (d)--(f) $A_{T=0.5t_1}^2(k,\omega)t_1$ for $U^{\prime}/t_1=4$ and $J_{\rm H}/t_1=2$ [(a), (d)]; $U^{\prime}/t_1=2$ and $J_{\rm H}/t_1=4$ [(b), (e)]; $U^{\prime}/t_1=0$ and $J_{\rm H}/t_1=6$ [(c), (f)]. The green lines indicate $\omega=0$.}
\label{fig:Akwntm}
\end{figure}

Although hole-doping-induced modes and temperature-induced modes from the LHB for the first orbital appear primarily in the momentum regime outside the Fermi sea [Figs. \ref{fig:Akw}(a), \ref{fig:Akw}(c), \ref{fig:Akw}(m), \ref{fig:Akw}(o), \ref{fig:AkwUp0}(a), and \ref{fig:AkwUp0}(c)], those for the second orbital appear predominantly in the opposite momentum regime (Fig. \ref{fig:Akwntm}) when the sign of the hopping parameter for the second orbital is opposite to that for the first orbital ($t_2/t_1=-0.5$). This feature holds true not only for the doping-induced electronic modes [Figs. \ref{fig:Akwntm}(a)--\ref{fig:Akwntm}(c)] but also for the temperature-induced modes [Figs. \ref{fig:Akwntm}(d)--\ref{fig:Akwntm}(f)]. 
\section{Electronic modes induced by spin and charge perturbations} 
\subsection{Spectral functions in charge-perturbed nonequilibrium states} 
\subsubsection{Spectral functions of nonzero $\eta$ states at half filling} 
\label{sec:eta}
Electronic modes can also be induced by spin or charge perturbations in the spin-1 antiferromagnetic insulators described by the KHM. 
Doping can be regarded as a charge perturbation that leads to an equilibrium state. The spectral features of a nonequilibrium charge-perturbed state at half filling can be discussed based on the results for the doped systems. 
In $\eta$-SU(2) symmetric systems \cite{Essler,TakahashiBook,YangEta,SO4}, the spectral function of a nonzero-$\eta$ state at half filling ($\eta^z=0$) can be obtained from those of hole-doped and electron-doped states \cite{KohnoDrivenMott} as follows: 
\begin{align}
A_{\eta^z=0}^m({\bm k},\omega)=&\frac{1}{2}\left[A_{\eta^z=n}^m({\bm k},\omega)+A_{\eta^z=n}^m({\bm \pi}-{\bm k},-\omega)\right]\nonumber\\
\overset{\text{Eq. (\ref{eq:AkwPH})}}{=}&\frac{1}{2}\left[A_{\eta^z=n}^m({\bm k},\omega)+A_{\eta^z=-n}^m({\bm k},\omega)\right],
\label{eq:Akweta}
\end{align}
where $A_{\eta^z=n}^m({\bm k},\omega)$ denotes the spectral function of the state with $\eta^z=n$ in the $\eta$-SU(2) multiplet for $\mu=H=0$. The relation for the particle-hole transformation [Eq. (\ref{eq:AkwPH})] is used in the second line. 
Here, $\eta$ and $\eta^z$ are defined as the total $\eta$ quantum number and the eigenvalue of ${\hat \eta}^z(=\sum_{i,m}\eta_i^{m,z})$, respectively; note that $\eta^z=-\frac{N_{\rm h}}{2}$ [Eq. (\ref{eq:spinEta})]. 
\par
In the KHM [Eq. (\ref{eq:Ham})], the system possesses $\eta$-SU(2) symmetry at $\mu=0$ for $U^{\prime}=\frac{J_{\rm H}}{2}$, because the $\eta$ operators commute with the spin operators, the $U$ term, and the hopping term on a bipartite lattice. 
Under the orbital rotational symmetry condition $U=U^{\prime}+J_{\rm H}$, $\eta$-SU(2) symmetry is present for $U^{\prime}=\frac{U}{3}$ and $J_{\rm H}=\frac{2U}{3}$. 
\par
In the case where hopping between different orbitals on neighboring sites is absent, the operators defined as ${\hat{\tilde \eta}}^{\gamma}=\sum_{i,m}{\tilde \eta}_i^{m,\gamma}$ for $\gamma=+, -, z$ also serve as the generators of the $\eta$-SU(2) symmetry, where 
\begin{equation}
\begin{array}{l}
{\tilde \eta}_i^{m,+}=(-1)^m(-1)^ic_{i,\uparrow}^{m\dagger}c_{i,\downarrow}^{m\dagger},\\
{\tilde \eta}_i^{m,-}=(-1)^m(-1)^ic_{i,\downarrow}^{m}c_{i,\uparrow}^m,\\
{\tilde \eta}_i^{m,z}=\eta_i^{m,z},
\end{array}
\label{eq:etatilde}
\end{equation}
including an orbital-dependent sign: $(-1)^m=-1$ and $+1$ for $m=1$ and $2$, respectively. 
The Hamiltonian of the KHM [Eq. (\ref{eq:Ham})] can be rewritten in terms of these operators as
\begin{align}
{\cal H}=&-\sum_{\langle i,j\rangle,m,\sigma}t_m\left(c_{i,\sigma}^{m\dagger}c_{j,\sigma}^m+{\rm H.c.}\right)\nonumber\\
&+U\sum_{i,m}\left(n_{i,\uparrow}^m-\frac{1}{2}\right)\left(n_{i,\downarrow}^m-\frac{1}{2}\right)\nonumber\\
&+U^{\prime}\sum_i\left(n_i^1-1\right)\left(n_i^2-1\right)\nonumber\\
&-J_{\rm H}\sum_i{\tilde {\bm \eta}}_i^1\cdot{\tilde {\bm \eta}}_i^2-J_{\rm H}\sum_i{\bm S}_i^1\cdot{\bm S}_i^2\nonumber\\
&-\mu\sum_{i,m}n_i^m-H\sum_{i,m}S_i^{m,z}.
\label{eq:Ham2}
\end{align}
Hence, the system has $\eta$-SU(2) symmetry for $U^{\prime}=0$ at $\mu=0$, where $\eta$ is defined in terms of the ${\tilde \eta}$ operators [Eq. (\ref{eq:etatilde})]. 
Under the orbital rotational symmetry condition $U=U^{\prime}+J_{\rm H}$, $\eta$-SU(2) symmetry is present for $U^{\prime}=0$ and $J_{\rm H}=U$. 
\par
\begin{figure}
\centering
\includegraphics[width=\linewidth]{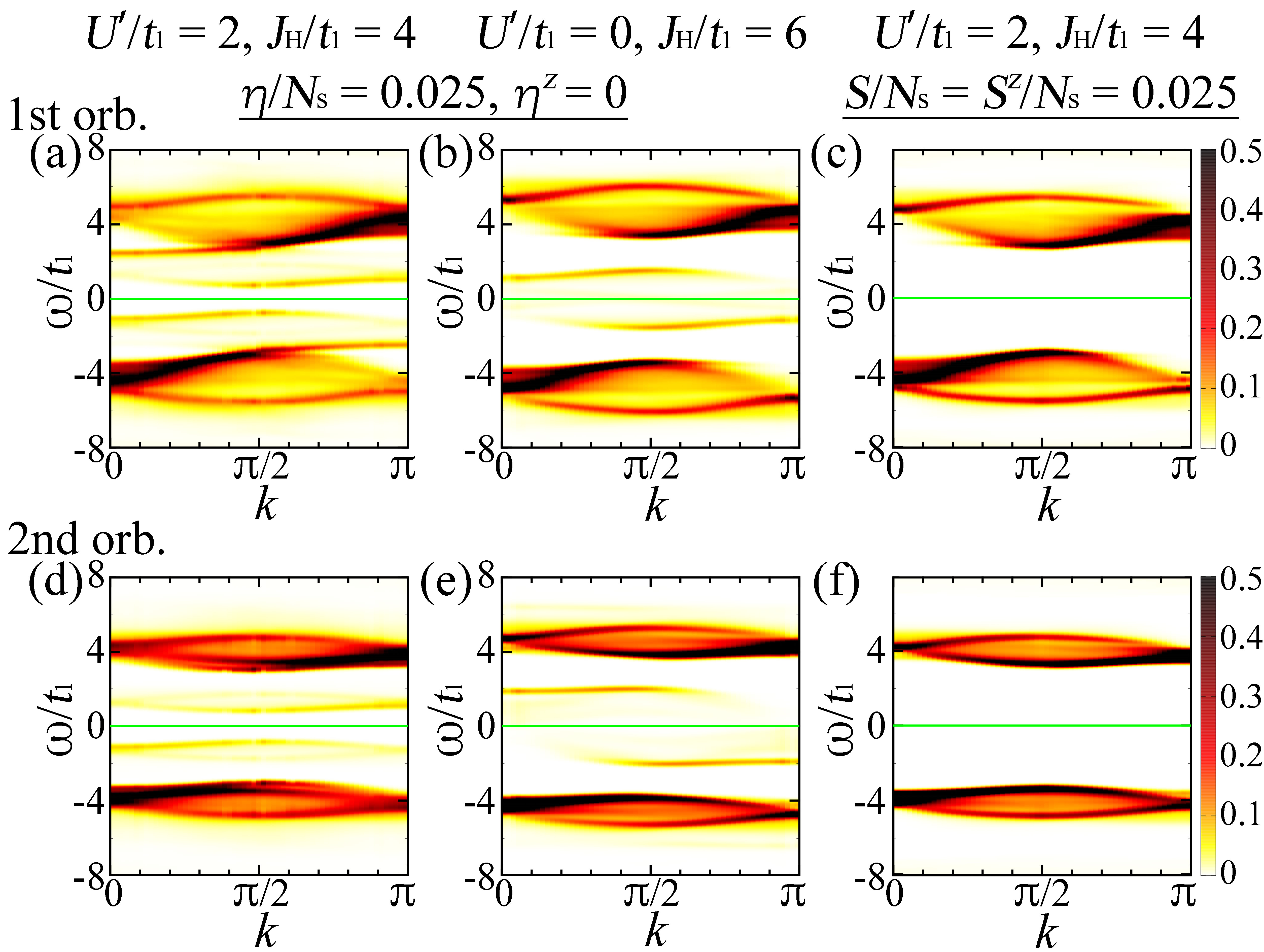}
\caption{Spectral functions of charge-perturbed states [(a), (b), (d), (e)] and a magnetized state [(c), (f)] at half filling ($\eta^z=0$) in the 1st orbital ($m=1$) [(a)--(c)] and 2nd orbital ($m=2$) [(d)--(f)] for $U^{\prime}/t_1=2$ and $J_{\rm H}/t_1=4$ [(a), (c), (d), (f)]; and $U^{\prime}/t_1=0$ and $J_{\rm H}/t_1=6$ [(b), (e)] with $U/t_1=6$ and $t_2/t_1=0.5$ in the 1D KHM. 
(a), (b), (d), (e) $A_X^m(k,\omega)t_1$ for $|X\rangle$ being the lowest-energy state with $\eta/N_{\rm s}=0.025$. 
(c), (f) $A_{\frac{S^z}{N_{\rm s}}=0.025}^{m,\uparrow}(k,\omega)t_1$ under a magnetic field. The green lines indicate $\omega=0$.}
\label{fig:etaSz}
\end{figure}

In $\eta$-SU(2) symmetric systems, the lowest-energy state with $\eta=n$ at half filling ($\eta^z=0$) belongs to the same $\eta$-SU(2) multiplet as the $2n$-hole-doped and $2n$-electron-doped ground states, if phase separation in the electron density does not occur at $|\delta|=\frac{2n}{N_{\rm s}}$. 
Hence, the spectral functions of this state, which is a nonequilibrium state, can be obtained using Eq. (\ref{eq:Akweta}) from those of the $2n$-hole-doped ground state, the $2n$-electron-doped ground state, or both [Fig. \ref{fig:etaSz}(a), \ref{fig:etaSz}(b), \ref{fig:etaSz}(d), and \ref{fig:etaSz}(e)]. 
In the nonequilibrium state in the parameter regime without pairing, electronic modes emerge from both band edges into the band gap, exhibiting momentum-shifted spin-mode dispersion relations in exactly the same way as in the hole- and electron-doped systems, as shown in Figs. \ref{fig:etaSz}(a) and \ref{fig:etaSz}(d). 
\subsubsection{Possible experimental realizations} 
Such a nonzero $\eta$ state at half filling can be generated by exciting particle-hole pairs whose excitation energies correspond to the band gap, i.e., the energy difference between the bottom of the UHB for the orbital with the lowest lower edge and the top of the LHB for the orbital with the highest upper edge. 
\par
To sustain the nonzero $\eta$ state as a nonequilibrium steady state under light irradiation, photons with energies comparable to the band gap must be continuously supplied at a rate sufficient to compensate for the relaxation of the particle-hole pairs. Since the spectral weights are essentially proportional to the density of the excited particle-hole pairs (as discussed in Sec. \ref{sec:SWdoping}), it is necessary to keep shining a macroscopic number $\mathcal{O}(N_{\rm s})$ of photons that actually excite particle-hole pairs (excluding transmitted or reflected photons that do not contribute to the excitations) during the measurement in order to observe the emergent electronic modes. 
It should be noted that the above discussion represents a necessary condition to observe the emergent modes. Specifying and modeling operational parameters, such as photon fluence, absorption cross section, and relaxation time, are beyond the scope of this study. 
\subsection{Spectral functions in spin-perturbed states} 
\subsubsection{Spectral functions in magnetized states} 
In spin-gapped strongly correlated insulators, electronic modes can be induced by magnetization \cite{KohnoDrivenMott}. 
In the 1D KHM, since the Haldane gap opens at half filling \cite{HaldaneGap}, electronic modes can in principle emerge upon magnetization. 
The emergent electronic modes can be explained as follows: 
\par
The ground state of $S^z=1$ can be effectively described [Eq. (\ref{eq:Spq})] as 
\begin{equation}
|{\rm GS}\rangle_{S^z=1}=\frac{1}{\sqrt{N_{\rm s}}}\sum_{i,l}e^{{\rm i}{\bm q}\cdot{\bm r}_i}{\bar \psi}({\cal C}^l)|{\rm T}^{\gamma_i^l}\rangle_i\prod_{j(\ne i)}|{\rm T}^{\gamma_j^l}\rangle_j,
\end{equation}
where ${\bm q}$ denotes the momentum of the lowest-energy excited state in the low-energy spin mode; ${\bm q}={\bm 0}$ or ${\bm \pi}$ in the KHM. The excitation energy from $|{\rm GS}\rangle_{N_{\rm h}=0}$ is given by ${\bar e}_{\bm q}^{S^+}-H=\Delta_{\rm s}-H$, where ${\bar e}_{\bm k}^{X}$ denotes $e_{\bm k}^{X}$ for $H=0$, and the spin gap at $H=0$ is denoted by $\Delta_{\rm s}$. 
\par
By adding a $\downarrow$-spin electron or removing an $\uparrow$-spin electron with momentum ${\bm k}$ in the $m$th orbital, the following states are obtained: 
\begin{align}
\label{eq:addGSSz1}
&c_{{\bm k},\downarrow}^{m\dagger}|{\rm GS}\rangle_{S^z=1}\nonumber\\
&=\frac{1}{N_{\rm s}}\sum_{l,i,j(\ne i)}e^{{\rm i}({\bm k}\cdot{\bm r}_j+{\bm q}\cdot{\bm r}_i)}{\bar \psi}({\cal C}^l)c_{j,\downarrow}^{m\dagger}|{\rm T}^{\gamma_j^l}\rangle_j|{\rm T}^{\gamma_i^l}\rangle_i\prod_{j(\ne i,j)}|{\rm T}^{\gamma_j^l}\rangle_j\nonumber\\
&+\frac{1}{N_{\rm s}}\sum_{i,l}e^{{\rm i}({\bm k}+{\bm q})\cdot{\bm r}_i}{\bar \psi}({\cal C}^l)c_{i,\downarrow}^{m\dagger}|{\rm T}^{\gamma_i^l}\rangle_i\prod_{j(\ne i)}|{\rm T}^{\gamma_j^l}\rangle_j,\\
\label{eq:rmvGSSz1}
&c_{{\bm k},\uparrow}^{m}|{\rm GS}\rangle_{S^z=1}\nonumber\\
&=\frac{1}{N_{\rm s}}\sum_{l,i,j(\ne i)}e^{{\rm i}(-{\bm k}\cdot{\bm r}_j+{\bm q}\cdot{\bm r}_i)}{\bar \psi}({\cal C}^l)c_{j,\uparrow}^{m}|{\rm T}^{\gamma_j^l}\rangle_j|{\rm T}^{\gamma_i^l}\rangle_i\prod_{j(\ne i,j)}|{\rm T}^{\gamma_j^l}\rangle_j\nonumber\\
&+\frac{1}{N_{\rm s}}\sum_{i,l}e^{{\rm i}(-{\bm k}+{\bm q})\cdot{\bm r}_i}{\bar \psi}({\cal C}^l)c_{i,\uparrow}^{m}|{\rm T}^{\gamma_i^l}\rangle_i\prod_{j(\ne i)}|{\rm T}^{\gamma_j^l}\rangle_j.
\end{align}
The first terms in Eqs. (\ref{eq:addGSSz1}) and (\ref{eq:rmvGSSz1}) essentially exhibit the same band structure as the unmagnetized system, whereas the second terms contribute to the emergent electronic modes. 
\par
The second terms in Eqs. (\ref{eq:addGSSz1}) and (\ref{eq:rmvGSSz1}) can be rewritten as 
\begin{align}
\label{eq:PGSSz1}
&\frac{1}{N_{\rm s}}\sum_{i,l}e^{{\rm i}({\bm k}+{\bm q})\cdot{\bm r}_i}{\bar \psi}({\cal C}^l)\alpha_{{\rm T}^{\gamma_i^l}}^{{\rm P}_{\sigma}^{{\bar m}}}|{\rm P}_{\sigma}^{{\bar m}}\rangle_i\prod_{j(\ne i)}|{\rm T}^{\gamma_j^l}\rangle_j,\\
\label{eq:RGSSz1}
&\frac{1}{N_{\rm s}}\sum_{i,l}e^{{\rm i}(-{\bm k}+{\bm q})\cdot{\bm r}_i}{\bar \psi}({\cal C}^l)\alpha_{{\rm T}^{\gamma_i^l}}^{{\rm R}_{\sigma}^{{\bar m}}}|{\rm R}_{\sigma}^{\bar m}\rangle_i\prod_{j(\ne i)}|{\rm T}^{\gamma_j^l}\rangle_j,
\end{align}
respectively, where $s^z=\gamma_i^l-\frac{1}{2}$. The state of Eq. (\ref{eq:PGSSz1}) can overlap with $|c_{\uparrow}^{m\dagger}\rangle_{{\bm k}+{\bm q}}$, for which the excitation energy from $|{\rm GS}\rangle_{N_{\rm h}=0}$ is ${\bar e}_{{\bm k}+{\bm q}}^{c^{m\dagger}}-\frac{H}{2}$. On the other hand, that of Eq. (\ref{eq:RGSSz1}) can overlap with $|c_{\downarrow}^m\rangle_{{\bm k}-{\bm q}}$, for which the excitation energy is ${\bar e}_{{\bm k}-{\bm q}}^{c^m}-\frac{H}{2}$. Since the value of $H$ for $S^z/S_{\rm max}\rightarrow +0$ is $\Delta_{\rm s}$, according to Eq. (\ref{eq:Akw}), the dispersion relations of the electron-addition and electron-removal emergent modes are given by 
\begin{align}
\label{eq:wAddSz1}
\omega&={\bar e}_{{\bm k}+{\bm q}}^{c^{m\dagger}}-\frac{\Delta_{\rm s}}{2},\\
\label{eq:wRmvSz1}
\omega&=-{\bar e}_{{\bm k}-{\bm q}}^{c^m}+\frac{\Delta_{\rm s}}{2},
\end{align}
respectively. 
\par
This implies that if a spin gap opens ($\Delta_{\rm s}>0$), electronic modes can emerge for each orbital. In the $\downarrow$-spin electron-addition spectral function, the emergent mode exhibits the dispersion relation of the UHB with the momentum shifted by ${\bm q}$ and the $\omega$ shifted downward by $\frac{\Delta_{\rm s}}{2}$ for the same orbital to which the electron is added. In the $\uparrow$-spin electron-removal spectral function, the emergent mode exhibits the dispersion relation of the LHB with the momentum shifted by $-{\bm q}$ and the $\omega$ shifted upward by $\frac{\Delta_{\rm s}}{2}$ for the same orbital from which the electron is removed. 
The spectral weights of the emergent modes are proportional to the magnetization density, as in the case of the doping-induced modes (Sec. \ref{sec:SWdoping}) \cite{KohnoDrivenMott}. 
\par
For the 1D KHM, the momentum $q=0$ or $\pi$ and $\Delta_{\rm s}\approx J_{\rm eff}\Delta$ [Eq. (\ref{eq:dispSb})]. In the large-$U/t$ regime, the value of $\Delta_{\rm s}$ is so small that the spectral functions appear almost unchanged from those of the unmagnetized system on the $\omega$ scale of $\mathcal{O}(U)$, as shown in Figs. \ref{fig:etaSz}(c) and \ref{fig:etaSz}(f). A high resolution in $\omega$ is required to confirm the emergent modes in experiments or numerical calculations. 
\par
In Figs. \ref{fig:etaSz}(c) and \ref{fig:etaSz}(f), the spectral functions of a magnetized system with $\eta$-SU(2) symmetry (Sec. \ref{sec:eta}) are obtained \cite{KohnoDrivenMott} via the Shiba transformation \cite{ShibaTrans,ShibaTrans_Heilmann} from those of the negative-$U$ model ($U/t_1=-6$, $U^{\prime}/t_1=-2$, $J_{\rm H}/t_1=-4$, and $t_2/t_1=0.5$) for $\delta=-0.05$. 
\subsubsection{Electronic modes induced by spin fluctuations} 
Electronic modes can emerge in nonequilibrium systems under spin fluctuations in strongly correlated insulators \cite{KohnoDrivenMott}. 
We consider the dominant eigenstate $|{\cal X}\rangle_{\bm q}$ of the state perturbed by the spin fluctuation operator ${\cal X}_{\bm q}$, ${\cal X}_{\bm q}|{\rm GS}\rangle_{N_{\rm s}=0}$, for $H=0$. The excitation energy from $|{\rm GS}\rangle_{N_{\rm s}=0}$ is $e_{\bm q}^{\cal X}$ [Eq. (\ref{eq:dominantState})], where ${\cal X}=S^{\pm}$ or $T^{\pm}$. 
The dominant eigenstate can be effectively expressed [Eqs. (\ref{eq:exStates}) and (\ref{eq:dominantState})] as 
\begin{equation}
|{\cal X}\rangle_{\bm q}
=\frac{1}{\sqrt{N_{\rm s}}}\sum_{i,l}e^{{\rm i}{\bm q}\cdot{\bm r}_i}\psi^{\prime}({\cal C}^l)|X\rangle_i\prod_{j(\ne i)}|{\rm T}^{\gamma_j^l}\rangle_j,
\end{equation}
where $|X\rangle_i$ is $|{\rm T}\rangle_i$ for ${\cal X}=S^+$, $|{\rm S}\rangle_i$ for $S^-$, $|{\rm D}^-\rangle_i$ for $T^+$, and $|{\rm D}^+\rangle_i$ for $T^-$ [Eqs. (\ref{eq:Snt0}) and (\ref{eq:Tpmt0})]. 
\par
By adding or removing an electron with momentum ${\bm k}$ for the $m$th orbital, the following states are obtained: 
\begin{align}
\label{eq:addSq}
&c_{\bm k}^{m\dagger}|{\cal X}\rangle_{\bm q}\nonumber\\
&=\frac{1}{N_{\rm s}}\sum_{l,i,j(\ne i)}e^{{\rm i}({\bm k}\cdot{\bm r}_j+{\bm q}\cdot{\bm r}_i)}\psi^{\prime}({\cal C}^l)c_{j}^{m\dagger}|{\rm T}^{\gamma_j^l}\rangle_j|X\rangle_i\prod_{j(\ne i,j)}|{\rm T}^{\gamma_j^l}\rangle_j\nonumber\\
&+\frac{1}{N_{\rm s}}\sum_{i,l}e^{{\rm i}({\bm k}+{\bm q})\cdot{\bm r}_i}\psi^{\prime}({\cal C}^l)c_{i}^{m\dagger}|X\rangle_i\prod_{j(\ne i)}|{\rm T}^{\gamma_j^l}\rangle_j,\\
\label{eq:rmvSq}
&c_{\bm k}^{m}|{\cal X}\rangle_{\bm q}\nonumber\\
&=\frac{1}{N_{\rm s}}\sum_{l,i,j(\ne i)}e^{{\rm i}(-{\bm k}\cdot{\bm r}_j+{\bm q}\cdot{\bm r}_i)}\psi^{\prime}({\cal C}^l)c_{j}^{m}|{\rm T}^{\gamma_j^l}\rangle_j|X\rangle_i\prod_{j(\ne i,j)}|{\rm T}^{\gamma_j^l}\rangle_j\nonumber\\
&+\frac{1}{N_{\rm s}}\sum_{i,l}e^{{\rm i}(-{\bm k}+{\bm q})\cdot{\bm r}_i}\psi^{\prime}({\cal C}^l)c_{i}^{m}|X\rangle_i\prod_{j(\ne i)}|{\rm T}^{\gamma_j^l}\rangle_j.
\end{align}
The first terms in Eqs. (\ref{eq:addSq}) and (\ref{eq:rmvSq}) essentially exhibit the same band structure as the unperturbed system, whereas the second terms contribute to the emergent electronic modes. 
\par
For the nonzero components of the second terms in Eqs. (\ref{eq:addSq}) and (\ref{eq:rmvSq}), according to Table \ref{tbl:states}, we have 
\begin{equation}
\begin{array}{lll}
\left.\begin{array}{c}
c_{i}^{m\dagger}|{\rm T}\rangle_i\propto|{\rm P}^{\bar m}\rangle_i\\
c_{i}^{m}|{\rm T}\rangle_i\propto|{\rm R}^{\bar m}\rangle_i
\end{array}\right\}&\text{for}&S^+,\\
\left.\begin{array}{c}
c_{i}^{m\dagger}|{\rm S}\rangle_i\propto|{\rm P}^{\bar m}\rangle_i\\
c_{i}^{m}|{\rm S}\rangle_i\propto|{\rm R}^{\bar m}\rangle_i
\end{array}\right\}&\text{for}&S^-,\\
\left.\begin{array}{c}
c_{i}^{m\dagger}|{\rm D}^-\rangle_i\propto|{\rm P}^{m}\rangle_i\\
c_{i}^{m}|{\rm D}^-\rangle_i\propto|{\rm R}^{m}\rangle_i
\end{array}\right\}&\text{for}&T^+,\\
\left.\begin{array}{c}
c_{i}^{m\dagger}|{\rm D}^+\rangle_i\propto|{\rm P}^{m}\rangle_i\\
c_{i}^{m}|{\rm D}^+\rangle_i\propto|{\rm R}^{m}\rangle_i
\end{array}\right\}&\text{for}&T^-.
\end{array}
\end{equation}
Hence, the second terms in Eqs. (\ref{eq:addSq}) and (\ref{eq:rmvSq}) can overlap with the dominant eigenstates $|c^{m\dagger}\rangle_{{\bm k}+{\bm q}}$ and $|c^{m}\rangle_{{\bm k}-{\bm q}}$ in the electron-addition and electron-removal excitation from $|S_{\bm q}^{\pm}\rangle$, respectively, as well as with 
$|c^{{\bar m}\dagger}\rangle_{{\bm k}+{\bm q}}$ and $|c^{\bar m}\rangle_{{\bm k}-{\bm q}}$ in the electron-addition and electron-removal excitation from $|T_{\bm q}^{\pm}\rangle$, respectively. 
\par
According to Eq. (\ref{eq:Akw}), the emergent electronic modes exhibit the following dispersion relations: 
\begin{equation}
\begin{array}{lll}
\left.\begin{array}{l}
\omega=e_{{\bm k}+{\bm q}}^{c^{m\dagger}}-e_{\bm q}^{S^{\pm}}\\
\omega=-e_{{\bm k}-{\bm q}}^{c^{m}}+e_{\bm q}^{S^{\pm}}
\end{array}\right\}&\text{for}&S^{\pm},\\
\left.\begin{array}{l}
\omega=e_{{\bm k}+{\bm q}}^{c^{{\bar m}\dagger}}-e_{\bm q}^{T^{\pm}}\\
\omega=-e_{{\bm k}-{\bm q}}^{c^{\bar m}}+e_{\bm q}^{T^{\pm}}
\end{array}\right\}&\text{for}&T^{\pm}.
\end{array}
\end{equation}
\par
\begin{figure*}
\centering
\includegraphics[width=\linewidth]{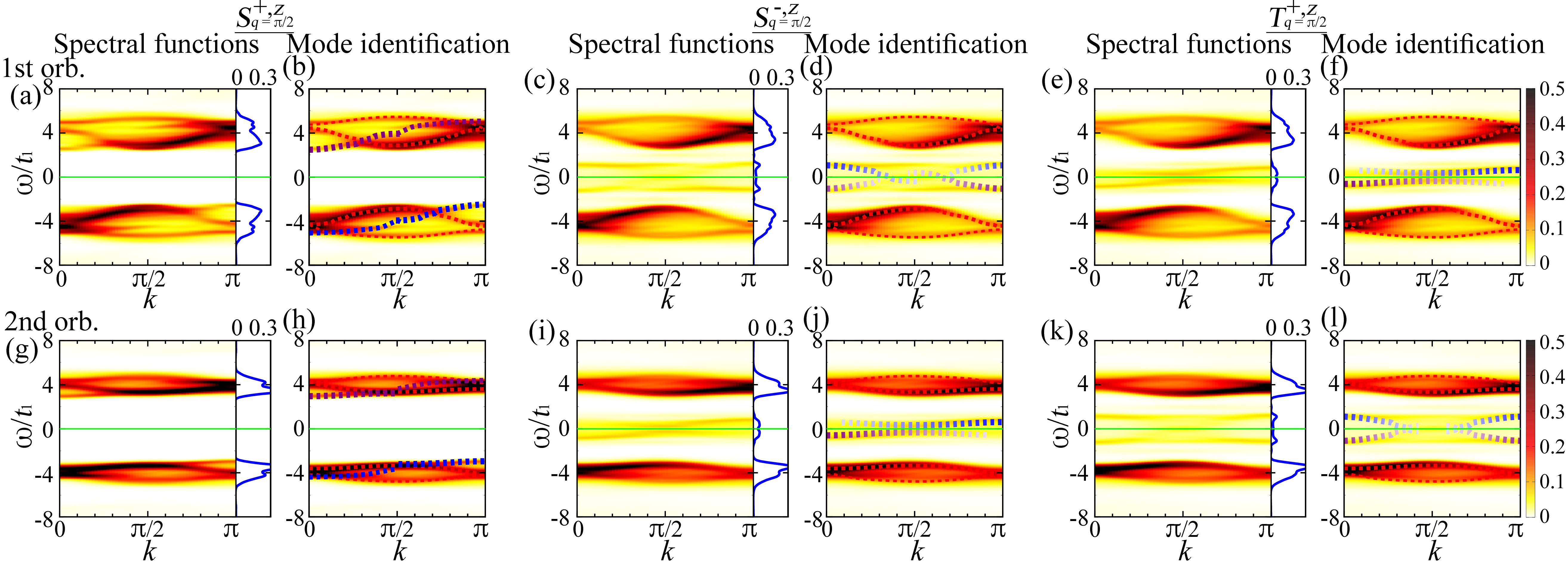}
\caption{Spectral functions and dispersion relations of the emergent modes under spin fluctuations in the 1st orbital ($m=1$) [(a)--(f)] and 2nd orbital ($m=2$) [(g)--(l)] for $U^{\prime}/t_1=2$, $J_{\rm H}/t_1=4$, $U/t_1=6$, and $t_2/t_1=0.5$ at half filling in the 1D KHM. 
$A_X^m(k,\omega)t_1$ for $|X\rangle$ corresponding to $\prod_iS_{q_i}^{+,z}|{\rm GS}\rangle$ [(a), (b), (g), (h)], $\prod_iS_{q_i}^{-,z}|{\rm GS}\rangle$ [(c), (d), (i), (j)], and $\prod_iT_{q_i}^{+,z}|{\rm GS}\rangle$ [(e), (f), (k), (l)], with $q_i=\frac{\pi}{2}$ and $S/N_{\rm s}=0.05$. The blue curves on the right-hand side show $\rho^m(\omega)t_1$ calculated from the spectral functions on the left in (a), (c), (e), (g), (i), and (k). 
In (b), (d), (f), (h), (j), and (l), $\omega=e_{k+\frac{\pi}{2}}^{c^{m\dagger}}-e_{\frac{\pi}{2}}^{X}$ (dashed purple curves) and $\omega=-e_{k-\frac{\pi}{2}}^{c^{m}}+e_{\frac{\pi}{2}}^{X}$ (dashed blue curves) for $X=S^{+}$ [(b), (h)] and $S^{-}$ [(d), (j)]; $\omega=e_{k+\frac{\pi}{2}}^{c^{{\bar m}\dagger}}-e_{\frac{\pi}{2}}^{X}$ (dashed purple curves) and $\omega=-e_{k-\frac{\pi}{2}}^{c^{\bar m}}+e_{\frac{\pi}{2}}^{X}$ (dashed blue curves) for $T^{+}$ [(f), (l)] are plotted in the regimes relevant for comparison with the spectral functions. The dispersion relations of the dominant modes in the unperturbed UHB ($\omega=e_k^{c^{m\dagger}}$) and LHB ($\omega=-e_k^{c^{m}}$), extracted from Figs. \ref{fig:Akw}(n) and \ref{fig:Akw}(t), are indicated by dashed red curves, and used for those of the emergent modes. The green lines indicate $\omega=0$.}
\label{fig:spinFluc}
\end{figure*}

This implies that in nonequilibrium states perturbed by spin fluctuations, the emergent electronic modes for $S_{\bm q}^{\tau}$ ($\tau=+,-$) exhibit the dispersion relations of the unperturbed bands for the same orbital $m$ as that of the added or removed electron [$e_{\bm k}^{c^{m\dagger}}$ (UHB), $-e_{\bm k}^{c^{m}}$ (LHB)] shifted by the momenta $\pm{\bm q}$ and the excitation energies $\mp e_{\bm q}^{S^{\tau}}$ [Figs. \ref{fig:spinFluc}(a)--\ref{fig:spinFluc}(d) and \ref{fig:spinFluc}(g)--\ref{fig:spinFluc}(j)]. In contrast, those for $T_{\bm q}^{\tau}$ exhibit the dispersion relations of the unperturbed bands for the opposite orbital ${\bar m}$ [$e_{\bm k}^{c^{{\bar m}\dagger}}$ (UHB), $-e_{\bm k}^{c^{\bar m}}$ (LHB)] shifted by the momenta $\pm{\bm q}$ and the excitation energies $\mp e_{\bm q}^{T^{\tau}}$ [Figs. \ref{fig:spinFluc}(e), \ref{fig:spinFluc}(f), \ref{fig:spinFluc}(k), and \ref{fig:spinFluc}(l)]. The emergence of the energy- and momentum-shifted unperturbed bands of the orbital opposite to that of the added or removed electron for the spin fluctuations $T_{\bm q}^{\pm}$ is a unique feature of multi-orbital strongly correlated systems. 
\par
The spectral weights of the emergent modes are proportional to the density of excited spins, as in the case of the doping-induced modes (Sec. \ref{sec:SWdoping}) \cite{KohnoDrivenMott}. 
\par
These emergent electronic modes may be induced by neutron irradiation; the sample must be continuously irradiated with a macroscopic number of neutrons that actually excite spins (excluding transmitted or reflected neutrons that do not contribute to the excitations) in order to maintain the nonequilibrium steady state during the measurement and to observe the emergent modes \cite{KohnoDrivenMott}. Instead of neutrons, photons may be used if they can generate spin excitations \cite{spinRIXS}. 
\section{Differences from noninteracting systems and the Hubbard model (HM)} 
\subsection{Differences from noninteracting systems} 
To clarify the peculiarity of the spectral features of orbitally degenerate strongly correlated systems described by the KHM, we compare their characteristic features with those of noninteracting systems and those of the HM. We first address the differences from noninteracting systems and point out that the emergence of the electronic modes is a distinctive feature of strong correlations, by noting that the band structure is identical for any eigenstate in noninteracting systems. 
\par
In noninteracting systems, any eigenstate can be expressed as $\prod_{i}c_{{\bm k}_i,\sigma_i}^{m_i\dagger}|0\rangle$, 
where $({\bm k}_i,\sigma_i,m_i)\ne({\bm k}_j,\sigma_j,m_j)$ for $i\ne j$. Here, $|0\rangle$ denotes the vacuum. 
If an electron with momentum ${\bm k}$ and spin $\sigma$ is added to or removed from the $m$th orbital, the resulting state is also an eigenstate of the Hamiltonian. This means that no additional spectral weight appears except for the conventional electronic modes, because other eigenstates are orthogonal to the electron-added or removed state and do not have any overlap with it. 
\par
The band structure does not depend on the electron occupation. In fact, if an electron is absent (present) at a given momentum, spectral weight appears in the electron-addition (electron-removal) spectral function at that momentum. The spectral functions, defined as the sums of the electron-addition and electron-removal spectral functions, always have values of one (per spin, per orbital) along the dispersion relations of the electronic modes. This means that the band structure is the same for any eigenstate of the Hamiltonian, although the Fermi level depends on the chemical potential or magnetic field. 
\par
In contrast, the emergent electronic modes in strongly correlated insulators discussed in this paper, as well as in Refs. \cite{Kohno1DHub,Kohno2DHub,KohnoDIS,KohnoRPP,KohnoHubLadder,KohnoKLM,KohnoGW,KohnoAF,KohnoSpin,Kohno1DtJ,Kohno2DtJ,KohnoMottT,KohnoTinduced,KohnoDrivenMott,KohnottpHub,KohnottpJ}, are associated with eigenstates that do not contribute to the unperturbed spectral function. In particular, the doping- or temperature-induced electronic modes within the band gap reflect the existence of spin excitations \cite{Anderson} in the energy regime lower than the band gap, i.e., spin-charge separation \cite{Kohno1DHub,Kohno2DHub,KohnoDIS,KohnoRPP,KohnoHubLadder,KohnoKLM,KohnoGW,KohnoAF,KohnoSpin,Kohno1DtJ,Kohno2DtJ,KohnoMottT,KohnoTinduced,KohnoDrivenMott,KohnottpHub,KohnottpJ}, which does not occur in noninteracting systems. 
\par
In mean-field approximations and ab initio calculations such as density functional theory, electrons are treated as moving in an effective potential, where spin-charge separation does not occur, so the resulting band structures are those of electronic quasiparticles, which do not reflect spin excitations, as in noninteracting systems. To describe emergent modes beyond these approximations, it is necessary to consider fluctuations \cite{KohnoAF}. 
\par
Thus, the emergence of electronic modes generated by perturbations such as doping, temperature, or magnetization is a remarkable effect of strong correlations. 
\subsection{Spin-triplet pairing from an antiferromagnetic insulator} 
\subsubsection{Spin-triplet pairing} 
One of the distinctive features of the KHM, compared to the HM and $t$-$J$ model, is the emergence of spin-triplet pairing upon doping in the small-$U^{\prime}$ regime (Sec. \ref{sec:tripletPairing}). This stands in marked contrast to the spin-singlet pairing in the attractive ($U<0$) HM, and the $d_{x^2-y^2}$-wave pairing observed in cuprate high-$T_{\rm c}$ superconductors, whose overall spectral features are governed by the 2D HM and $t$-$J$ model \cite{Kohno2DHub,Kohno2DtJ,KohnoRPP,KohnottpHub,KohnottpJ}. 
\par
The spin-triplet pairing in the KHM is an onsite pairing that appears in doped systems, because the energy of a two-electron pair on a single site is lower than that of two single electrons on different sites in the small-$U^\prime$ regime (Sec. \ref{sec:tripletPairing}). This characteristic does not depend on the geometry of the lattice or the spatial dimension, as it is an onsite effect. 
\par
In contrast, the spin-singlet pairing in the attractive ($U<0$) HM occurs regardless of the doping level; even at half filling, electrons form pairs. In cuprate high-$T_{\rm c}$ superconductors, the $d_{x^2-y^2}$-wave pairing is sensitive to the lattice geometry and the spatial dimension, as it is an offsite effect. 
\subsubsection{Pair-breaking gap different from the band gap}
In the doped KHM in the small-$U^{\prime}$ regime, electronic excitations exhibit a nonzero pair-breaking gap (Sec. \ref{sec:tripletPairing}) [Figs. \ref{fig:AkwUp0}(a), \ref{fig:AkwUp0}(d), \ref{fig:AkwUp0}(g), and \ref{fig:AkwUp0}(j)]. This gap is approximately $\frac{J_{\rm H}}{2}-U^{\prime}$, which is distinct from the band gap at half filling [Figs. \ref{fig:AkwUp0}(b), \ref{fig:AkwUp0}(e), \ref{fig:AkwUp0}(h), and \ref{fig:AkwUp0}(k)]; the band gap is primarily determined by $U$ in the large-$U$ regime ($U\gg |t|$). 
\par
In contrast, in the attractive HM, the electronic excitations exhibit a single pair-breaking gap of $\mathcal{O}(|U|)$, regardless of whether the system is doped or not \cite{KohnoDrivenMott}. In the 1D and 2D repulsive HMs, the electronic excitations exhibit a Mott gap of $\mathcal{O}(U)$ at half filling, and the gapless and dispersive electronic mode emerges upon doping \cite{Kohno1DHub,Kohno2DHub,KohnoRPP}. Even if $d_{x^2-y^2}$-wave superconductivity occurs, the electronic excitations are gapless in the nodal directions in 2D. 
\subsubsection{Pairing coexisting with low-energy spin excitations} 
The triplet-pairing state with a pair-breaking gap in the KHM can coexist with low-energy spin excitations, which can be effectively described by the spin-1 Heisenberg model with mobile vacancies [Fig. \ref{fig:nkSk}(i)], because onsite triplet pairs formed at each undoped site effectively generate spin-1 moments. 
\par
In contrast, in the attractive HM, there are no low-energy spin excitations; the lowest spin-excitation energy is the same as the pair-breaking gap. In cuprate high-$T_{\rm c}$ superconductors, the pairing state coexists with low-energy spin excitations via $d_{x^2-y^2}$-wave symmetry, which possesses nodes and is compatible with antiferromagnetic fluctuations; both electronic and spin excitations are gapless. 
\par
Such coexistence of onsite spin-triplet pairs with low-energy spin excitations is a distinctive characteristic of Hund-coupled orbitally degenerate systems, and is absent in the HM. 
\subsection{Orbital-selective emergent electronic modes} 
\subsubsection{Orbital-selective Mott transition (OSMT)} 
A significant spectral feature of the KHM, distinct from those of the HM, PAM, and KLM \cite{Kohno1DHub,Kohno2DHub,KohnoDIS,KohnoRPP,KohnoHubLadder,KohnoKLM,KohnoGW,KohnoAF,KohnoSpin,Kohno1DtJ,Kohno2DtJ,KohnoMottT,KohnoTinduced,KohnoDrivenMott,KohnottpHub,KohnottpJ}, is the emergence of orbital-selective electronic modes. Although the PAM and KLM also involve orbital degrees of freedom, their orbital characters are mixed due to hybridization. In contrast, the KHM, without hybridization, exhibits well-defined and characteristic orbital-dependent features. 
\par
Traditionally \cite{ImadaRMP}, the Mott transition has been considered to be characterized either by the loss of carriers, as in a band insulator or the antiferromagnetic mean-field theory \cite{SlaterAF}, or by the divergence of the effective mass of electronic quasiparticles \cite{BrinkmanRice}, as in the Fermi liquid theory \cite{LandauFL}. However, the essential nature of the Mott transition is that the electronic mode above the Fermi level in hole-doped systems gradually loses its spectral weight, while exhibiting a low-energy spin-mode dispersion relation shifted by the Fermi momentum. From the insulating side, low-energy spin excited states emerge as electronic excited states in the band gap from the band edge, exhibiting the spin-mode dispersion relation shifted by the Fermi momentum \cite{Kohno1DHub,Kohno2DHub,KohnoDIS,KohnoRPP,KohnoHubLadder,KohnoKLM,KohnoGW,KohnoAF,KohnoSpin,Kohno1DtJ,Kohno2DtJ,KohnoMottT,KohnoTinduced,KohnoDrivenMott}. 
\par
In multi-orbital systems, the OSMT usually refers to the insulator--metal transition where the Fermi level enters the LHB of one orbital; the electrons in that orbital become metallic, whereas the electrons in other orbitals, whose LHBs are still below the Fermi level, remain insulating [Figs. \ref{fig:Akw}(g), \ref{fig:Akw}(j), \ref{fig:Akw}(s), and \ref{fig:Akw}(v)]. 
This transition appears similar to that of the band-insulator--metal transition. However, this study shows that only for the doped orbital, the aforementioned nature of the Mott transition appears [Sec. \ref{sec:holedoping}; Figs. \ref{fig:Akw}(a), \ref{fig:Akw}(d), \ref{fig:Akw}(m), \ref{fig:Akw}(p), \ref{fig:compDoping}(a), \ref{fig:compDoping}(b), \ref{fig:compDoping}(i), and \ref{fig:compDoping}(j)]; specifically, an electronic mode that exhibits the low-energy spin-mode dispersion relation shifted by the Fermi momentum emerges from the top of the LHB into the band gap and evolves into a gapless mode carrying significant spectral weight which may be regarded as a quasiparticle of a Fermi liquid as doping concentration increases. 
\par
Thus, the OSMT, particularly the emergence of an electronic mode associated with low-energy spin excitations only in the doped orbital, and not in other orbitals, constitutes a key feature of orbitally degenerate strongly correlated systems described by the KHM. 
\subsubsection{Emergent electronic modes originating from inter-orbital spin excitations} 
Although an electronic mode associated with low-energy spin excitations emerges only in the doped orbital, electronic modes also emerge in undoped orbitals, originating from distinct types of spin excitations. 
\par
In the KHM, there exist spin excitations in which an electron in one orbital hops to another orbital with its spin flipped [Eqs. (\ref{eq:Tkwpm}) and (\ref{eq:Ti})]. These types of spin excitations possess the same quantum number $S=1$ as the conventional spin excitations [Eq. (\ref{eq:Skwpm})]. 
Upon doping one of the orbitals, electronic modes associated with these types of spin excitations emerge in undoped orbitals, exhibiting the dispersion relations shifted by the Fermi momentum [Secs. \ref{sec:holedoping} and \ref{sec:eledoping}; Figs. \ref{fig:Akw}(g), \ref{fig:Akw}(j), \ref{fig:Akw}(s), \ref{fig:Akw}(v), \ref{fig:compDoping}(e)--\ref{fig:compDoping}(h), and \ref{fig:compDoping}(m)--\ref{fig:compDoping}(p)]. 
\par
Thus, the emergence of electronic modes originating from unconventional inter-orbital spin excitations, as well as their emergence in undoped orbitals, also constitutes a distinctive feature of the KHM. 
\subsubsection{Emergent electronic modes in different momentum regimes depending on orbitals}
Not only do the origins of the emergent modes depend on orbitals, but the momentum regimes in which these modes primarily appear can also vary with orbitals (Sec. \ref{sec:momentum}; Fig. \ref{fig:Akwntm}). 
If the momentum regimes of the Fermi sea differ between orbitals---as can occur when the signs of the hopping parameters are different---the momentum regimes where the emergent modes appear may also differ from orbital to orbital. 
\par
This is also a characteristic of strongly correlated multi-orbital systems described by the KHM. 
\subsubsection{Distinct band structures between doped and nonzero-temperature cases}
Upon increasing temperature, electronic modes associated with low-energy spin excitations emerge from the band edges of all the orbitals, exhibiting the momentum-shifted spin-mode dispersion relations. This contrasts with the above-described orbital-selective features of the doping-induced electronic modes. These differences become evident in multi-orbital systems; in the HM, the temperature- and doping-induced modes exhibit essentially the same features \cite{KohnoMottT,KohnoTinduced}. 
\subsubsection{Emergence of electronic modes reflecting band structures of other orbitals} 
Emergent electronic modes induced by conventional spin fluctuations in one orbital can exhibit the momentum- and energy-shifted band structure of that orbital. In contrast, those induced by inter-orbital spin fluctuations can exhibit that of another orbital. This is also a distinctive feature of multi-orbital systems. 
\subsection{Implications}
The emergence of electronic states upon doping Mott insulators has been recognized since the early 1990s \cite{Eskes,DagottoDOS}. Although various interpretations of the doping-induced states have been proposed \cite{Kohno1DHub,Kohno2DHub,KohnoDIS,KohnoRPP,KohnoHubLadder,KohnoKLM,KohnoGW,KohnoAF,KohnoSpin,Kohno1DtJ,Kohno2DtJ,KohnoMottT,KohnoTinduced,KohnoDrivenMott,KohnottpHub,KohnottpJ,SakaiImadaPRL,ImadaCofermionPRL,PhillipsRPP,EderOhta2DHub,EderOhtaIPES,Eskes,DagottoDOS,Hundband,HDDMRG,dopeHDDMFT,undopeHDDMFT,splitHDJDMFT}, the view presented in this paper and in Refs. \cite{Kohno1DHub,Kohno2DHub,KohnoDIS,KohnoRPP,KohnoHubLadder,KohnoKLM,KohnoGW,KohnoAF,KohnoSpin,Kohno1DtJ,Kohno2DtJ,KohnoMottT,KohnoTinduced,KohnoDrivenMott,KohnottpHub,KohnottpJ} provides a unified explanation for the doping-, temperature- and perturbation-induced states not only in single-orbital or orbital-hybridized systems, such as the HM, PAM, and KLM, but also in orbitally degenerate systems described by the KHM. 
\par
Although the two-orbital KHM has been considered in this paper, the results can be generalized to models with more orbitals. For example, if the top of the LHB of the $m$th orbital has the highest upper edge, hole-doping-induced electronic modes for a different $m^{\prime}$th orbital should emerge along $\omega=e_{{\bm k}-{\bm k}_{\rm F}^-}^{T^{m,m^{\prime};\pm}}$ with the Fermi momentum ${\bm k}_{\rm F}^-$, where the inter-orbital spin operator between the $m$th and $m^{\prime}$th orbitals is denoted by $T_{\bm k}^{m,m^{\prime};\pm}$. 
\section{Summary} 
In this paper, using the effective theory for weak inter-site hopping together with a quantum-number analysis (selection rules) and numerical calculations (non-Abelian DDMRG method and CPT) for the 1D and 2D KHMs, the following features have been clarified: the emergence of electronic modes originating from spin excitations upon doping; their orbital-specific correspondence with distinct types of spin excitations; the emergence of electronic modes reflecting spin excitations at nonzero temperatures; the emergence of electronic modes in nonequilibrium states under spin or charge perturbations; and the formation of onsite spin-triplet pairs mediated by Hund's coupling. 
\par
Specifically, upon doping the spin-1 antiferromagnetic insulator described by the KHM in a typical parameter regime, electronic modes emerge in the band gap, exhibiting spin-mode dispersion relations shifted by the Fermi momentum. In the doped orbital, the emergent modes reflect conventional spin excitations, whereas those in the undoped orbital reflect inter-orbital spin excitations. In particular, the low-energy emergent mode appears only in the doped orbital and evolves into a gapless mode carrying significant spectral weight which may be regarded as a quasiparticle of a Fermi liquid as doping concentration increases. 
These emergent modes constitute an essential feature of the OSMT, which is absent in band-insulator--metal transitions. 
\par
At nonzero temperatures, electronic modes reflecting the low-energy spin excitations emerge from the band edges into the band gap in all orbitals, exhibiting the momentum-shifted spin-mode dispersion relations, and alter the band structure, in contrast to conventional band insulators. 
\par
In nonequilibrium states under spin fluctuations or the charge perturbation that generates the lowest-energy particle-hole excitations, electronic modes can be induced, whose dispersion relations can be expressed as combinations of unperturbed electronic and spin-mode dispersion relations. 
\par
In the regime of small inter-orbital repulsion, onsite spin-triplet pairs can be formed upon doping; superconductivity is expected to occur in dimensions higher than one in specific parameter regimes. 
\par
These findings contribute to a comprehensive understanding of unconventional features associated with large-spin antiferromagnetic insulators, which are relevant to a wide class of transition-metal oxides including Ni, Co, Fe, and Mn-based materials, as well as other orbitally degenerate strongly correlated systems. 
\begin{acknowledgments} 
This work was supported by JSPS KAKENHI Grant No. JP25K07160 and World Premier International Research Center Initiative (WPI), MEXT, Japan. 
Numerical calculations were partly performed on the Numerical Materials Simulator at NIMS.
\end{acknowledgments}
\appendix
\section{Finite-size, temperature, and doping effects} 
\label{sec:AppendixA}
\subsection{Computational methods in Appendix \ref{sec:AppendixA}} 
In Appendix \ref{sec:AppendixA}, spectral functions at different hole densities and temperatures, as well as dynamical spin structure factors at half filling and zero temperature, are investigated for clusters with different sizes from those of the main text. 
\par
In the 1D KHM, spectral functions and dynamical spin structure factors were calculated at zero temperature on 40-site clusters using the non-Abelian DDMRG method, keeping 120 eigenstates of the density matrix. 
\par
In the 2D KHM, spectral functions were calculated at $T=0.5t_1$ and $0.25t_1$ at half filling using CPT with full diagonalization of $2\times 2$-site clusters, instead of the low-temperature Lanczos method. 
\par
Gaussian broadening with a standard deviation of $0.1t_1$ was used after deconvolution from Lorentzian broadening of a half width at half maximum of $0.16t_1$ (non-Abelian DDMRG) and $0.1t_1$ (CPT). 
\begin{figure}
\centering
\includegraphics[width=\linewidth]{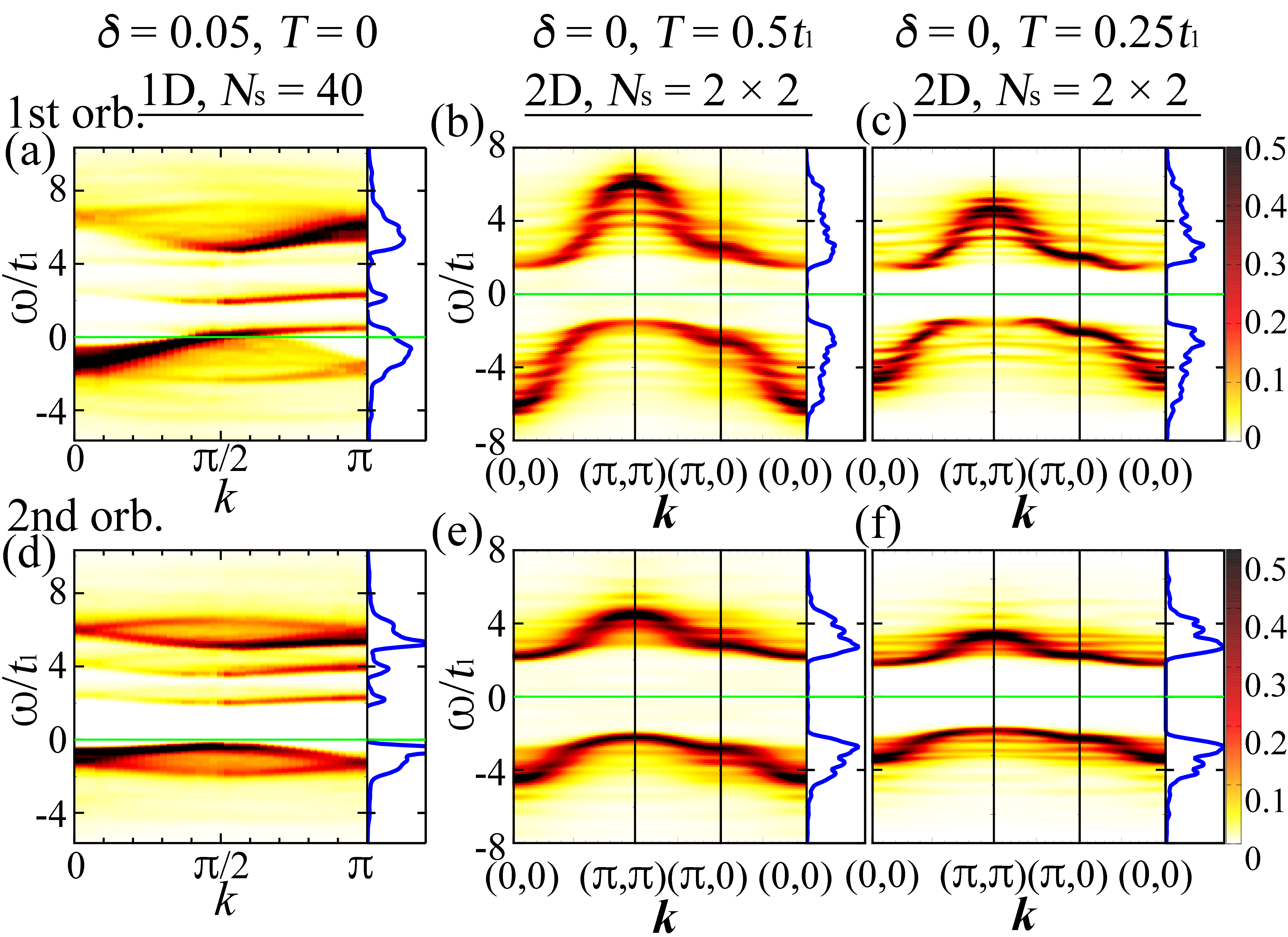}
\caption{Spectral functions for the 1st orbital ($m=1$) [(a)--(c)] and 2nd orbital ($m=2$) [(d)--(f)] at $U^{\prime}/t_1=4$, $J_{\rm H}/t_1=2$, $U/t_1=6$, and $t_2/t_1=0.5$. 
(a), (d) $A_{\delta=0.05}^m(k,\omega)t_1$ on the 40-site cluster in the 1D KHM. 
(b), (c), (e), (f) $A_{T=0.5t_1}^m({\bm k},\omega)t_1$ [(b), (e)] and $A_{T=0.25t_1}^m({\bm k},\omega)t_1$ [(c), (f)] using the $2\times 2$-site cluster in the 2D KHM. The blue curves on the right-hand side show $\rho^m(\omega)t_1$ calculated from the spectral functions on the left. The green lines indicate $\omega=0$.}
\label{fig:AkwSizeT}
\end{figure}
\subsection{Finite-size effects} 
\label{sec:sizeEffects}
To investigate finite-size effects, spectral functions calculated for clusters with different sizes are compared. 
\par
In the 1D KHM, the overall spectral features obtained on the 40-site cluster [Figs. \ref{fig:AkwSizeT}(a) and \ref{fig:AkwSizeT}(d)] are almost the same as those on the 60-site cluster [Figs. \ref{fig:Akw}(a) and \ref{fig:Akw}(g)], and no significant finite-size effects are observed for $U/t_1=6$, $U^{\prime}/t_1=4$, $J_{\rm H}/t_1=2$, and $t_2/t_1=0.5$ at $\delta=0.05$ and $T=0$. This implies that the system size ($N_{\rm s}=60$), as well as the number of eigenstates of the density matrix kept in non-Abelian DDMRG ($240$), used in the main text is sufficiently large to discuss the overall spectral features. 
Quantitative comparisons of excitation energies and spectral weights of the emergent modes also indicate that the finite-size effects are not significant (Fig. \ref{fig:dopingDpn}). 
\par
In the 2D KHM, the overall spectral features obtained using the $2\times 2$-site cluster [Figs. \ref{fig:AkwSizeT}(b) and \ref{fig:AkwSizeT}(e)] are in reasonable agreement with those obtained using the $3\times 2$-site cluster [Figs. \ref{fig:Akw}(f) and \ref{fig:Akw}(l)] for $U/t_1=6$, $U^{\prime}/t_1=4$, $J_{\rm H}/t_1=2$, and $t_2/t_1=0.5$ at $\delta=0$ and $T=0.5t_1$. Although extra stripes due to discrete energy levels are observed in the case of the $2\times 2$-site cluster, basic spectral features, such as the emergence of electronic modes within the band gap from the band edges of the UHB and LHB, are well captured for both orbitals. This implies that the results obtained using $3\times 2$-site clusters, which are presented in the main text, are reliable for discussing the overall spectral features. Furthermore, these comparisons indicate that the rectangular geometry of $3\times 2$-site clusters, as well as the use of the low-temperature Lanczos method, does not significantly affect the overall spectral features. 
\subsection{Temperature effects} 
\label{sec:temperatureEffects}
To investigate temperature effects, spectral functions at different nonzero temperatures are compared. 
The spectral weights around the top (bottom) of the emergent electronic modes from the LHB (UHB) at a temperature comparable to the highest excitation energy of the low-energy spin mode [$T=0.5t_1$; Figs. \ref{fig:AkwSizeT}(b) and \ref{fig:AkwSizeT}(e)] become suppressed at a lower temperature [$T=0.25t_1$; Figs. \ref{fig:AkwSizeT}(c) and \ref{fig:AkwSizeT}(f)].
\par
This can be understood based on the arguments in Sec. \ref{sec:nonzeroT} \cite{KohnoMottT,KohnoTinduced} as follows: As the temperature is lowered, the number of thermally excited spins with excitation energies near the top of the low-energy spin mode decreases. As a result, the spectral weights around the band edges of the emergent electronic modes, which primarily originate from these spin-excited states, are reduced. The numerical results shown in Figs. \ref{fig:AkwSizeT}(b), \ref{fig:AkwSizeT}(c), \ref{fig:AkwSizeT}(e), and \ref{fig:AkwSizeT}(f) are consistent with this interpretation. 
\subsection{Doping effects} 
\label{sec:dopingEffects}
\begin{figure}
\centering
\includegraphics[width=\linewidth]{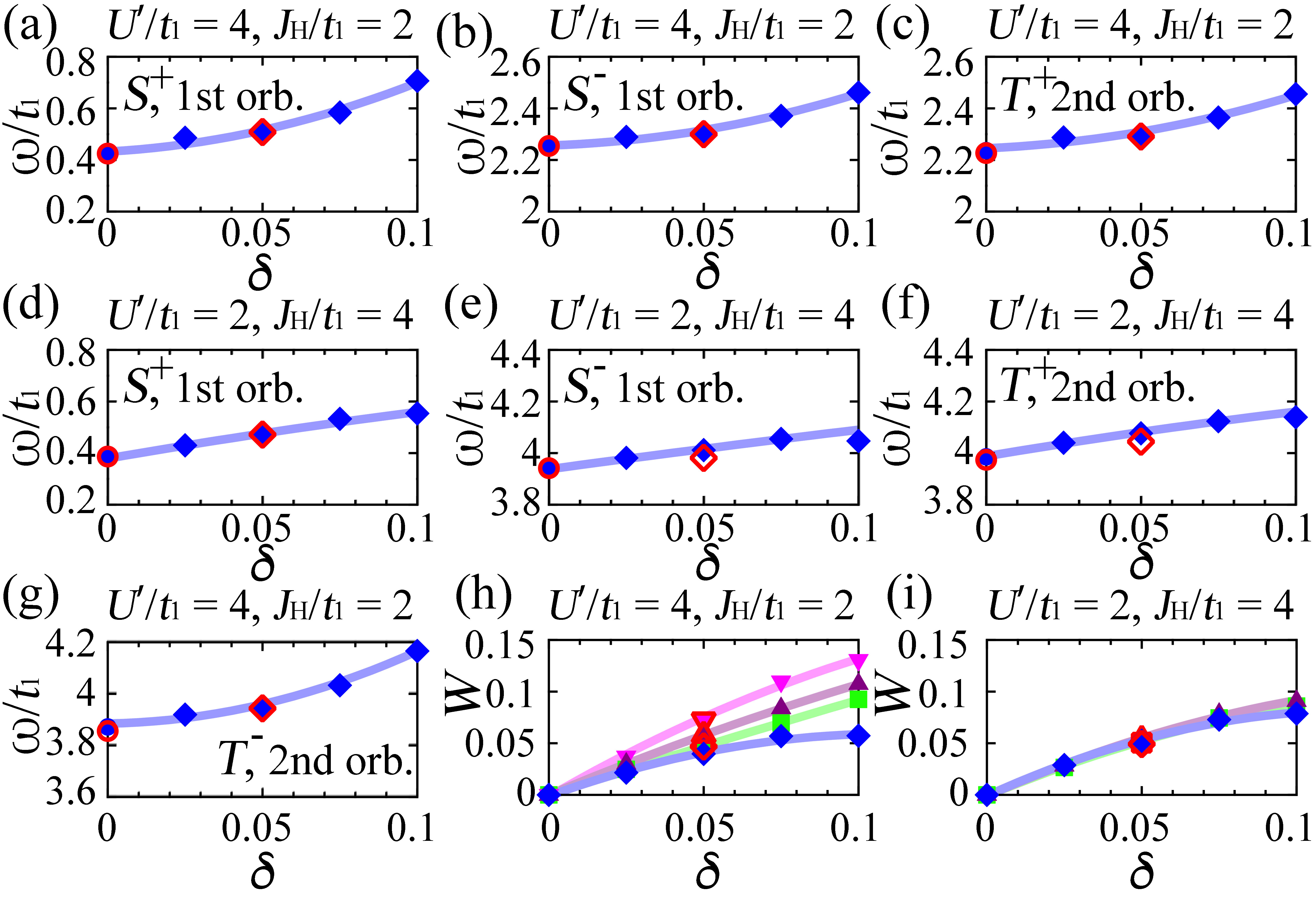}
\caption{Excitation energies [(a)--(g)] and spectral weights [(h), (i)] of hole-doping-induced modes as a function of doping, and excitation energies of spin modes at half filling [(a)--(g)], for $U^{\prime}/t_1=4$, $J_{\rm H}/t_1=2$ [(a)--(c), (g), (h)] and $U^{\prime}/t_1=2$, $J_{\rm H}/t_1=4$ [(d)--(f), (i)], with $U/t_1=6$ and $t_2/t_1=0.5$, on clusters with $N_{\rm s}=40$ (solid symbols) and $60$ (open red symbols) in the 1D KHM. 
(a), (d) $\omega/t_1$ of the low-energy emergent modes for the 1st orbital at $k=\pi$ (diamonds) and $e_{\frac{\pi}{2}}^{S^+}/t_1$ at half filling (circles). 
(b), (c), (e), (f) $\omega/t_1$ of the intermediate-energy emergent modes for the 1st orbital [(b), (e)] and 2nd orbital [(c), (f)] at $k=\pi$ (diamonds), and $e_{\frac{\pi}{2}}^{S^-}/t_1$ [(b), (e)] and $e_{\frac{\pi}{2}}^{T^+}/t_1$ [(c), (f)] at half filling (circles). 
(g) $\omega/t_1$ of the high-energy emergent mode for the 2nd orbital at $k=\pi$ (diamonds) and $e_{\frac{\pi}{2}}^{T^-}/t_1$ at half filling (circles). 
(h), (i) Spectral weights of the low-energy emergent modes for the 1st orbital (diamonds), intermediate-energy emergent modes for the 1st orbital (upward triangles) and 2nd orbital (squares), and high-energy emergent modes (downward triangles). 
The curves are guides for the eye.}
\label{fig:dopingDpn}
\end{figure}

To investigate doping effects, the highest excitation energy and spectral weight of each hole-doping-induced mode are shown as a function of doping in Fig. \ref{fig:dopingDpn}. 
\par
All the highest excitation energies of the emergent electronic modes approach those of the corresponding spin modes at half filling as doping concentration decreases [Figs. \ref{fig:dopingDpn}(a)--\ref{fig:dopingDpn}(g)]. These results strongly support the validity of the theoretical argument presented in Sec. \ref{sec:holedoping} and indicate that the emergent electronic modes originate from the spin modes. 
\par
The spectral weights of the emergent electronic modes increase in proportion to doping concentration from the low-doping limit [Figs. \ref{fig:dopingDpn}(h) and \ref{fig:dopingDpn}(i)], as expected from the effective theory described in Sec. \ref{sec:SWdoping}. 
\section{Gauge transformation} 
\label{sec:gaugeTrans}
We consider a $U(1)$ gauge transformation defined as 
\begin{equation}
c_{i,\sigma}^m\rightarrow (-1)^ic_{i,\sigma}^m
\label{eq:gaugeTrans2}
\end{equation}
for all $\sigma$ and sites $i$ on a bipartite lattice. 
This transformation is described by the following unitary operator: 
\begin{equation}
U=\prod_{\sigma}\exp\left(-{\rm i}\sum_i{\bm \pi}\cdot{\bm r}_i n_{i,\sigma}^m\right).
\label{eq:transOperator}
\end{equation}
The Hamiltonian [Eq. (\ref{eq:Ham})], which is represented by ${\cal H}(t_m,t_{\bar m},U,U^{\prime},J_{\rm H},\mu,H)$, is transformed to ${\cal H}(-t_m,t_{\bar m},U,U^{\prime},J_{\rm H},\mu,H)$; only the sign of $t_m$ is changed. As a shorthand notation, ${\cal H}(t_m,t_{\bar m},U,U^{\prime},J_{\rm H},\mu,H)$ is denoted by ${\cal H}(t_m)$. 
The eigenstate $|X\rangle$ for ${\cal H}(t_m)$ is transformed as
\begin{equation}
|X\rangle\rightarrow|X^{\prime}\rangle=U|X\rangle, 
\end{equation}
where $|X^{\prime}\rangle$ is an eigenstate for ${\cal H}(-t_m)$ with the same norm and energy; $\langle X|X\rangle=\langle X^{\prime}|X^{\prime}\rangle$ and $E_X=E_{X^{\prime}}$, because this transformation is unitary. 
An operator $O$ is transformed as 
\begin{equation}
O\rightarrow UOU^{\dagger}. 
\end{equation}
In particular, $c_{i,\sigma}^m$ is transformed as shown in Eq. (\ref{eq:gaugeTrans}) [$(-1)^i=e^{{\rm i}{\bm \pi}\cdot{\bm r}_i}$], which leads to 
\begin{equation}
c_{{\bm k},\sigma}^m\rightarrow c_{{\bm k}-{\bm \pi},\sigma}^m.
\end{equation}
\par
Thus, the spectral function of $|X^{\prime}\rangle$ for ${\cal H}(-t_m)$ can be expressed using that of $|X\rangle$ (the eigenstate with the same energy) for ${\cal H}(t_m)$ as
\begin{equation}
A_{X^{\prime}}^m({\bm k},\omega)=A_X^m({\bm \pi}-{\bm k},\omega) 
\label{eq:Akwntm2}
\end{equation}
in inversion symmetric systems; $A_X^m({\bm k},\omega)=A_X^m(-{\bm k},\omega)$. 

\end{document}